\documentclass[12pt]{article}
\usepackage{amssymb,amsmath,amsthm,mathrsfs,latexsym,amsxtra,graphicx,appendix,hyperref,setspace,subfig,fix-cm,color,cite,dsfont,adjustbox,tabularx,comment,afterpage, bbold, amsfonts}
\numberwithin{equation}{section}

\usepackage{titlesec}
\titleformat{\paragraph}
{\normalfont\normalsize\bfseries}{\theparagraph}{1em}{}
\titlespacing*{\paragraph}{0pt}{3.25ex plus 1ex minus .2ex}{1.5ex plus .2ex}

\usepackage{tikz}
\usetikzlibrary{positioning}
\usetikzlibrary{intersections} 
\newlength{\PicScale}
\definecolor{indiagreen}{rgb}{0.07, 0.53, 0.03}

\newcommand{\op}{\hspace{1pt}}

\newcommand{\nn}{\nonumber}

\newcommand{\ttf}{\mathtt{f}}
\newcommand{\ttw}{\mathtt{w}}
\newcommand{\ktl}{\Gamma_{\!\!(19,3)}}
\newcommand{\rkt}{\text{K3}}
\newcommand{\zp}{Z'}
\newcommand{\opm}{\hspace{-3pt}}

\usepackage{longtable, multirow, array}

\newcolumntype{M}[1]{>{\centering\arraybackslash}m{#1}}
\hypersetup{
    pdftitle={},
    pdfauthor={},
    pdfsubject={}
}

\begin{document} 

\newcommand{\bea}{\begin{eqnarray}}
\newcommand{\eea}{\end{eqnarray}}
\newcommand{\be}{\begin{equation}}
\newcommand{\ee}{\end{equation}}
\newcommand{\eq}[1]{(\ref{#1})}

\newcommand{\del}{\partial}
\newcommand{\delbar}{\overline{\partial}}
\newcommand{\zbar}{\overline{z}}
\newcommand{\wbar}{\overline{w}}
\newcommand{\vbar}{\overline{\varphi}}
\newcommand{\tm}{{\text{-}}}
\newcommand{\hf}{\frac{1}{2}}
\newcommand{\qrt}{\frac{1}{4}}
\newcommand{\bz}{{\mathbb Z}}
\newcommand{\R}{{\mathbb R}}
\newcommand{\C}{{\mathbb C}}
\newcommand{\A}{{\mathbb A}}
\newcommand{\N}{{\mathbb N}}
\newcommand{\bH}{{\mathbb H}}
\renewcommand{\P}{{\mathbb{CP}}}
\newcommand{\Q}{{\mathbb Q}}
\newcommand{\tX}{\widetilde{X}}
\newcommand{\mO}{\Omega}
\newcommand{\mJ}{{\mathbb J}}
\def\taubar{\overline{\tau}}
\def\Tr{{\rm Tr}}
\def\qhat{\hat{q}_0}
\def\ap{\alpha^{\prime}}
\def\eg{{\emph{e.g.}~}}
\def\a{\alpha} 
\def\b{\beta} 
\def\g{\gamma} 
\def\G{\Gamma}
\def\e{\epsilon}
\def\h{\eta}
\def\th{\theta} 
\def\Th{\Theta}  
\def\k{\kappa}
\def\la{\lambda} 
\def\L{\Lambda} 
\def\m{\mu}
\def\n{\nu}
\def\r{\rho} 
\def\s{\sigma} 
\def\t{\tau}
\def\f{\phi} 
\def\F{\Phi} 
\def\w{\omega}
\def\W{\Omega} 
\def\v{\varphi} 
\def\z{\zeta}
\def\ts{\textstyle}

\newcommand{\sA}{{\mathscr A }}  
\newcommand{\sR}{{\mathscr R }}  
\newcommand{\cN}{{\mathcal N }}            
\newcommand{\cZ}{\mathcal Z}
\newcommand{\cT}{\mathcal T}
\newcommand{\cE}{\mathcal E}
\newcommand{\cG}{\mathcal G}
\newcommand{\cZh}{\hat{\mathcal Z}}
\newcommand{\bt}{{\mathbb T}}
\newcommand{\bone}{{\mathbb 1}}
\newcommand{\tf}{\textstyle\frac}
\newcommand{\etc}{{\it etc.~}}
\newcommand{\tv}{\text{v}}
\newcommand{\tw}{\text{w}}

\newcommand{\one}{\mathbf{1}}
\newcommand{\two}{\mathbf{2}}
\newcommand{\three}{\mathbf{3}}
\newcommand{\eo}{\epsilon_1}
\newcommand{\et}{\epsilon_2}
\newcommand{\bp}{\mathbf{+}}
\newcommand{\bm}{\mathbf{-}}

\newcommand{\wb}{{\bar w}}
\newcommand{\zb}{{\bar z}}
\newcommand{\xb}{{\bar x}}
\newcommand{\hb}{{\bar h}}
\newcommand{\qb}{{\bar q}}

\begin{titlepage}
\begin{center}

\hfill \\
\hfill \\
\vskip 0.75in
\begin{flushright}
KCL-PH-TH/2022-26
\end{flushright}
{\Large 
{\bf Heterotic Strings on ${\bt^3}/{\bz_2}$, \\[3mm]
Nikulin involutions and M-theory}
}\\

\vskip 0.4in

{\large Bobby Samir Acharya${}^{a,b}$, Gerardo Aldazabal${}^{c,d,e}$, Anamar\'ia Font${}^{f,g}$,\\Kumar Narain${}^{a}$, and Ida G.~Zadeh${}^{a}$
}\\
\vskip 4mm

${}^{a}$
{\it International Centre for Theoretical Physics, Strada Costiera 11, 34151 Trieste, Italy} \vskip 1mm
${}^{b}$
{\it Department of Physics, Kings College London, London, WC2R 2LS, UK} \vskip 1mm
${}^{c}$
{\it G. F\'isica CAB-CNEA, Centro At\'omico Bariloche, Av. Bustillo 9500, Bariloche, Argentina} \vskip 1mm
${}^{d}$
{\it Consejo Nacional de Investigaciones Cient\'ificas y T\'ecnicas {\rm(}CONICET{\rm)}} \vskip 1mm
${}^{e}$
{\it Instituto Balseiro, Universidad Nacional de Cuyo {\rm(}UNCUYO{\rm)}, Av. Bustillo 9500, R8402AGP,  Bariloche, Argentina} \vskip 1mm
${}^{f}$
{\it Fac. de Ciencias, Universidad Central de Venezuela, A.P.20513, Caracas 1020-A, Venezuela} \vskip 1mm
${}^{g}$
{\it Max-Planck-Institut f\"ur Gravitationsphysik, Albert-Einstein-Institut, 14476 Golm, Germany} \vskip 1mm

\end{center}

\vskip 0.35in

\begin{center} {\bf Abstract } \end{center}
We first describe the low energy dynamics of ten dimensional heterotic supergravity compactified on the smooth, flat 3-manifold
${\bt^3}/{\bz_2}$, without supersymmetry, and explain how it arises from flat heterotic gauge fields. 
The semi-classical theory has both Coulomb and Higgs branches of non-supersymmetric vacua. 
We then give an exact worldsheet description as asymmetric orbifolds of $\bt^3$, where the orbifold generator involves
a Nikulin non-symplectic involution $\theta$ of the even self-dual lattice $\ktl$. 
Along the way we briefly compare our findings with M-theory on $\text{K3}/\theta$.
Our construction gives a novel CFT description of the semi-classical field theory moduli space. In particular, the Wilson line 
parameters in the lattice $I\subset \ktl$ of signature 
$(19-s,1)$ which is invariant under $\theta$, and in its orthogonal complement $N$, correspond respectively to Coulomb
and Higgs branch moduli. 
There is a rich pattern of transitions amongst Higgs and Coulomb branches which we describe using the worldsheet theory.
\vfill


\end{titlepage}

\setcounter{page}{1}
\setcounter{tocdepth}{2}

\tableofcontents

\newpage
\section{Introduction and summary}\label{section_intro}

Non-supersymmetric vacua arising from compactifications of superstrings have only received limited attention throughout the years. Although these solutions are typically unstable, their study could still give new insights on properties of theories that include gravity 
at the quantum level. With this goal in mind, in this work we will explore a class of compactifications of the heterotic string on 
$\bt^3/\bz_2$ in which supersymmetry is broken despite the internal manifold being
Ricci-flat. Aspects of non-supersymmetric heterotic compactifications have been considered before by several authors,
see for instance \cite{Dixon:1986iz, AlvarezGaume:1986jb, Seiberg:1986by, Nair:1986zn, Narain:1986gd, Ginsparg:1986wr, Itoyama:1986ei, Blum:1997gw, Blum:1997cs, Font:2002pq, Faraggi:2009xy, Blaszczyk:2014qoa, Angelantonj:2014dia, Blaszczyk:2015zta, Abel:2015oxa, Aaronson:2016kjm, Basile:2018irz, Itoyama:2020ifw, Faraggi:2020wej, Faraggi:2020hpy, Basile:2021vxh, Itoyama:2021itj, Cribiori:2021txm} and references therein.
In the present paper we are additionally motivated to extend our previous investigation \cite{Acharya:2020hsc} about type II string 
compactifications on Ricci-flat manifolds without supersymmetry. The concrete problem that we pose is to describe
the heterotic compactification on $\bt^3/\bz_2$ at the string worldsheet level. We will explain how it can be solved using the 
formalism of asymmetric orbifolds \cite{Narain:1986qm, Narain:1990mw} and as a byproduct will uncover interesting phenomena 
such as properties of their moduli spaces.

Before entering into asymmetric orbifolds we will discuss the supergravity limit of the 10-dimensional heterotic 
string on $\bt^3/\bz_2$. Understanding the resulting low-energy field theory will be rather instructive.
We will develop a physical interpretation of the moduli space and we will eventually learn how to define the orbifold action on 
the gauge degrees of freedom. In the field theory setup the latter 
corresponds to specifying a flat connection on the heterotic group $E_8 \times E_8$ or $Spin(32)/\bz_2$,
which amounts to finding a set of holonomies, $g_i$ for each of the three translations along $\bt^3$ and $g_\theta$ for the
orbifold generator, satisfying the defining relations of the fundamental group. Instead of looking for general solutions we will
consider two classes of flat connections referred to as Higgs and Coulomb branch, according to the nature of their moduli.
At generic points in the Higgs branch the gauge group is completely broken whereas it is only broken to the maximal torus in the Coulomb branch. At the origin of the Higgs branch the two branches intersect and a transition can occur because moduli 
of either branch could be switched on. 

For the string worldsheet picture it will be crucial to discuss first the conformal field theory realization of the holonomies, which we develop in some detail, especially in the case of gauge group $SU(2)$. Using this realisation we are able to describe the Higgs and Coulomb branch vacua as well as transitions amongst them from the worldsheet viewpoint.

The duality between the heterotic string on $\bt^3$ and M-theory on K3  \cite{Hull:1994ys,Witten:1995ex} suggests that the resulting non-supersymmetric vacua can also be regarded as compactifications of M-theory on $\bz_2$ orbifolds of K3 surfaces. In both frameworks a key factor is the even self-dual lattice of signature (19,3), denoted $\ktl$.
In the heterotic compactification on $\bt^3$, modular invariance requires
the internal left and right-moving momenta to lie on $\ktl$ \cite{Narain:1985jj,Narain:1986am}. 
On the other hand, the second cohomology group of K3, with the intersection form of K3, is isometric to $\ktl$. 
The quotient of K3 is by a non-symplectic involution, say $\theta$,  
 that inverts the holomorphic 2-form but leaves a K\"ahler form invariant. Such involutions have been classified by Nikulin \cite{Nikulin80,Nikulin83,Nikulin86} in terms of the sublattice of $\ktl$, called $I$, which is left invariant by $\theta$.
 This invariant lattice $I$ has rank $r$, signature $(r-1,1)$, satisfies $I^*/I=\bz_2^a$, and turns out to be
 completely specified up to isomorphisms by the three invariants $(r,a,\delta)$, where $\delta$ equals zero if all elements 
 of the dual $I^*$ have integer norm, and equals one otherwise. 
 All 75 possibilities for $(r,a,\delta)$ are shown in figure \ref{figureNikulin}.
 The sublattice orthogonal to $I$ in $\ktl$, denoted $N$, which
 has rank $s+2$ with $s:=20-r$, signature $(s,2)$, and satisfies $N^*/N=\bz_2^a$, is also uniquely determined by the triple $(r,a,\delta)$.
 
In the heterotic $\bt^3$ orbifold the quotient is by an action that involves a reflection
of $s$ left-moving and two right-moving directions, and is realized on the momentum lattice $\ktl$ by one
of the Nikulin involutions $\theta$ specified by a given $(r,a,\delta)$.
In general $\theta$ is accompanied by a translation in the invariant lattice $I$, the so-called shift $v$,
satisfying $4v\in I$, so that the orbifold generator, denoted $g$, is order four acting on both fermions and bosons. 
The doubling of the order is an expected  consequence of modular invariance and consistency of the Hilbert space interpretation
\cite{Narain:1990mw, Aoki:2004sm,Harvey:2017rko}.
We will show that level matching in the $g$-twisted sector in fact constrains the norm of $v$ in terms of the number $s$ of left-moving reflected directions.
We further find that in the definition of the orbifold generator it is essential to include an additional phase that depends on the normal lattice $N$. More precisely, the phase in $g^2$ turns out to be $e^{2\pi i P^2_N}$, with $P_N \in N^*$, which can take 
values $1$  or $-1$ depending on whether $2P_N^2$ is even or odd. The necessity
of this phase is suggested by the conformal field theory realization of $SU(2)$ flat connections on $\bt^3/\bz_2$, 
where it happens that in the Higgs branch $g_\theta^2$ acts as $+1$ on the root lattice and as $-1$ on the
conjugacy class of the fundamental weights.
In practice the phase can be conveniently written as $e^{2\pi i P_I \cdot w}$, $\forall P_I \in I^*$, where $w \in I^*/I$
always exists and satisfies the conditions required for level matching and consistent operator interpretation
in the $g^2$-twisted sector.

One of the main results of our work is to have revealed and characterized flows in the moduli space of the 
heterotic orbifold theory. Indeed, the continuous Wilson line parameters in the $N$ lattice of signature $(s,2)$ 
and the $I$ lattice of signature $(19-s,1)$ correspond respectively to Higgs and Coulomb branch moduli.
By deforming $N$ we can go to a point with enhanced $SU(2)$ and then switch moduli along $I$,
implying a transition $s \to (s-1)$. This is the building block of transitions between models and is referred to as 
the $s$-transition. Multiple applications of the $s$-transition then connect models with different $(r,a,\delta)$.
In fact, we will show that all models in figure \ref{figureNikulin} may be connected in this way.
This result motivates the interesting question of whether there exists a unique grand moduli space containing each asymmetric model 
$(r,a,\delta)$?

To answer the above question we need to recall that the heterotic asymmetric orbifolds are determined by the triple 
$(r,a,\delta)$ together with the shift vector $v$ and that for each triple there can be several $v$'s which satisfy the level matching condition. We will argue that for some models all allowed shift vectors can be connected to one another by
automorphisms of the invariant lattice, or differ by an element of the dual invariant lattice.
However, this is not always the case. 
This implies that orbifold models with such inequivalent shift vectors do not sit in the same moduli space, which in turn seems 
to contradict the existence of a unique grand moduli space.
On the other hand, we will see in particular examples that a given model with seemingly inequivalent shift vectors 
can be connected after an $s$-transition to another model with only one $v$.
Since we do not yet have a full understanding of the structure of the moduli space of the asymmetric orbifold models 
the existence of a unique grand moduli space remains as a conjecture at this point.

To our knowledge the construction of heterotic orbifolds including Nikulin non-symplectic involutions
has not appeared in the literature. One could have certainly thought that asymmetric orbifolding was required 
and expected that conceptual as well as technical challenges would be surmounted. But the occurrence of 
transitions between models with different invariant lattices, i.e. corresponding to different triples $(r,a,\delta)$,
was perhaps less predictable.
These transitions manifest naturally in the heterotic string because the moduli of the theory are massless states 
which can acquire vacuum expectation values (vevs) in a controlled way.
In M-theory there are instead branes wrapping the cycles of the K3 surface and the corresponding states are non-perturbative.
As such, assigning vevs to them and following their paths is a more complicated problem open to investigation.

The paper is organised as follows. In section \ref{section_nikulin} we will review the basic features of Nikulin non-symplectic 
involutions of $\ktl$.
In section \ref{section_ft} we will study various aspects of the supergravity limit of the heterotic string theory on a Ricci-flat 3-manifold 
$\bt^3/\bz_2$. 
In section \ref{section_cft} we will analyse the worldsheet conformal field theory realisation of the $SU(2)$ flat connections in
the Higgs and Coulomb branches discussed in the supergravity limit.
In section \ref{section_orbifold} we will construct the asymmetric orbifolds describing the quotient of the heterotic string 
by the non-symplectic involutions. After defining the orbifold action involving the shift vector $v$, we will compute the 
partition function, check consistency of the operator interpretation, and obtain the modular invariance condition on $v$. We will then determine solutions of this condition and address the question of whether orbifold models with different shift vectors sit in the same moduli space. We will also survey generic properties of the spectrum of the heterotic orbifolds, 
such as existence of tachyonic and massless states, illustrating the salient points in various examples.
In section \ref{sec_motion} we will discuss transitions between models with different $(r,a,\delta)$. 
Some final remarks and a list of open questions are collected in section \ref{sec_final}.
In Appendix \ref{app_lattices} we provide explicit realisations of the involution $\theta$, and the corresponding invariant lattice, 
for all the 75 Nikulin triples $(r,a,\delta)$.

\section{Nikulin non-symplectic involutions of $\ktl$}\label{section_nikulin}

\begin{figure}[htb!]
\centering
\begin{tikzpicture}[scale=0.55]
    \draw[gray] (0, -0.5) grid (20.5, 11.5);    \draw[thin, black, ->] (0, -0.5) -- (20.5, -0.5);
      \foreach \x in {0,2,4,6,8,10,12,14,16,18,20}
      \draw[thin] (\x, -0.6) -- (\x, -0.6)        node[anchor=north] {\(\x\)};
    \foreach \x in {10}      \draw[thin] (\x, -1.7) -- (\x, -1.7)        node[anchor=north] {$r:=20-s$};    \draw[thin, black, ->] (0,-0.5) -- (0, 11.5);
    \foreach \y in {0,1,2,3,4,5,6,7,8,9,10,11}      \draw[thick] (-0.0, \y) -- (0.0, \y)        node[anchor=east] {\(\y\)};
    \foreach \y in {5}      \draw[thick] (-1.0, \y) -- (-1.0, \y)        node[anchor=east] {\large$a$};
    
    \draw[thick, black, fill=black] (1,1) circle (1.2mm);    
    \draw[thick, black] (2,0) circle (2.5mm);
    \draw[thick, black] (2,2) circle (2.5mm);
    \draw[thick, black, fill=black] (2,2) circle (1.2mm);

    \draw[thick, black, fill=black] (3,1) circle (1.2mm);
    \draw[thick, black, fill=black] (3,3) circle (1.2mm);

    \draw[thick, black, fill=black] (4,2) circle (1.2mm);
    \draw[thick, black, fill=black] (4,4) circle (1.2mm);

    \draw[thick, black, fill=black] (5,3) circle (1.2mm);
    \draw[thick, black, fill=black] (5,5) circle (1.2mm);

    \draw[thick, black] (6,2) circle (2.5mm);
    \draw[thick, black] (6,4) circle (2.5mm);
    \draw[thick, black, fill=black] (6,4) circle (1.2mm);
    \draw[thick, black, fill=black] (6,6) circle (1.2mm);

    \draw[thick, black, fill=black] (7,3) circle (1.2mm);
    \draw[thick, black, fill=black] (7,5) circle (1.2mm);
    \draw[thick, black, fill=black] (7,7) circle (1.2mm);

    \draw[thick, black, fill=black] (8,2) circle (1.2mm);
    \draw[thick, black, fill=black] (8,4) circle (1.2mm);
    \draw[thick, black, fill=black] (8,6) circle (1.2mm);
    \draw[thick, black, fill=black] (8,8) circle (1.2mm);

    \draw[thick, black, fill=black] (9,1) circle (1.2mm);
    \draw[thick, black, fill=black] (9,3) circle (1.2mm);
    \draw[thick, black, fill=black] (9,5) circle (1.2mm);
    \draw[thick, black, fill=black] (9,7) circle (1.2mm);
    \draw[thick, black, fill=black] (9,9) circle (1.2mm);

    \draw[thick, black] (10,0) circle (2.5mm);
    \draw[thick, black] (10,2) circle (2.5mm);
    \draw[thick, black, fill=black] (10,2) circle (1.2mm);
    \draw[thick, black] (10,4) circle (2.5mm);
    \draw[thick, black, fill=black] (10,4) circle (1.2mm);
    \draw[thick, black] (10,6) circle (2.5mm);
    \draw[thick, black, fill=black] (10,6) circle (1.2mm);
    \draw[thick, black] (10,8) circle (2.5mm);
    \draw[thick, black, fill=black] (10,8) circle (1.2mm);
    \draw[thick, black] (10,10) circle (2.5mm);
    \draw[thick, black, fill=black] (10,10) circle (1.2mm);

    \draw[thick, black, fill=black] (11,1) circle (1.2mm);
    \draw[thick, black, fill=black] (11,3) circle (1.2mm);
    \draw[thick, black, fill=black] (11,5) circle (1.2mm);
    \draw[thick, black, fill=black] (11,7) circle (1.2mm);
    \draw[thick, black, fill=black] (11,9) circle (1.2mm);
    \draw[thick, black, fill=black] (11,11) circle (1.2mm);
    
     \draw[thick, black, fill=black] (12,2) circle (1.2mm);
     \draw[thick, black, fill=black] (12,4) circle (1.2mm);
    \draw[thick, black, fill=black] (12,6) circle (1.2mm);
    \draw[thick, black, fill=black] (12,8) circle (1.2mm);
    \draw[thick, black, fill=black] (12,10) circle (1.2mm);

     \draw[thick, black, fill=black] (13,3) circle (1.2mm);
    \draw[thick, black, fill=black] (13,5) circle (1.2mm);
    \draw[thick, black, fill=black] (13,7) circle (1.2mm);
    \draw[thick, black, fill=black] (13,9) circle (1.2mm);

    \draw[thick, black] (14,2) circle (2.5mm);
    \draw[thick, black] (14,4) circle (2.5mm);
    \draw[thick, black, fill=black] (14,4) circle (1.2mm);
    \draw[thick, black] (14,6) circle (2.5mm);
    \draw[thick, black, fill=black] (14,6) circle (1.2mm);
    \draw[thick, black, fill=black] (14,8) circle (1.2mm);

     \draw[thick, black, fill=black] (15,3) circle (1.2mm);
    \draw[thick, black, fill=black] (15,5) circle (1.2mm);
    \draw[thick, black, fill=black] (15,7) circle (1.2mm);

     \draw[thick, black, fill=black] (16,2) circle (1.2mm);
    \draw[thick, black, fill=black] (16,4) circle (1.2mm);
    \draw[thick, black, fill=black] (16,6) circle (1.2mm);

     \draw[thick, black, fill=black] (17,1) circle (1.2mm);
    \draw[thick, black, fill=black] (17,3) circle (1.2mm);
    \draw[thick, black, fill=black] (17,5) circle (1.2mm);

    \draw[thick, black] (18,0) circle (2.5mm);
    \draw[thick, black] (18,2) circle (2.5mm);
    \draw[thick, black, fill=black] (18,2) circle (1.2mm);
    \draw[thick, black] (18,4) circle (2.5mm);
    \draw[thick, black, fill=black] (18,4) circle (1.2mm);

     \draw[thick, black, fill=black] (19,1) circle (1.2mm);
    \draw[thick, black, fill=black] (19,3) circle (1.2mm);

     \draw[thick, black, fill=black] (20,2) circle (1.2mm);

\node[above,black] at (3.5,8.9) {\large ${\delta=1}$};
\node[above,black] at (3.5,10) {\large$\delta=0$};
    \draw[thick, black] (1.5,10.5) circle (2mm);
    \draw[thick, black, fill=black] (1.5,9.5) circle (1.2mm);
  \end{tikzpicture}
\caption{Points $(r,a,\delta)$ determining all 75 invariant lattices of signature 
$(r-1,1)$ which are embedded primitively in the K3 lattice $\ktl$.}
\label{figureNikulin}
\end{figure}
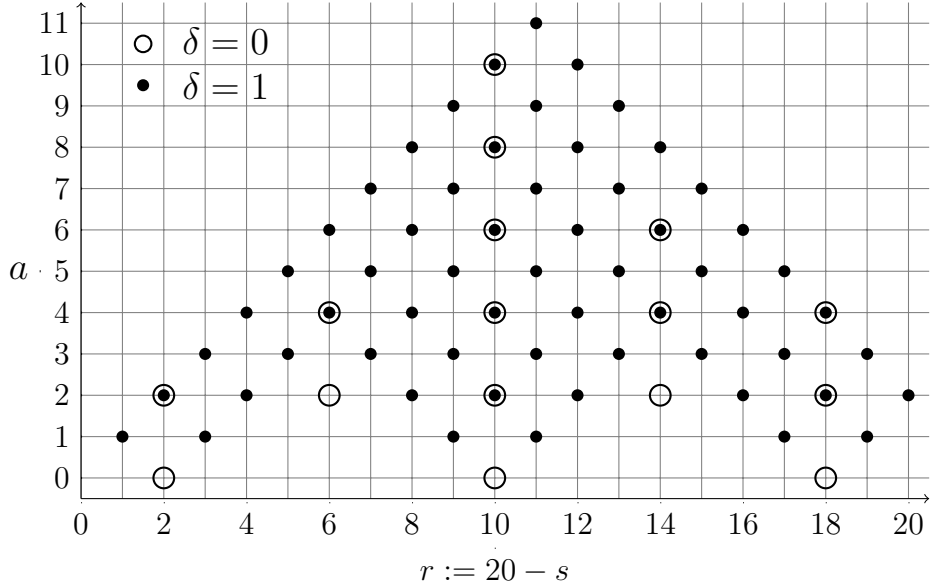

The even self-dual lattice $\ktl$ plays a central role in our discussion.
In M-theory compactification on K3, it is isometric to the second cohomology group of
K3, with the intersection form of K3. In the heterotic compactification on $\bt^3$, modular invariance forces
the internal left and right-moving momenta to live on $\ktl$ \cite{Narain:1985jj,Narain:1986am}. We are interested in a $\bz_2$ involution of
$\ktl$, denoted $\theta$, that reflects $s$ left movers and two right movers. Equivalently, $\theta$ leaves invariant
$(19-s)$ left movers and one right mover. On the K3 side  this means that $\theta$ leaves invariant a K\"ahler
2-form while it acts by $(-1)$ on the holomorphic 2-form. 
Such non-symplectic involutions were classified by Nikulin \cite{Nikulin80}, see also \cite{Nikulin83} and \cite{Nikulin86}. 
They turn out to be completely characterized by a triple $(r,a,\delta)$, where $r=20-s$. There are 75 allowed triples depicted in Figure
\ref{figureNikulin}.
Below we briefly sketch the significance of this classification in terms of the lattices that enter in the construction of
asymmetric orbifolds \cite{Narain:1986qm}.

The $\bz_2$ involution $\theta$ leaves invariant a lattice, denoted $I$, of rank $r$ and signature $(r-1,1)$. 
The normal lattice in $\ktl$, denoted $N$, has rank $(2+s)$ and signature $(s,2)$. Both $I$ and $N$ are even
sub-lattices of $\ktl$, their duals are denoted $I^*$ and $N^*$ respectively.
As explained in \cite{Narain:1986qm}, any vector $P \in \ktl$ can be written as
$(P_N,P_I)$ with $P_N\in N^*$ and $P_I \in I^*$. Moreover, $I^*/I=N^*/N$ so that
there is a definite one to one pairing of conjugacy classes in the cosets. 
The rotation $\theta$ acts on lattice vectors as $\theta (P_N,P_I)= (-P_N,P_I)$. Therefore $(1+\theta)(P_N,P_I)= (0,2 P_I) \in \ktl$. 
Since $P_I$ is an arbitrary element of $I^*$, this implies that $2 I^* \subset I$. Similarly $2 N^*\subset N$. 
Thus, necessarily
\be\label{gendisc}
I^*/I =N^*/N = (\bz_2)^a \ ,
\ee
for some integer $a\ge 0$. Notice also that from $2 I^* \subset I$ it follows that for all $P_I \in I^*$, $2 P_I^2$ is integer.
The invariant $\delta$ indicates whether $P_I^2$ itself is integer or not. Concretely,
\be\label{deldef}
\delta_I = \delta =\left\{
\begin{array}{ll}
0 & \quad {\rm if} \ P_I^2 \in \bz \ \ \forall \op P_I \in I^*\ , \\
1 & \quad {\rm otherwise}\ .
\end{array}
\right.
\ee
Notice also that for all $P_N \in N^*$, $2 P_N^2$ is integer and $\delta_N=\delta_I=\delta$.
This means that for each point in Figure \ref{figureNikulin} there is a lattice $N$ with the same $a$ and $\delta$.

According to Theorem 3.6.2 in \cite{Nikulin80}, the invariants $(r,a,\delta)$ determine the lattice $I$ uniquely up to
isomorphisms, i.e. up to boosts in $SO(r-1,1)$. At special points in moduli space, $I$ will be given by orthogonal 
direct sums of root lattices of $A_1$, $D_{2j}$, $E_7$, $E_8$, and the lattice $U$, 
which stands for the unique even self-dual lattice of signature $(1,1)$. Other lattices that will appear
are $U(2)$, $A_1(-1)$ and $E_8(2)$, where $L(n)$ denotes the lattice 
obtained by multiplying the Gram matrix of $L$ by $n$. For example, for the point $(10,10,0)$, 
$I \simeq U(2) + E_8(2)$ and $N \simeq U + U(2) + E_8(2)$. These results can be obtained 
by starting at a particular point in the moduli space of $\ktl$ in which it is given by 
$U+U+U+E_8+E_8=\oplus_i \Lambda_i$ and taking 
$\theta(\Lambda_1,\Lambda_2,\Lambda_3,\Lambda_4,\Lambda_5)=(-\Lambda_1, \Lambda_3,\Lambda_2,\Lambda_5,\Lambda_4)$.

In subsection \ref{sec_U3E8} and appendix \ref{app_lattices} we will discuss possible involutions $\theta$ for all points $(r,a,\delta)$,
and will identify the corresponding lattices $I$ and $N$.
In table \ref{tab_IN} we collect invariant and normal lattice representatives for all $(r,a,\delta)$.
We stress that the given lattices are unique up to boosts. This means that there can be equivalences 
such as $I\simeq U(2) + A_1^7 \simeq A_1(-1) + A_1^8 \simeq A_1(-1) + E_8(2)$, for the point $(9,9,1)$.
In subsection \ref{subsec_LMC} we will describe the moduli of $I$ in more detail and will indicate how such equivalences 
can be proven.

For future purposes we also define $N^*_e$ and $N^*_o$ to be the subsets of $N^*$ with $2P_N^2$ even and 
odd respectively, namely
\be\label{nenodef}
N^*_e = \left\{P_N \in N^* || 2 P^2_N \in 2\bz  \right\}, \quad 
N^*_o = \left\{P_N \in N^* || 2 P^2_N \in 2\bz +1 \right\}.
\ee
The subsets $I^*_e$ and $I^*_o$ are defined analogously.
It is easy to show that $I^*_e$ is a sublattice of $I^*$ and it is also clear that $I \subset  I^*_e$, and similarly replacing 
$I$ by $N$.
If we shift $P_I \in I^*$ by an arbitrary vector $P'\in I$, then $(P_I+P')^2=P_I^2\, \mathrm{mod}\, 2$,
therefore  the classes $I^*/I$ split into $(I^*/I)_e$ and $(I^*/I)_o$. The same holds also for $N^*/N$ classes. 
Since $\ktl$ is even, the $(I^*/I)_e$ classes are paired one to one with  $(N^*/N)_e$ classes and  $(I^*/I)_o$ are paired one to one with  $(N^*/N)_o$ classes, as  $(P_N,P_I)^2= P_N^2+P_I^2$ should be even.
It also follows that if $P_1 \in I^*_e$ and $P_2,P_3 \in I^*_o$, then  $P_1+P_2 \in I^*_o$ whereas 
$P_2+P_3 \in I^*_e$.  Thus, if $P \in I^*_o$ then $I^*_o= P+ I^*_e$.

To conclude this brief review we collect some additional results.
The parameters $(r,a,\delta)$ are also related to the K3 locus left fixed by the involution $\theta$ \cite{Nikulin80}. 
For $(r,a,\delta) \not=(10,10,0), (10,8,0)$, the fixed locus is a disjoint union of a curve 
of genus $\frac12(22-r-a)$, and $\frac12(r-a)$ rational curves, i.e. ${\mathbb {CP}}^1$'s. For $(r,a,\delta) =(10,8,0)$,
the fixed locus is the union of two elliptic curves. For $(r,a,\delta)=(10,10,0)$, there are no fixed points.
In the latter case the quotient of the K3 by the involution is an Enriques surface. Thus, in all cases except that of the Enriques surface, the quotient $Y$ has real codimension two orbifold singularities along the fixed point set. However, all of the $Y$ may be considered smooth as complex surfaces. The corresponding compact 4-manifolds are simply connected, apart from the case of the Enriques surface.

An important physical question is whether $Y$ is a spin manifold or not.
The answer is relevant for the existence of fermions in M-theory or type IIA compactifications on $Y$. We now show that only in a handful of cases, $Y$ is spin. 
Rokhlin's theorem states that smooth compact spin 4-manifolds have
signature divisible by 16 (see Theorem 1.2.6 in \cite{DonaldsonKronheimer}). 
Besides, in many quotients the singularities of $Y$ can be resolved to give a smooth manifold  \cite{NikulinSaito, AlexeevNikulin}.
Since the signature of the quotient is $(r-2)$, i.e. equal to that of the
invariant lattice, only the cases $r=2$ and $r=18$ are not excluded by the above argument. For instance, for $(r,a,\delta)=(10,10,0)$ and
$(r,a,\delta)=(1,1,1)$, the quotients are respectively Enriques and $\P^2$, which are not spin. 
For $r=2$ it has been found \cite{Morrison:1996pp, NikulinSaito} that the cases with $\delta=0$ are spin, concretely for 
$a=2$, $Y=\P^1 \times \P^1$, whereas for $a=0$, $Y={\mathbb F}_4$, where ${\mathbb F}_n$ is a Hirzebruch ruled surface. For $n$ even Hirzebruch surfaces are spin and in fact are topologically equivalent to $\P^1 \times \P^1$. For $n$ odd they are not spin and as topological 4-manifolds are equivalent to $\P^2$ blown up at a point.
For $(r,a,\delta)=(2,2,1)$ the quotient is instead $Y={\mathbb F}_1$. For $r=18$, the manifolds can also not be spin since a compact, simply connected, spin topological 4-manifold must be the connected sum of some number of K3's and $S^2 \times S^2$'s, all of which have the wrong Betti numbers.

In the physics literature Nikulin non-symplectic involutions first appeared in \cite{Aspinwall:1995mh, Morrison:1996pp} where
the so-called Voisin-Borcea \cite{Voisin93,  Borcea97} Calabi-Yau threefolds were considered. 
These are special Calabi-Yau manifolds constructed as orbifolds of the form 
$({\text {K3}}\times \bt^2)/\bz_2$, where the $\bz_2$ changes the sign of
the $\bt^2$ complex coordinate and acts on K3 by one of the Nikulin involutions characterized by $(r,a,\delta)$.

In \cite{Nikulin80b} Nikulin has also classified Abelian symplectic automorphisms of $\ktl$ which 
in the K3 picture leave the holomorphic 2-form invariant, thus maintaining supersymmetry.
The order 2 case, i.e. the symplectic involution, corresponds to exchange of the two $E_8$ factors. 
Symplectic automorphisms can be used to build 7-dimensional supersymmetric asymmetric heterotic orbifolds, 
as well as their M-theory and type IIA duals, characterized by having reduced rank of the gauge group \cite{Mikhailov:1998si, deBoer:2001wca, Fraiman:2021soq}. Compactifying on an additional circle gives 6-dimensional theories 
\cite{Schwarz:1995bj, Chaudhuri:1995dj, Mikhailov:1998si, Fraiman:2021hma} such as the CHL model
first obtained in the fermionic formulation \cite{Chaudhuri:1995fk} and later as an asymmetric orbifold \cite{Chaudhuri:1995bf}.
Lower dimensions have been studied likewise \cite{Chaudhuri:1995bf, Mikhailov:1998si}.

\section{Field Theory}\label{section_ft}

We would like to consider vacuum solutions of ten dimensional heterotic or Type I supergravity theory which are compactifications to seven dimensional Minkowski space.
The simplest such smooth solutions are spacetimes of the form $M^{9,1} = M^3 \times {\R^{6,1}}$ with a product metric
\be
g(M^{9,1}) = \left(\!\!\begin{array}{cc}
g(M^3) & 0 \\
0 & \eta 
\end{array}
\!\! \right),
\ee
where $\eta$ is the Minkowski metric in seven dimensions and $g(M^3)$ is a Ricci flat metric on $M^3$. In such solutions, both the NS-NS 3-form field strength $H$ and the non-Abelian gauge field strength $F$ are zero.

Ricci flatness in dimension three is equivalent to local flatness i.e. the Riemann tensor is zero, hence $M^3$ is locally isometric to Euclidean space. Since we are interested only in compact $M^3$ this implies that $M^3 = {\bt^3}/{\Gamma}$ for a freely acting finite group $\Gamma$. If we assume that $M^3$ is orientable, then there are only six possibilities for $\Gamma$: ${\mathds1, \bz_2, \bz_3, \bz_4, \bz_6, \bz_2 \times \bz_2}$ \cite{McInnes:1998wq,Pfaeffle}. The elements of these groups act on ${\bt^3}$ as a finite order rotation in a plane combined with a translation along the orthogonal circle.

In order to fully specify the heterotic supergravity background, one has to specify additional data. First, one must specify a spin structure since the theory has fermions. Second, although the gauge field strength is zero, since $M^3$ is non-simply connected there can be non-trivial flat gauge connections. In total there are 28 spin structures on these six compact manifolds and only one of these is supersymmetric: the 3-torus with the totally periodic spin structure. The remaining 27 are non-supersymmetric backgrounds of the heterotic string theory. We will focus on the ${\bt^3}/{\bz_2}$ compactifications here. Note that all 27 of these backgrounds have been shown to have generalised Witten bubble-of-nothing instabilities \cite{Acharya:2019mcu,GarciaEtxebarria:2020xsr}. 

We will now discuss the flat gauge connections and their relation to the low energy effective field theory in seven dimensions. As we will see, even though the theory is not supersymmetric, the low energy dynamics bears similarities to $N=2$ theories in four dimensions. Later on in the paper we will give a complete worldsheet description of these semi-classical vacua.

\subsection{Flat Connections on ${\bt^3}/{\bz_2}$}\label{subs_ft_het}
In standard Euclidean coordinates $(x_1, x_2, x_3)$ we can describe the generators of the fundamental group of ${\bt^3}/{\bz_2}$ as the three commuting translations of ${\bt^3}$, $g_1, g_2, g_3$ and the generator $g_\theta$ which is order two on ${\bt^3}$.
We take the translations to act as
\be
g_i: x_i \longrightarrow x_i + 1
\ee
and the fourth generator acts as
\be
g_\theta: (x_1, x_2, x_3) \longrightarrow (-x_1, -x_2, x_3 + {\tfrac12})\ .
\ee
The fundamental group can be described abstractly as having four generators subject to the relations
\be\label{flatcond}
\begin{split}
g_i g_j &= g_j g_i, \quad \forall i,j=1,2,3 , \\
g_\theta g_1 g_\theta^{-1} &= g_1^{-1}, \\
g_\theta g_2 g_\theta^{-1} &= g_2^{-1}, \\
g_\theta g_3 g_\theta^{-1} &= g_3, \\
g_\theta^2&=g_3\, ,
\end{split}
\ee

A flat connection on the heterotic $E_8 \times E_8$ or $Spin(32)/{\bz_2}$ gauge bundle
is specified by a set of four Wilson lines, one for each generator, obeying these relations.
In other words we look for homomorphisms from $\pi_1(M^3)$ to the gauge group. As we will see, there are different classes of solutions. 

One possible strategy to find solutions consists of first finding a flat connection on ${\bt^3}$ which would give candidate group elements for $g_1$, $g_2$,  and $g_3$ satisfying the first of the above relations (and is a problem which is completely solved 
\cite{Witten:1997bs, Keurentjes:1998uu, Keurentjes:1999qf, Kac:1999gw, Keurentjes:1999mv}) 
and then trying to find a $g_\theta$ representative in the gauge group which satisfies the remaining relations. From now on, by abuse of notation, we will denote the gauge group representatives by $g_i, g_\theta$. We will discuss two main classes of flat connections which we refer to as Higgs and Coulomb branch solutions following their descriptions in the low energy effective theory.

\subsubsection{Higgs branch solutions}

Let us first consider the flat connection to be restricted to an $SU(2)$ subgroup of the gauge group. Then, if we take $g_1$ and $g_2$ diagonal, it is straightforward to see that the flat connections take the form
\be\label{higgsbranch}
g_1 =
\begin{pmatrix}
\lambda_1 & 0 \\
0 & \bar\lambda_1
\end{pmatrix},
\quad
g_2 =
\begin{pmatrix}
\lambda_2 & 0 \\
0 & \bar\lambda_2
\end{pmatrix},
\quad g_\theta =
\begin{pmatrix}
0 & \lambda_\theta \\
-\bar\lambda_\theta & 0
\end{pmatrix},
\quad
g_3 =
\begin{pmatrix}
-1 & 0 \\
0 & -1
\end{pmatrix},
\ee
where the $\lambda$'s are all unitary numbers. Note also that $g_\theta$ in these solutions obeys $g_\theta^4=\bone$ so one can use the gauge symmetry to fix $\lambda_\theta = 1$.
If we view ${\bt^3}/{\bz_2}$ as $({\bt^2} \times S^1)/{\bz_2}$, one can regard this type of solution as a generic flat connection on ${\bt^2}$ which descends to the quotient.
Clearly such solutions generalise to higher rank subgroups since we have, up to a discrete factor, that $SU(2)^{16} \subset E_8 \times E_8$ (or $Spin(32)/{\bz_2}$). Hence we can embed the above solution into any of the sixteen $SU(2)$ factors. 

These solutions have a moduli space which is the moduli space of flat $SU(2)$-connections on ${\bt^2}$. Hence the low energy field theory will contain two light scalars, which will naturally form a complex scalar field. Notice that at the origin of the moduli space, when $\lambda_{1,2} = 1$, there is an $SO(2)$ subgroup of $SU(2)$ which commutes with the flat connection; these are the $SU(2)$ matrices with real entries.
Therefore the 7d theory has an enhanced $SO(2)$ gauge symmetry at that point, broken for generic values of the $\lambda_i$. Our proposal for the low energy effective theory is an $SO(2)$ gauge theory coupled to a complex field in the fundamental representation of $SO(2)$. The potential for this theory arises from the reduction of $SU(2)$ Yang-Mills on the 3-manifold. The generic vacuum expectation values for these charged scalars which minimise the potential break $SO(2)$ completely, leaving behind two massless scalars without a potential. These are identified with the $\lambda_i$.

If we now take block diagonal sums of these solutions we can obtain flat connections for e.g. $SU(2N)$ by considering an $SU(2)^N$ subgroup. Now $g_1$ and $g_2$ each have $N$ moduli subject to one overall constraint since the determinant is one, so the low energy theory contains at least $(N-1)$ complex scalars. At the origin an enhanced gauge symmetry appears. In this case $g_\theta$ is the block diagonal sum of $N$ copies of our two-by-two example. In other words, $g_\theta$ is the standard $Sp(N, {\C})$-invariant symplectic form. 
By definition this group is formed by $2N \times 2N$ matrices satisfying
\be
M \cdot g_\theta \cdot M^T = g_\theta
\ee
and $M$ is complex. 

In our problem, we must determine the subgroup of $SU(2N)$ matrices which commute with $g_\theta$. This determines the unbroken gauge symmetry at the origin of the Higgs branch. This is the condition:
\be
M \cdot g_\theta \cdot M^\dagger = g_\theta
\ee
where $M \subset SU(2N)$. Clearly by restricting to real $SU(2N)$ matrices we obtain a subgroup of the symplectic group. The set of real $SU(2N)$ matrices is isomorphic to $SO(2N)$, but clearly not all of them preserve $g_\theta$. However, a well known fact about real symplectic transformations is that the intersection of the symplectic group with $SO(2N)$ is $U(N)$. Hence, at the origin of the Higgs branch at least an enhanced $U(N)$ gauge symmetry appears. However, the actual symmetry group at the origin turns out to be $G_{max} = (SU(N) \times SU(N) \times U(1))/{\bz_N}$. To see this, one can diagonalise $g_\theta$ at the origin of the Higgs branch. This gives the diagonal sum of $N$ copies of $i \sigma_3$. The commutant of this form of $g_\theta$ is $G_{max}$. Therefore, in this branch the low energy effective theory would be a $G_{max}$ gauge theory coupled to complex scalars transforming in some representation, ${\bf r}$ of $G_{max}$. The higher dimensional origin of the theory restricts ${\bf r}$ as these fields originate for the adjoint representation of the gauge group, in this case $SU(2N)$. Since, the adjoint of $SU(2N)$ decomposes under $G_{max}$ as the adjoint plus $(N,\bar{N})$ plus complex conjugate, the matter in the low energy theory must be bifundamentals under $G_{max}$. In fact there is one complex bifundamental and one complex anti-bifundamental. The symmetry breaking pattern which occurs can be understood as follows. One can give a vev to one of the bifundamentals which breaks $G_{max}$ to $SU(N)$, the diagonal of the two $SU(N)$ groups. This leaves adjoint fields plus one singlet in the spectrum arising from the other bifundamental. Giving a vev to these breaks $SU(N)$ to its maximal torus $U(1)^{N-1}$. The singlet plus the diagonal components of the $SU(N)$ adjoint fields remain massless, giving $2N$ moduli, which we can identify with the $2N$ moduli of our flat connection.

Of course, we are ultimately interested in the case of $E_8 \times E_8$ and $Spin(32)/{\bz_2}$. In these cases there are other embeddings of the $SU(2)$ solutions that we can use. For instance, if we consider the the fact that $E_8$ contains a rank eight subgroup isomorphic to $Spin(4)^4 /\bz_2 = (SU(2)^8)/{\bz_2}$ we obtain a solution with eight copies of the $SU(2)$ Higgs solution and has eight complex moduli. In this case, it may be shown that the unbroken gauge symmetry at the origin of the moduli space is $Spin(16)/{\bz_2}$ and contains a massless complex scalar in the 128-dimensional spinor representation. A generic vev for this field breaks $Spin(16)/{\bz_2}$ completely but leaves eight complex scalars massless and these can be identified with the eight complex moduli. An analogous model can also be obtained in the $Spin(32)/{\bz_2}$ theory.

\subsubsection{Coulomb branch solutions}
Another family of solutions, which {\it is} identity connected, are the following:
\be\label{coulombbranch}
g_1 =
\begin{pmatrix}
1 & 0 \\
0 & 1
\end{pmatrix},
\quad g_2 =
\begin{pmatrix}
1 & 0 \\
0 & 1
\end{pmatrix},
\quad g_\theta =
\begin{pmatrix}
\lambda_\theta & 0 \\
0 & \bar{\lambda}_\theta
\end{pmatrix},
\quad g_3 =
\begin{pmatrix}
\lambda_\theta^2 & 0 \\
0 & \bar{\lambda}_\theta^2
\end{pmatrix}
\ee
These simply reflect the fact that $b_1({\bt^3}/{\bz_2}) = 1$ corresponding to the $x_3$ direction and are simply Wilson lines along that circle in ${\bt^3}$ which are ${\bz_2}$-invariant. Clearly, if one takes $g_\theta$ to be any element of the maximal torus of the full gauge group, these solutions break the gauge symmetry down to the maximal torus generically. In this case the low energy theory in seven dimensions will have an $E_8 \times E_8$ gauge symmetry with a real adjoint scalar field. Diagonalising the field minimises the potential and hence we have a sixteen dimensional moduli space of vacua with 
$E_8 \times E_8$ unbroken at the origin. We refer to these solutions as Coulomb branch vacua since the gauge group is the maximal torus at generic points. Notice that, at the origin of the Higgs branch solution above, the solution is gauge equivalent to a particular Coulomb branch solution. Hence these two types of branches of moduli space intersect there.

\subsubsection{Mixed Higgs-Coulomb solutions}
Clearly one can combine the Higgs and Coulomb branch solutions. In the $SU(2)^{16} \subset E_8 \times E_8$ one can pick $n_V$ Coulomb $SU(2)$'s and $n_H=16-n_V$ Higgs $SU(2)$'s to obtain vacua whose low energy effective theory contains both types of moduli and gauge fields.

There are additional massless fields which arise from the metric, Kalb-Ramond $B$-field and dilaton. One can understand these by examining the ${\bz_2}$ action on the theory on ${\bt^3}$, regarded as ${\bt^2} \times S^1$. In the theory on ${\bt^3}$, one has 22 $U(1)$ gauge fields generically, however only 19 of these have associated scalar moduli. In fact for each of these 19 gauge fields there are three associated moduli, generating the familiar Grassmanian
$Gr(19,3)$ moduli space locally. Of the 22 gauge fields, four are always odd under the ${\bz_2}$, and two of these four have associated moduli. The maximal rank of the gauge group is thus 18 from this point of view. 
This corresponds to a situation in which the Wilson lines are all of the Coulomb type and the sixteen $U(1)$'s are the maximal torus of 
$E_8 \times E_8$ 
plus two Kaluza-Klein gauge fields from the metric and $B$-field respectively. 
In these vacua the moduli space is locally
$Gr(2,2) \times Gr(17,1)$ corresponding to the moduli of ${\bt^2}$ (all of which are invariant)
and the radius plus 16 Wilson line moduli along the $S^1$ direction.

We can now consider switching on Higgs solutions for some of the directions in the gauge group. 
This then gives vacua with local moduli space $Gr(2+n_H,2) \times Gr(17-n_H,1)$, or equivalently
$Gr(18-n_V,2) \times Gr(n_V+1,1)$.

Note that there is a universal $U(1)$ gauge field (one of the three without associated moduli) that is also present.
Thus, including this universal ``graviphoton", the gauge group at a generic point in the moduli space has rank $n_V +2$.
The maximal rank is achieved when $n_V=16$ as explained above.

\subsubsection{Other Branches}
The above vacua are by no means all since we have not classified flat connections on ${\bt^3}/{\bz_2}$. In the closely related problem of flat connections on $\bt^3$ a complete solution has been obtained and the independent components of moduli space are in one to one correspondence with flat connections distinguished by the Chern-Simons invariant \cite{Borel:1999bx, deBoer:2001wca}. 
We can utilise some of these to describe other vacua. 
For instance for gauge group $SO(7)$ there is a flat connection on the 3-torus generated by diagonal order two elements of $SO(7)$: $a=(--++--+), b=(++----+)$ and $c=(-+-+-+-)$. These three elements cannot belong to the same maximal torus of $SO(7)$ so cannot arise in the identity connected component of moduli space. Taking $g_1=a$, $g_2=b$, $g_3=\bone$, and $g_\theta = c$,  
we obtain a flat connection of ${\bt^3}/{\bz_2}$. This example can naturally be embedded into the heterotic field theories.

\subsubsection{Quantum Moduli Space}
Since all of the moduli fields discussed so far are Wilson line moduli, they all enjoy shift symmetries, remnants of the underlying gauge symmetry. Hence there can be no polynomial terms in the effective potential on the moduli space. Usually such axion-like-fields in four and lower dimensions can have non-perturbative effective potentials generated by instantons. In the seven dimensional theory we are discussing here, however, the gauge dynamics is infrared free, so it might be the case that there are no such non-perturbative corrections and that the effective potential vanishes identically in the quantum theory.

\subsection{Duality with M-theory on $\rkt/{\bz_2}$}\label{subs_ft_M}
At the classical level one can view the heterotic string vacua discussed above as arising from M-theory on Nikulin orbifolds of the form $\rkt/{\bz_2}$ and briefly described in the previous section. The K3 surface comes equipped with a Ricci flat metric which is equivalent to a hyperKahler triple of harmonic 2-forms, $\omega_i$ which combine into a holomorphic 2-form $\omega^{2,0}=\omega_1+i\omega_2$ and Kahler form $\omega=\omega_3$. 
Any given Nikulin involution acts as ${\bz_2}:(\omega, \omega^{2,0}) \rightarrow (\omega, - \omega^{2,0})$. Since it preserves the K\"ahler form, it preserves the Ricci flat metric on that given K3 surface. Hence we obtain a Ricci flat metric on $\rkt/{\bz_2}$ which breaks supersymmetry. This is therefore a classical solution of M-theory.
The moduli space of fixed volume Ricci flat metrics on K3 is obtained by taking the periods of the three self-dual forms $\omega_i$ and evaluating them against the 19 orthogonal 2-cyles in the K3 lattice.
One can think of these 19 directions as the anti self-dual (ASD) harmonic 2-forms in $H^2(K3, {\R})$ and the periods as the integrals over K3 of the wedge product of the $\omega_i$ with these ASD forms.
The eigenvalues of the Nikulin involution on the ASD forms will have $(r-1)$ plus and $s=(20-r)$ minus signs. 
The even directions give moduli by taking periods with $\omega$ giving an $(r-1)$-dimensional moduli space. 
This is the K\"ahler moduli space of $\rkt/{\bz_2}$. The odd directions give moduli by taking periods with $\omega^{2,0}$ (which is also odd under the involution). These are the complex structure moduli of $\rkt/{\bz_2}$.

The original moduli space of K3, which is {\it locally} the Grassmanian $Gr(19,3)$, decomposes into 
$Gr(r-1,1) \times Gr(20-r,2)$. 
Each of the $r$ even two forms also gives rise to a $U(1)$ gauge field in seven dimensions from the harmonic expansion of the $C$-field in M-theory. Hence, we can identify the K\"ahler moduli space $Gr(r-1,1)$ with the Coulomb branch moduli space discussed in the heterotic discussion above. Similarly, we can identify the complex structure moduli space with the Higgs branch, since generically there is no gauge symmetry arising at generic points there because the would be $U(1)$'s from the $C$-field are odd under the involution.

\subsubsection{Non-supersymmetric BPS States}

When $\rkt/{\bz_2}$ develops $ADE$ singularities, the $U(1)^r$ gauge symmetry enhances to corresponding non-Abelian groups. Note that: although the background is non-supersymmetric, $M2$-branes wrapping the exceptional divisors of the $ADE$-singularity are {\it exactly} massless, because the exceptional divisors are calibrated submanifolds, with volume calibrated by the Kahler form $\omega$, hence there is a BPS bound on the masses of such states. This shows that BPS states in the supersymmetric theory on K3 which are 
${\bz_2}$-invariant are non-supersymmetric BPS states in the theory on $\rkt/{\bz_2}$. This will presumably also be true on the heterotic string side.

\section{CFT of $SU(2)$ holonomies}\label{section_cft}

In this section we want to analyze the worldsheet realization of the flat connections on $T^3/\bz_2$ considered in 
section \ref{section_ft}, focusing on gauge group $SU(2)$. We are mostly interested in
the holonomy $g_\theta$ corresponding to the $\bz_2$ generator.
Understanding $g_\theta$ will lead us to the proper definition of the orbifold action in the heterotic string context.
For convenience we now use the gauge freedom to choose $\lambda_\theta=i$ in eq. \eqref{higgsbranch} so that
$g_\theta= e^{i \frac{\pi}{2} \sigma_1 }=i \sigma_1$. To simplify typing we will write $g=g_\theta$ in most places.
We will discuss two possible formulations. 

\medskip
\noindent
{\bf{Formulation 1}}

The  weight vectors $ |\ttw\rangle$ will be eigenfunctions in a given representation of the Cartan generator
chosen to be $\sigma_1/\sqrt2$. Therefore,
$g |\ttw\rangle = e^{i \frac{\pi}{2} \sqrt{2} \ttw} |\ttw\rangle$, where the factor $\sqrt{2}$ appears because we 
use roots of norm 2 and fundamental weights of norm $1/2$.
Thus, the action of $g$ can be defined as a shift vector in the weight space.

\medskip
\noindent
{\bf{Formulation 2}}

The Cartan generator is chosen to be $\sigma_3/\sqrt{2}$. 
Let us denote by $L$ and $L_R$  the weight
and root lattice of $SU(2)$ respectively. $L/L_R$ has two elements $(0,\ttf)$ where  $\ttf$ is a weight of the fundamental rep 
(i.e. spin $\frac12$) of $SU(2)$ then
$L$ consists of $L_R$ and vectors of type $L_R+\ttf \equiv L_F$. 
Since $g= i \sigma_1$, it acts as $|\ttf\rangle \rightarrow i |-\ttf\rangle$ and $|-\ttf\rangle \rightarrow  i|\ttf\rangle$. On the adjoint
representation $g$ acts as $(\sigma_1,\sigma_2,\sigma_3 ) \rightarrow ( \sigma_1,-\sigma_2,-\sigma_3 )$.  
It follows that $g^2$ which is the non-trivial element of the center of $SU(2)$, acts as the identity on the adjoint representation
and all integer spin weights (i.e. in the root lattice $L_R$) and as $-1$ on the fundamental representation and indeed on all the 
weights in  $L_F$.
Thus, in formulation 2 the action of $g$ includes a reflection of the weights together with a phase. Since all other representations can be obtained by tensoring spin $\frac12$ representations
we can deduce the action of $g$ in all the representations. We note that  the phase does not depend only on the weight. For example consider the product of two spin 1/2 states $(|+ \frac12\rangle |- \frac12\rangle \pm |- \frac12\rangle |+ \frac12\rangle)$. 
For both relative signs the total weight is zero but for $+$ sign it is in a spin 1 representation while for $-$ sign it is in the singlet representation.   
The action of $g$ on this state gives $\mp (|+ \frac12\rangle |- \frac12\rangle \pm |- \frac12\rangle |+ \frac12\rangle)$. 
Thus, this phase cannot be expressed as a shift vector on the weight space.

\medskip 

In the following we will look into the CFT realization of the two formulations.

\subsection{Holonomies and Kac-Moody currents}\label{subsec_km}

Let us consider the $SU(2)$ Kac-Moody algebra
\begin{equation}\label{su2km1}
  [ J^a_n, J^b_m] = i \epsilon_{abc} J^c_{m+n} + \frac{1}{2} k m \delta^{ab} \delta_{m,-n}
\end{equation}
where $k$ is the level. Setting $n=m=0$ above gives an $SU(2)$ subalgebra 
with the identification $J^a_0 \rightarrow \sigma_a/2$.
The states are built by starting from a highest weight vector which is annihilated by all the positive frequency modes 
$J^a_n$, $n > 0$, and is labelled by the highest weight with respect to the $SU(2)$ subalgebra.
In turn this highest weight is specified by the
eigenvalue with respect to the Cartan generator, say $J^3_0$, together with the condition that it is annihilated by the raising operator 
$J^1_0+i J^2_0$. All the other states are obtained by applying creation operators involving negative frequency modes 
$J^a_{-n}$, $n>0$, and the lowering operator $J^1_0-i J^2_0$. 

We want to study the adjoint action of 
$g= e^{\frac{i\pi}{2} \sigma_1} \equiv e^{i\pi J^1_0} $ on the Kac-Moody currents and on all the states.
To begin notice that
\begin{equation}\label{su2km2}
  [J^1_0, J^b_m]= i\epsilon_{1bc} J^c_m ,
\end{equation}
as the central term drops out. 
Let us now see how $g$ acts on the highest weight states and determine the adjoint action of $g$ on the oscillator modes of the 
currents. The $J^1_n$ are clearly invariant under the adjoint action of $g$. For $b\neq 1$, 
\begin{equation}\label{su2km3}
  e^{i \theta J^1_0} J^b_n   e^{-i \theta J^1_0} = \cos \theta  J^b_n - \sin \theta \epsilon_{1bc}  J^c_n ,
\end{equation}
where we used \eqref{su2km2}.
Since $g$ is obtained by setting $\theta=\pi$, it follows that for all $n \in \bz$,  $Ad_g J^1_n = J^1_n$ and for $b \neq 1$,   
$Ad_g J^b_n =- J^b_n$. We still need to know how
$g$ acts on the highest weight states. In general this can be worked out for arbitrary level $k$, but we will 
consider level $k= 1$, in which case there are only two highest weight states: 
the ground state $|0\rangle$ which is singlet of $SU(2)$ and a spin
$1/2$ representation $|\pm \frac12\rangle$ where
$\pm \frac12$ denote the eigenvalues with respect to $J^3_0$. On the ground state $g$ acts as the identity and we already saw 
above that $g|\pm \frac12\rangle=i |\mp \frac12\rangle$.

\subsection{Free Field Realization}\label{subsec_ff}

We are interested in $k=1$ which can be realized in terms of a free scalar CFT.
Denoting this scalar by $Y$, we can write one of the currents as $\partial Y$ 
(we take $SU(2)$ to be left-moving  so $Y$ depends on ${z}$).
For the remaining currents we consider two choices according to the two formulations described before.

In formulation 1 we take $ J^1(z)= \partial Y(z)$. The other two currents are 
$J^2(z)\pm i J^3(z) = e^{\pm i \sqrt{2} Y}$,  with the correlator 
$\langle Y( z) Y( w)\rangle = -\log( z- w)$ so that all the currents have dimension 1. 
Since  $g$ leaves $J^1$ invariant, $Y$ is not reflected and can at most be shifted by $g$. 
Since $J^2$ and  $J^3$ pick up a minus sign under $g$ we can determine
this shift to be $Y \rightarrow Y+2 \pi/(2\sqrt{2})$. This also gives the correct action of $g$ on the highest weight states,
now taken to be eigenstates of $J_0^1$, which we denoted before by $|\ttw\rangle$.
Indeed these states correspond to $ e^{\pm \frac{i}{ \sqrt{2}} Y}$ and hence the above shift gives the phase 
$e^{\pm \frac{i \pi}{2}}= \pm i$. 
It is easy to see that this can be extended to all the states of the theory
and agrees with the action of $g$ on all the states in the representations of the Kac-Moody algebra. 
Thus, in this formulation the orbifold is very simple, it is just generated by an order 4 shift. 

In formulation 2 we bosonize $J^3(z)= \partial Y'( z)$ (we use a different variable $Y'$, related to the above $Y$ in a non-local way). 
Then $J^{\pm}( z)=J^1( z)\pm i J^2( z)=e^{\pm i \sqrt{2} Y'}$. Since $J^3$ picks up a sign under $g$, we conclude that all the non-zero modes of $Y'$ must be reflected. Furthermore, $J^+$ and  $J^-$ are exchanged under $g$, 
which means that also the zero mode of $Y'$ gets a minus sign. We then conclude that $g: Y' \rightarrow -Y'$. 
Let us now look at the highest weight states of the Kac-Moody algebra. On the spin 0 state, of course $g$ is the identity. 
On the spin 1/2 states 
$|\pm \frac12\rangle$, we already know that $g: |\pm \frac12\rangle \rightarrow i |\mp \frac12\rangle$ 
(recall that now these states are the eigenvectors of $J^3_0$, unlike in the formulation 1 where they are eigenvectors of $J^1_0$).

In formulation 2 the correspondence of states in the $SU(2)$ Kac-Moody algebra (i.e. states that are obtained by applying creation operators of $J^a$ oscillators on the highest weight state) with the states that are naturally defined in the $Y'$ CFT  (namely states obtained by 
applying $Y'$ creation operators on a vacuum carrying momentum i.e. weight) is quite involved. 
To see this, consider  $J^+_{-n}|\frac12\rangle = \oint d z  z^{-n}J^+( z) |\frac12\rangle = \oint d z  z^{-n} e^{i \sqrt{2} Y'( z)} e^{\frac{i}{ \sqrt{2}} Y'(0)} =\oint d z  z^{1-n} e^{\frac{i}{ \sqrt{2}}(2 Y'( z)+ Y'(0))}$. 
In the exponential, $Y'( z)$  can be expanded  
near $ z=0$ as $Y'( z)= Y'(0) +  z \partial Y'(0)+ \frac12  z^2 \partial^2 Y'(0)+...$ to obtain a sum of terms 
$\sum_m C(n_j)(\prod_{j=1}^m \partial^{n_j} Y'(0) ) e^{i\frac{3}{ \sqrt{2}} Y'(0)}$ with various complicated coefficients $C(n_j)$  such that $\sum_{j=1}^{m} n_j= n-2$.
Note that the weight $\frac{3}{ \sqrt{2}}$  of this state is just the sum of the weight of $J^+$ and that of $|\frac12\rangle$.
Now, $\partial^{n_j} Y'(0)$ is proportional to the oscillator mode $Y'_{-n_j}$.  Thus, a single $J^+$ creation operator would correspond to a polynomial of $Y'$ oscillators. We know how $g$ acts on the oscillators of the Kac-Moody currents. The question is whether we obtain the same result by reflecting $Y'$ when the oscillators of Kac-Moody currents are expressed as polynomials of $Y'$ oscillators. 
The answer is yes. 
This is because $J^1 \propto \big(e^{i \sqrt{2} Y'( z)} + e^{-i \sqrt{2} Y'( z)}\big)$, implying that the expansion of $J^1$ oscillators will involve only  polynomials having even powers of  $Y'$ oscillators and therefore reflecting $Y'$ leaves $J^1$ oscillators invariant as expected. Similarly, the oscillator expansion of $J^2 \propto \big(e^{i \sqrt{2} Y'(z)} - e^{-i \sqrt{2} Y'(z)}\big)$ will involve only polynomials with odd powers of  $Y'$ oscillators and therefore reflecting $Y'$ will give a minus sign to  $J^2$ oscillators as required. 

We also remark that the precise phases associated to the action of $g$ on  various weight states is basis dependent. 
For example, we could change the basis  $\{|+\frac12\rangle, |-\frac12\rangle \}$ to a new basis 
$\{|+\frac12\rangle', |-\frac12\rangle'\}=\{a |+\frac12\rangle,b |-\frac12\rangle\}$. 
Then $g$  on $\{|+\frac12\rangle', |-\frac12\rangle'\} $ gives  
$\{i\frac{a}{b} |-\frac12\rangle',i\frac{b}{a} |+\frac12\rangle'\}$. However, the statement that is invariant under the change of basis is 
that $g^2$ acts as the identity on integer spin states and as minus the identity on half-integer spin states. Indeed, the action of $g$ on 
$\{i\frac{a}{b}|-\frac12\rangle',i\frac{b}{a} |+\frac12\rangle'\}$ gives back $-\{|+\frac12\rangle', |-\frac12\rangle'\}$.

The upshot is that the action of $g$ in formulation 2 is a reflection of $Y'$ together with a phase which is  order 4  on the Kac-Moody representation built on the spin $1/2$ highest weight state, while it is order 2 on the  Kac-Moody representation built on the spin 
$0$ highest weight state.

\subsection{Characters}\label{subsec_ch}

Formulations 1 and 2 must be equivalent since they are just two different ways of expressing the $SU(2)$ Kac-Moody currents in terms 
of a free scalar. Let us explicitly check that the characters agree.
In the $(1,1)$ sector the characters in both formulations are
\begin{equation}\label{z11}
Z^{\epsilon}_{(1,1)} = \frac{1}{\prod_{n=1}^{\infty} (1- q^n)} \sum_{m\in \bz}  q^{(m+\epsilon)^2},
\end{equation}
where $\epsilon=0$ and $\epsilon=1/2$ correspond respectively to
the character of the representation based on the trivial and spin 1/2 highest weight states.  
In these two cases the lattice sum is the same as summing over $L_R$ and $L_F=L_R+ \ttf$ respectively.
In the $(1,g^2)$ sector, in both formulations we have
\begin{equation}
Z^{\epsilon}_{(1,g^2)} = (-1)^{2\epsilon}  Z^{\epsilon}_{(1,1)},
\label{su21}
\end{equation}
since $g^2$ acts as +1 on $L_R$ and as (-1) on $L_R+\ttf$. 
 
In the $(1,g)$ sector, in formulation 1 we get
\begin{equation}\label{z1gf1}
  Z^{\epsilon}_{(1,g)} =  \frac{1}{\prod_{n=1}^{\infty} (1- q^n)} \sum_{m\in \bz} 
 q^{(m+\epsilon)^2}e^{-2\pi i (m+\epsilon)/2}
\end{equation}
Note that this vanishes for $\epsilon=1/2$   as  under   $m\rightarrow -m-1$ the lattice sum picks up an overall minus sign. 
For $\epsilon=0$ this can be rewritten as
\begin{equation}
  Z^{0}_{(1,g)} =   \frac{1}{\prod_{n=1}^{\infty} (1+ q^n)}.
\end{equation}
On the other hand, in formulation 2 we first note that for $\epsilon=1/2$ the invariant lattice is null because none of the lattice vectors is invariant under $g$ due to the reflection of $Y'$. Distinguishing the character with a prime,
that means $\zp^{1/2}_{(1,g)}=0$, precisely as in the first formulation. For $\epsilon=0$ the invariant lattice is just one point, namely $m=0$ and moreover all the oscillator modes of $Y'$ pick up a minus sign due to the action of $g$. Hence
$\zp^{0}_{(1,g)}$ agrees exactly with  $Z^{0}_{(1,g)}$ in \eqref{z1gf1}. 
Since by modular transformations we can get other sectors, we conclude that the characters in the two formulations are the same.

\subsection{Moduli Deformation}\label{subsec_mod}

Even though the two formulations are the same at the $SU(2)$ point, formulation 1 is more convenient to 
describe the Coulomb branch whereas formulation 2 is more convenient for the Higgs branch.
By convenient we mean that the effect of moduli deformations can be realized as Lorentz boosts of appropriate lattices.

The key to understand the difference is to look at massless states, which depend on the action of $g$.
Recall that $g$ acts simultaneously on the right-moving $\bt^2 \times S^1$ 
(part of the fermionic string)  as rotation on $\bt^2$. Let us call $(X_1,X_2)$  the coordinates
of $\bt^2$ and $X_3$  that of $S^1$.  Then $g: (X_1,X_2,X_3)\rightarrow (-X_1,-X_2,X_3)$ modulo shifts.  

The $g$-invariant massless states in the first formulation will include
$e^{\pm i \sqrt{2} Y} \bar\partial \bar X_1$, $e^{\pm i \sqrt{2} Y}\bar\partial \bar X_2$ and ${\partial}Y\bar\partial \bar X_3$. 
We cannot give expectation values to all of these massless states simultaneously. 
If we deform by giving a vev to ${\partial}Y\bar\partial \bar X_3$ then $e^{\pm i \sqrt{2} Y}\bar\partial \bar X_1$ and  
$e^{\pm i \sqrt{2} Y} \bar\partial \bar X_2$ will become massive. 
This means that the latter are frozen to a zero vev. This defines the branch of moduli deformation dubbed Coulomb branch. 
Indeed, in the Coulomb branch holonomies given in eq. \eqref{coulombbranch}, the moduli are encoded in $g_3$, while
$g_1$ and $g_2$ are fixed.
On the other hand, if we give a vev to say \mbox{$(e^{ i \sqrt{2} Y}+ e^{- i \sqrt{2} Y})\bar\partial \bar X_1$} then  the only 
invariant state among the above  
which remains massless is \mbox{$( e^{ i \sqrt{2} Y}+ e^{- i \sqrt{2} Y})\bar\partial \bar X_2$}, the others become massive. 
Thus, in this branch, called Higgs branch, there are two moduli. These two moduli are continuous Wilson lines $A_1$ and $A_2$
encoded in the Higgs branch holonomies $g_1=e^{i A_1 \sigma_3}$ and $g_2=e^{i A_2 \sigma_3}$ given in 
eq. \eqref{higgsbranch}.

If we want to study the Coulomb branch clearly the first formulation is 
very convenient since giving vev to   
${\partial}Y\bar\partial \bar X_3$ is equivalent to boosting the lattice along $(1;1)$ directions in the $(Y;X_3)$ plane. 
On the other hand, studying the Higgs branch will be very difficult because it involves exponential operators.

In the second formulation, it is exactly the opposite. The Coulomb branch modulus is 
$(e^{ i \sqrt{2} Y'}+ e^{- i \sqrt{2} Y'})\bar\partial \bar X_3$, while the Higgs branch moduli are 
${\partial}Y'\bar\partial \bar X_1 $ and ${\partial}Y'\bar\partial \bar X_2 $. Therefore turning on the Higgs moduli corresponds to boosting along the $(1;2)$ directions $(Y';X_1,X_2)$.  

In the Higgs branch we can go to the origin of moduli space $A_1=A_2=0$, which corresponds to the $SU(2)$ point.
At this point $g_1$ and $g_2$ become the identity and we can turn on deformations along $g_3$,
i.e. $g_3=e^{i A_3 \sigma_1}$ so that $g=g_\theta=e^{i \frac{A_3}2 \sigma_1}$. 
Now we are in the Coulomb branch but in a basis where $g_\theta$ is not diagonal.
We can diagonalize so that  $\sigma_1 \rightarrow \sigma_3$ to get back  eq. \eqref{coulombbranch}.
In the CFT this diagonalization corresponds to rebosonization, i.e. using $Y$ instead of $Y'$ to describe the
Coulomb branch. In terms of the $s$ transitions discussed in great detail in section \ref{sec_motion},
this rebosonization corresponds to $s \to (s-1)$.

Starting from the Coulomb branch, with diagonal $g_\theta=e^{i \frac{A_3}2 \sigma_3}$  and $g_3=e^{i A_3 \sigma_3}$, 
we may also go to the special point $A_3= \pi$. The resulting $g_\theta$ gives a minus sign to $J^\pm$ and $\pm i$ to the
fundamental, i.e. it corresponds to a shift by half the fundamental weight. Now we can change the basis so that
$\sigma_3 \rightarrow \sigma_1$, such that $g_\theta$ and $g_3$ are as in eq.~\eqref{higgsbranch} (with $\lambda_\theta=i$). 
We can further turn on $A_1$ and $A_2$ to obtain $g_1=e^{i A_1 \sigma_3}$ and
$g_2=e^{i A_2 \sigma_3}$.  This change of basis is again accomplished by rebosonization. 
Finally, now there is a transition 
$s \to  (s+1)$ but, as remarked above, for this transition to work there must be a shift in the $SU(2)$ equal to half the fundamental weight.

\subsection{Combining with the rest of the String CFT}\label{subsec_rest}

In the full heterotic string theory, $SU(2)$ characters will be coupled with the rest of the CFT consisting of the right-moving superstring and the left-moving bosonic string with central charge $25$. 
The generator of the orbifold group $g$ will also simultaneously act on the rest of the CFT. 
In particular, on the right-moving fermionic string compactified on $\bt^2 \times S^1$ it will act as reflection on $\bt^2$ and possibly a shift on $S^1$. 
Similarly on the left-moving $c=25$ CFT which is compactified on $\bt^{18}$, $g$ acts as a reflection on $s$ compact directions, and some suitable shift $v_1$. 
The lattice of the left-right momenta will be defined by some $(18,3)$ lattice which will couple with the $SU(2)$ weight lattice in a particular way so that the combined lattice is even-self dual. 

Thus the full even self dual lattice  $\Gamma_{\!\!(19,3)}$ lattice can be expressed as the union of 
$\Gamma^0_{\!\!(18,3)} \times L_R$ and
$\Gamma^1_{\!\!(18,3)} \times L_F$ where $L_R$ is the $SU(2)$ root lattice and $L_F=L_R+ \ttf$ is set of the half integer spin weights of $SU(2)$. 
As an example, we could think of $\Gamma^0_{\!\!(18,3)}$ as the as even  self dual lattice 
$\Gamma_{\!\!(3,3)}$ times $ E_8 \times E_7$ root lattice. 
$\Gamma_{\!\!(18,3)}^1$ is then given by $\Gamma_{\!\!(18,3)}^0+ \ttw_{56}$ where $\ttw_{56}$ is a weight of the $56$ dimensional representation of $E_7$ (where $E_7$ roots are normalized so that their length squares are 2). 
Let us denote by $\hat{Z}^{\epsilon}_{(1,1)}$ the  partition function of the rest of the CFT with $\Gamma_{(18,3)}^0$ for $\epsilon=0$ and the partition function of the rest of the CFT with $\Gamma_{(18,3)}^1$ for $\epsilon=1$. 
Then the full partition function in the $(1,1)$ sector is $Z_{(1,1)}=\sum_{\epsilon=0}^1\hat{Z}^{\epsilon}_{(1,1)} Z^{\epsilon}_{(1,1)}$ where $Z^{\epsilon}_{(1,1)}$ is defined in eq. (\ref{z11}). 
Similarly $Z_{(1,g^n)}=\sum_{\epsilon=0}^1\hat{Z}^{\epsilon}_{(1,g^n)} Z^{\epsilon}_{(1,g^n)}$ with $Z^{\epsilon}_{(1,g)}$ defined in eq. (\ref{z1gf1}). 
The rest of the sectors can be obtained by modular transformations so that $Z_{(g^m,g^n)}=\sum_{\epsilon=0}^1\hat{Z}^{\epsilon}_{(g^m,g^n)} Z^{\epsilon}_{(g^m,g^n)}$ which would give a modular invariant theory if the level matching  condition is satisfied.

In the first formulation, where $g$ is realized as a shift $v$ equal to half of the fundamental weight in the $SU(2)$ part, the resulting theory will have a moduli space given by the deformation of the lattices $(s,2)$ and $(19-s,1)$. 
On the other hand, in the second formulation, since $g$ acts as reflection in the $SU(2)$ part, the moduli space will be given by the  deformations of the lattices $(s+1,2)$ and $(18-s,1)$. 
Thus, starting from a consistent orbifold model for a given $s$, by going to a point with  $SU(2)$ enhancement, we can move to a different branch of the moduli space with $s$ shifted by $1$.

In section \ref{section_orbifold}, we will describe the general construction for all $s$, without referring to $SU(2)$ enhanced points, 
but the method is inspired by the discussion above.

\section{Partition function and spectrum of asymmetric Nikulin orbifolds}\label{section_orbifold}

The starting point is the 10-dimensional supersymmetric heterotic string.
In our conventions the left-moving sector is the bosonic string while the right-moving sector is
the superstring. For the right-moving worldsheet fermions we will use
the description in terms of $SO(8)$ bosons and take the GSO projection to select weights in
the vector class ($V$) and the spinor class ($Sp$). 

The next step is to compactify on $\bt^3$ and mod out by an action that involves a rotation
by $\pi$ of $s$ left-moving and two right-moving directions. Since the action on left and right movers is different,
we will apply the formalism of asymmetric orbifolds \cite{Narain:1986qm, Narain:1990mw}. 
To this end we need to specify the action on the left and right-moving momenta living on the even self-dual lattice
$\ktl$ of the $\bt^3$ compactification. This is precisely the involution $\theta$,
characterized by the triples $(r,a,\delta)$ in Figure \ref{figureNikulin}, as explained in section \ref{section_nikulin}.

We will begin by specifying the orbifold generator in subsection \ref{subsec_generator}.
We will then write down the partition function and determine the conditions for modular invariance in subsection \ref{subsec_pf}. 
We will also verify the validity of the operator interpretation. 
In subsection \ref{subsec_LMC} we will solve the level matching condition and derive the constraints it puts on the moduli space of the invariant lattices.
Finally, in subsection \ref{subsec:spectrum} we will discuss generic features of the spectrum of tachyonic and massless states and will also present some examples.

\subsection{Orbifold generator}\label{subsec_generator}

In order to construct the heterotic asymmetric orbifolds based on Nikulin involutions we first need to define how
the orbifold generator acts on the worldsheet. The action on $\ktl$ includes the involution $\theta$ and in general also
a lattice shift. In fact, we will find that a shift is required by modular invariance.

To be more precise let us denote the orbifold generator by $g$. The action on the worldsheet fermions is
\be\label{defvf}
g: |p \rangle \to e^{-2\pi i p \cdot v_f} |p \rangle\ ,
\ee
where $p$ is an $SO(8)$ weight and $v_f=(\frac12,0,0,0)$.
Notice that $g^2$ acts by $(-1)$ on space-time fermions that have $p \in Sp$. In other words, $g^2$ acts as
$(-1)^{F_S}$, where $F_S$ is space-time fermion number. The generator $g$ also reflects $s$ left and two right-moving bosonic
oscillators.  

The action of $g$ on the $\ktl$ lattice momenta is most conveniently given in terms of the projections $(P_N,P_I)$ 
along normal and invariant directions that transform as $\theta(P_N,P_I)=(-P_N, P_I)$. We then define
\begin{equation}
  g: |P_N,P_I\rangle  \to f(P_N)e^{2\pi i P_I.v} |-P_N,P_I\rangle\ .
  \label{defg1}
\end{equation}
Here $v$ is a constant shift that can be taken along the invariant directions without loss of generality.
Moreover, we impose $4 v \in I$ to also have $g^4$ acting as the identity on $\ktl$.
The additional $\bz_4$ phase $f(P_N)$ is such that  $f(0)=1$ and further satisfies
\be\label{defg2}   
  f(P_N)f(-P_N) = \left\{\!\!
\begin{array}{ll}
\phantom{-}1 & {\text{if}}\ \, P_N \in N^*_e \\[1mm]
-1 & {\text{if}}\ \, P_N \in N^*_o
\end{array}
\right.\ ,
\ee
where $N^*_e$ and $N^*_o$ were defined in \eqref{nenodef}. In fact we can shorten the above equation as
\begin{equation}
  f(P_N)f(-P_N)= e^{2\pi i P_N^2} =  e^{2\pi i P_I^2}\ ,
  \label{defg3}
\end{equation}
where we used the fact that $P_N^2+P_I^2=P^2$ is even.

The definition of the phase $f$ is inspired by the $SU(2)$ discussion in section \ref{section_cft}. $f$ itself is basis dependent 
since the $g$ action takes one state to another and
therefore it depends on the relative normalizations of the two states. But $g^2$ takes a state to itself and in the $g^2$ action the phase that appears is the combination $f(P_N)f(-P_N)$.
The prescription is therefore basis independent and it basically says that, setting $v=0$,  if $ P_N \in N^*_e$ then $g^2=1$ and 
if $P_N \in N^*_o$ then $g^2=-1$.
This is analogous to the $SU(2)$ discussion where $g^2=1$ on the $SU(2)$ root lattice (whose length squares are all integers - actually even integers, and therefore they are part of $N^*_e$ in our present notation) 
while $g^2=-1$ on the conjugacy class of the fundamental representation (whose length squares are all 1/2 mod integers and are therefore  part of $N^*_o$).
In the case of $SU(2)$, $g$ is a $\bz_4$ element in the global $SU(2)$ subgroup generated by zero modes of the Kac-Moody currents $J^a_0$ and therefore it is an exact symmetry of the theory. 
In particular it means that $g$ is also a symmetry of the OPE. This can be generalized to other level one Kac-Moody algebras, e.g. $SO(4)$ which can be written as $SU(2)\times SU(2)$ and has representations $Sc=(R,R)$, $V=(F,F)$, $Sp=(F,R)$ and $Sp'=(R,F)$,
where $R$ and $F$ are the adjoint and fundamental representations of $SU(2)$. If in both $SU(2)$ factors
we define the action of $g$ in identical way then $g^2$ acting on $Sc$ and $V$ gives $+1$, while acting on $Sp$ and $Sp'$ gives $(-1)$. Indeed this is consistent with our prescription above since $Sc$ and $V$ weights have integer length square and are therefore 
in $N^*_e$, while $Sp$ and $Sp'$ weights have half-integer length square and belong in $N^*_o$. This is also true for any $SO(8n+4)$. On the other hand, for $SO(8n)$ all the four classes have integer length square and therefore $g^2$ acts as the identity on the entire weight lattice.

The important point  is that when there is an underlying Kac-Moody algebra, the phase $f$ can be realized as an action by a  symmetry generated by the zero modes of the Kac-Moody current, so it is an
exact symmetry of the CFT. In particular this means that OPE's will also respect this symmetry. In the present case, we are discussing an arbitrary $\ktl$ and the corresponding
$N^*$, which need not have any non-abelian Kac-Moody algebra (another way to say this is that by boosting the lattices we may have broken all the non-abelian symmetries). In this case it is
not obvious that the above definition of $g$ involving $f$ is an exact symmetry of the CFT and would respect the OPE's.
However, we can still argue that we could have started from the non-abelian points, do the orbifolding there, including $f$, and then we could deform those points by boosting via marginal operators that are $g$-invariant and therefore would not break the $g$-symmetry.

Having defined the action of $g$ we now turn to writing down the partition function for the asymmetric orbifold.

\subsection{Partition function}\label{subsec_pf}

Since the orbifold generator $g$ has order 4, the full partition function $Z$ takes the form
\be\label{fullZ}
Z=\frac{1}{4}\sum_{m=0}^{3} \sum_{n=0}^{3}\int_{\mathcal F}  \frac{d^2\t} {\tau_2^2} \, Z_{m,n}\ ,
\ee
where $\cal F$ is the $SL(2,\bz)$ fundamental domain.
The sum in $m$ is over twisted sectors while the sum in $n$ enforces the projection over the $g$ action.
In operator language, $\int\! \frac{d^2\tau}{\tau_2^2} Z_{m,n}$ corresponds to
${\rm Tr}_{{\cal H}_m} g^n q^{L_0} \bar{q}^{\bar{L}_0}$, where ${\cal H}_m$ is the $g^m$  twisted Hilbert space. 
The untwisted sector contributions $Z_{0,n}$ can be calculated by inserting in the trace the known action of
$g^n$ on the Hilbert space ${\cal H}_0$.  
The remaining $Z_{m,n}$ can be derived by applying the transformations $S: \tau \rightarrow-\tf{1}{\tau}$ and  
$T: \tau \rightarrow{\tau+1}$. In the end we have to check that the operator interpretation is valid.

For $m=n=0$ we just have the toroidal partition function
\be\label{Z00}
  Z_{0,0}=\frac{1}{\tau_2^{\frac{5}{2}}\eta^{24}\bar{\eta}^{12}}
              \left(\sum_{p \in V}-\sum_{p \in Sp}\right)\opm \bar q^{\frac12 p^2} \,
              \sum_{P\in \Gamma} q^{\frac{1}{2}P_L^2}\bar{q}^{\frac{1}{2}P_R^2}\ ,
\ee
where we defined $\Gamma = \ktl$ to simplify expressions. The left and right components of $P \in \Gamma$ are denoted
$P_L$ and $P_R$ respectively. Our convention for the Lorentzian metric is $P^2=P_L^2-P_R^2$.
For definitions of Dedekind and Jacobi functions used below we refer to \cite[appendix A.1]{Acharya:2020hsc}.

Let us now consider the sector $(1,g)$. Inserting $g$ in the trace will only allow states in the invariant lattice $I$, i.e. it annihilates all 
states with $P_N \neq 0$. Since $f(P_N)=1$ for $P_N=0$, the phase $f$ will not affect the $(1,g)$ sector, nor any
of the sectors that are in the same modular orbit as $(1,g)$. 
The partition function for the $(1,g^{2n+1})$ sector is then
\begin{equation} \label{Z01}    
\begin{split}
  Z_{0,2n+1}=&\frac{1}{\tau_2^{\frac{5}{2}}\eta^{24}\bar{\eta}^{12}}\!
            \left(\frac{2\eta^3}{\vartheta_2}\right)^{\!\! s/2} \!\! \left(\frac{2\bar{\eta}^3}{\bar{\vartheta}_2}\right) \\[2mm]
            &\times \,   \left(\sum_{p \in V}-\sum_{p \in Sp}\right)\opm \bar q^{\frac12 p^2}e^{-2\pi  i p\cdot (2n+1)  v_f} \, 
              \sum_{P\in I} q^{\frac{1}{2}P_L^2}\bar{q}^{\frac{1}{2}P_R^2} e^{2\pi iP\cdot  (2n+1) v}
 \end{split}        
\end{equation}                 
where $n=0,1$. Note that $Z_{0,4+k}=Z_{0,k}$ and $Z_{0,1}=Z_{0,3}$. 

In the $(1,g^2)$ sector there is an important effect due to the phase $f$. We now have
\begin{equation}
  Z_{0,2}=\frac{1}{\tau_2^{\frac{5}{2}}\eta^{24}\bar{\eta}^{12}}
  \left(\sum_{p \in V}-\sum_{p \in Sp}\right)\opm\bar q^{\frac12 p^2}e^{-2\pi i p\cdot 2 v_f}
  \sum_{P\in \Gamma} e^{2\pi i P\cdot 2v}e^{2\pi i P_I^2} q^{\frac{1}{2}P_L^2}
  \bar{q}^{\frac{1}{2}P_R^2}\ , 
  \label{Z02}
  \end{equation}
where the extra term $e^{2\pi i P_I^2}$ is $f(P_N)f(-P_N)$ using \eqref{defg3}. 
The sectors $(g^2,1)$ and $(g^2,g^2)$, that are in the same modular orbit as $(1,g^2)$, will also feel the 
effect of the phase $f$.

The partition functions $Z_{1,n}$ for $g$-twisted sectors are obtained from $Z_{0,1}$ applying first $S$ and then $T$ transformations.
In this way we find
\be \label{Z1n}
\begin{split}
  Z_{1,n}=&\frac{1}{\tau_2^{\frac{5}{2}}\eta^{24}\bar{\eta}^{12}} \frac{1}{\sqrt{[I^*/I]}}
  \! \left(\frac{2\eta^3}{\vartheta(n)}\right)^{\!\! s/2}\!\! \left(\frac{2\bar{\eta}^3}{\bar{\vartheta}(n)}\right)
e^{\pi i  n \left(\frac{s-2}{8}-1\right)} \\[2mm]
&\times \,    \left(\sum_{p \in V}-\sum_{p \in Sp}\right)\opm 
\bar q^{\frac12(p+v_f)^2}e^{-\pi i  n(p+v_f)^2}
    \sum_{P\in I^*}e^{\pi i n(P+v)^2} q^{\frac{1}{2}(P+v)_L^2}\bar{q}^{\frac{1}{2}(P+v)_R^2}\ ,
 \end{split}  
\ee
where $\vartheta(n)$ stands for $\vartheta_4$ for even $n$ and $\vartheta_3$ for odd $n$.
The phase in the first line arises from the $T$ transformation of oscillator modes, i.e. from $\eta$ and $\bar{\eta}$. 
The phases in the second line come instead from the $SO(8)$ and $I$ lattice modes.
This clearly has a well defined operator interpretation for the action of $g^n$ on the $g$-twisted Hilbert space provided
that for $n=4$ the overall phase is one. Imposing the constraint 
\begin{equation}
  e^{2\pi i\left(\frac{s-2}{4}-2(p+v_f)^2+2(P+v)^2\right)} = 1
  \label{level1}
\end{equation}
gives the level matching condition
\be\label{LMC}
2 v^2 +\frac{s}{4}  \in {\bz}
\ee
after using that $p^2$, $4 p.v_f$, $4 v_f^2$, $4P.v$ and $2P^2$ are all integers, the last two facts following from 
$4v\in I$, $P\in I^*$ and $2I^*\subset  I$.

To evaluate $Z_{2,0}$ we can start from $Z_{0,2}$ given in eq.\eqref{Z02} and do an $S$ transformation but  the presence of the 
phase $e^{2\pi i P_I^2}$ in $Z_{0,2}$ complicates Poisson resummation. This difficulty can be overcome by
rewriting this phase in a way which simplifies Poisson resummation and makes the $g^2$-twisted spectrum more transparent. 
To this end, first observe that this phase basically gives $(+1)$ for $P_I \in I^*_e$ and $(-1)$ for $P_I \in I^*_o$.  If $I^*_o$ 
is null, which occurs when $\delta=0$, then this phase is always 1 and Poisson resummation can be easily done. The complication 
arises when $I^*_o$ is not null. In this case, we can ask if there exists a vector $w$ along the invariant directions such that 
$w \cdot P_I$  is an integer for all $P_I \in I^*_e$ and is a half integer for all  $P_I\in I^*_o$. 
If such a $w$ exists then we can write  
\be\label{wcond}
e^{2\pi i P_I^2} = e^{2\pi i P_I \cdot w} \ \ \forall \ \ P_I   \in I^*
\ee
and Poisson resummation becomes easy because the phase is linear in $P_I$.

We now address the questions whether $w$ satisfying condition \eqref{wcond} exists, and how it could be determined.
Similar to $v$ we choose $w$  purely along $I$ directions.
The condition  $w \cdot P_I \in \bz$ for all $P_I \in I^*_e$ implies that $w \in (I^*_e)^*$. 
Now $I^*_e \subset I^*$, therefore $(I^*_e)^* \supset I$. Since we are assuming that $I^*_o$ is not null,
these latter relations mean proper subsets, i.e. $(I^*_e)^* \neq I$. 
We can take $w$ to be a non-trivial element in  $(I^*_e)^* / I$, i.e. $w\in (I^*_e)^*$ but 
$w \notin I$. If $P_I \in I^*_o$ then $P_I \cdot w$ cannot be integer,
because if it was then $w$ dotted with any vector of $I^*_o$ would be integer since any vector of $I^*_o$ can be expressed as 
$P_I$ plus an element of $I^*_e$ (as follows from the discussion at the end of section \ref{section_nikulin}).  That would mean that
$w$ dotted with any vector of $I^*$ is an integer but that would imply that $w \in I$, a contradiction. 
We also know that  $2 P_I \in I$ which implies that $2 P_I \in I^*_e$ and therefore  $2 P_I\cdot w$ is an integer. This proves
that $w\cdot P_I$ is a half integer for all $P_I \in I^*_o$. Note also that $I^*_e \supset I$, therefore $(I^*_e)^* \subset I^*$. This means that $w$ is a nontrivial element of $I^*/I$. 

Let us consider some examples to illustrate the above discussion. Suppose that $I=U+D_{4n+2}$ and neglect $U$ because 
it is self-dual. Then, in terms of $SO(8n+4)$ classes, $I^*_e$ is formed by scalar
$Sc$ and vector $V$ classes,  while $I^*_o$ includes the spinor classes $Sp$ and $Sp'$. 
In this case $(I^*_e)^*= I^*_e$ and $w$ is in the $V$ conjugacy class. If $I=U+A_1$,  then $I^*_e =I$ and $w$ is an element of the fundamental class of $SU(2)$.

The great advantage of introducing $w$ is that $Z_{0,2}$ in eq.\eqref{Z02} can be rewritten as 
\begin{equation}
 Z_{0,2}=\frac{1}{\tau_2^{\frac{5}{2}}\eta^{24}\bar{\eta}^{12}}
  \left(\sum_{p \in V}-\sum_{p \in Sp}\right)\opm\bar q^{\frac12 p^2} e^{-2\pi i p\cdot 2 v_f}
 \sum_{P\in \Gamma} e^{2\pi i P \cdot (2v+w)} q^{\frac{1}{2}P_L^2}
 \bar{q}^{\frac{1}{2}P_R^2}\ .
 \label{Z02p}
  \end{equation}  
The $S$ transformation can then be easily performed to obtain
\be\label{Z20}
 Z_{2,0}=  \frac{1}{\tau_2^{\frac{5}{2}}\eta^{24}\bar{\eta}^{12}}
  \left(\sum_{p \in V}-\sum_{p \in Sp}\right)\opm\bar q^{\frac12(p+2v_f)^2}
  \sum_{P\in \Gamma} q^{\frac{1}{2}(P+2v+w)_L^2}
            \bar{q}^{\frac{1}{2}(P+2v+w)_R^2}\ .
\ee  
Applying $T$ transformations then yields
\be\label{Z22n}
\begin{split}
 Z_{2,2n}= &\frac{1}{\tau_2^{\frac{5}{2}}\eta^{24}\bar{\eta}^{12}}
  \left(\sum_{p \in V}-\sum_{p \in Sp}\right)\opm\bar q^{\frac12(p+2v_f)^2} e^{-2\pi i n(p\cdot 2v_f)}\\[2mm]
&\times \,   \sum_{P\in \Gamma} q^{\frac{1}{2}(P+2v+w)_L^2}
        \bar{q}^{\frac{1}{2}(P+2v+w)_R^2}e^{2\pi i n P\cdot (2v+w)}e^{\pi i n (1+ (2v+w)^2)}
\end{split}
\ee    
after some rearrangements.
      
The level matching condition in the $g^2$-sector is $Z_{2,0}=Z_{2,4}$. From \eqref{Z22n} this implies that $(2v+w)^2$ must be an integer. 
Now, we already found from the level matching condition (\ref{LMC}) in the $g$ sector that $2v^2=-\frac{s}4 \, {\mathrm{mod}}\, \bz$,
and we also know that $4v\cdot w \in \bz$ as $4v \in I$ and $w\in I^*$. Hence, the condition 
$(2v+w)^2 \in \bz$ amounts to
\begin{equation}
    w^2+\frac{s-2}{2} \in \bz
\label{level2}  
\end{equation}
It is instructive to look at some examples to verify that the $w$'s determined according to our previous analysis
do satisfy this condition. Consider first $I=U+E_7$, with $s=11$. Then, omitting the self-dual $U$ for simplicity,
$I^*$ is the weight lattice of $E_7$ composed of the adjoint class (i.e. the root lattice itself) plus the
class of the fundamental ${\bf{ 56}}$, dubbed $F$. Hence, $I^*_e=E_7$, $I^*_o=F$, and $w$ can be chosen to
be a weight of the ${\bf{ 56}}$ with $w^2=\frac32$ so that \eqref{level2} holds.
For an example with $s$ even, take $I=U+D_6$. In this case we already argued that $w$ is in the $V$ class of $SO(12)$,
therefore $w^2 \in \bz$ and \eqref{level2} is again fulfilled. In fact, notice that 
$(w^2+\frac{s-2}{2})$  is actually an even integer in both examples. We will show shortly that $w$ satisfies this
latter condition by construction.
 
We still need to show that the operator interpretation is valid, namely we must prove that the
actions of $g$ deduced from $Z_{2,1}$ and $Z_{2,2}$ are mutually consistent.
In fact, our first problem is whether $Z_{2,1}$ can be understood as ${\mathrm {Tr}}\, g$ in the $g^2$ twisted Hilbert space 
that can be deduced from $Z_{2,0}$. In order to obtain $Z_{2,1}$  we start from $Z_{1,2}$ in eq. \eqref{Z1n}
and perform an $S$ transformation to obtain
\be\label{Z21}
\begin{split}
  Z_{2,-1}=&\frac{1}{\tau_2^{\frac{5}{2}}\eta^{24}\bar{\eta}^{12}}
\left(\frac{2\eta^3}{\vartheta_2}\right)^{\!\! s/2}\!\! \left(\frac{2\bar{\eta}^3}{\bar{\vartheta_2}}\right)
e^{2\pi i  ( \frac{s-4}{8}+2v_f^2-v^2-v.w)} \\[2mm]
& \times \,   \left(\sum_{p \in V}-\sum_{p \in Sp}\right)\opm\bar q^{\frac12(p+2v_f)^2} e^{2\pi i   p\cdot v_f}
          \,    \sum_{P\in I}e^{-2\pi i P\cdot v} q^{\frac{1}{2}(P+2v+w)_L^2}\bar{q}^{\frac{1}{2}(P+2v+w)_R^2}.
\end{split}
\ee
To arrive at this result, in $Z_{1,2}$ we used that $e^{2\pi i P^2} = e^{2\pi i P.w}$ for all $P \in I^*$, in order to
do directly the Poisson resummation.

By comparing the above $Z_{2,-1}$ with $Z_{2,0}$ in eq. \eqref{Z20}, we see that $Z_{2,-1}$ can be obtained inserting 
$g^{-1}$ in the trace over the $g^2$-twisted Hilbert space, where $g$ acts as a rotation along $N$ directions. 
The reason is that only $P_N=0$ will contribute in the trace and as a result the lattice states that will survive are precisely 
$|P+2v+w\rangle$, where $P\in I$. Thus,  we pass the basic test that in $Z_{2,m}$ the $g^m$ are all playing on the same ground, 
i.e. they all act on the spectrum that appears in $Z_{2,0}$. 
Now we come to the more detailed question of matching the phases. 
From the phases in $Z_{2,2}$ and $Z_{2,-1}$, cf. eqs. \eqref{Z22n} and \eqref{Z21},
we deduce the action of $g^2$ and $g^{-1}$ on the momenta to be
\begin{align} 
 g^2 | p+2v_f,  P+2v+w \rangle &=e^{2\pi i  (\frac{1+w^2}{2}+ 2v.w+2v^2)}e^{-4\pi i (p+2v_f).v_f} 
e^{2\pi i P.(2v+w)}| p+2v_f,  P+2v+w \rangle ,
\nonumber  \\[2mm] 
g^{-1} | p+2v_f,  P+2v+w \rangle &= e^{2\pi i  ( \frac{s-4}{8}-v^2-v.w)} e^{2\pi i   (p+2v_f).v_f}e^{-2\pi i P.v}
  | p+2v_f,  P+2v+w \rangle .
\label{ging2}
\end{align}  
Consistency then requires that the phase in the first line times the square of the phase in the second line must be one. 
Using $v_f^2=-\frac14$, and the fact that $P \cdot w \in \bz$ for all $P\in I$ as $w \in I^*$, leads to the condition
\begin{equation}
w^2+\frac{s-2}{2} \in 2 \bz
\label{21operatorint}
\end{equation}
Notice that this is a stronger constraint than the level matching condition in the $g^2$-twisted sector, 
i.e. eq.~\eqref{21operatorint} implies eq.~\eqref{level2} but not vice versa.

Interestingly enough, condition \eqref{21operatorint} always holds.
It can be shown from the identity
\be\label{milgram}
C=\sum_{P\in I^*/I} e^{i\pi P^2} = \sqrt{\left|I^*/I\right|} e^{-i\pi\frac{(s-2)}{4}}
\ee
which follows from Milgram's Theorem proven e.g. in Appendix 4 of \cite{Milnor73}.
Since $w \in I^*$, the sum $C$ can as well be written as
\be
C=  \sum_{P \in I^*/I}e^{\pi i (P-w)^2} 
= e^{i\pi w^2} \sum_{P \in I^*/I}e^{-i \pi  P^2} =\bar{C} e^{i \pi w^2}
\ee
where in the second equality we used that $e^{-2\pi i P \cdot w}=e^{-2\pi i P^2}$ for all $P \in I^*$. 
Substituting  \eqref{milgram} then gives
\begin{equation}
  e^{-i\pi \frac{(s-2)}{2}} = e^{i\pi w^2} .
\end{equation}
This proves that $w$ satisfies eq.~\eqref{21operatorint} by construction.
Indeed, we have previously explained, and provided examples, how we can always find $w \in I^*/I$ such that
$w^2= (1 -\frac{s}2)\, {\rm{mod}}\, 2$.

Summarizing, in this section we have shown that imposing level matching in the $g$-twisted sector is sufficient to ensure a modular invariant theory endowed with a consistent operator interpretation.

\subsection{Solving the level matching condition and connecting different shifts by boosting}\label{subsec_LMC}

In order to analyze the spectrum of tachyonic and massless states for a given triple $(r,a,\delta)$ we need to specify the
lattice $I$ and choose a shift $v$ satisfying the level matching condition \eqref{LMC}. Actually, for two lattices
of small rank there is no solution of this condition. These correspond to the triples $(1,1,1)$ and $(2,2,1)$. 
In the case $(2,2,1)$, the generic invariant lattice can be obtained applying a boost in $SO(1,1)$ to a vector of $A_1 + A_1(-1)$, 
namely to $(\sqrt2 \ell_1; \sqrt2 \ell_2)$, $\ell_1, \ell_2 \in \bz$.
A generic shift $v$, with $4v \in I$, then has $v^2=\frac18(\ell_1^2- \ell_2^2)$ and \eqref{LMC} cannot
be fulfilled for $s=18$. The same occurs for the triple $(1,1,1)$ having $I=A_1(-1)$, $v=\sqrt2 \ell/4$, $v^2=-\ell^2/8$, and $s=19$.
For other triples there are always solutions. In the following we will classify the $v$'s that solve the level matching condition.

Before proceeeding let us clarify the problem.
In general we begin by fixing $I$.
Boosting $I$, i.e. applying a transformation in $SO(r-1,1)$,  will change $I$ continuosly. 
But $I$ could be invariant for some discrete boosts which are therefore
part of the $T$-duality group $\cT(I)$ that leaves $I$ invariant. 
An element $\cE \in \cT(I)$ maps $I$ to $I$, i.e. for all $P \in I$, $\cE(P) \in I$.
Consider now a shift $v$, with $4 v \in I$, and satisfying level matching.
A discrete boost could change $v$, but not its squared norm. Thus, $\cE(v)$ also satisfies level matching
and defines the same theory as $v$.
Note however that $\cE(v)$ and $v$ do not give the same spectrum.
If a state $|P\rangle$ is invariant under $v$, i.e. $e^{2\pi i v\cdot P}=1$, then
$|\cE(P)\rangle$ is invariant under $\cE(v)$ because  $\cE(v)\cdot \cE(P) = v\cdot P$.
But in general $\cE(P)$ is not invariant under $v$ since  $v \cdot \cE(P)$ needs not be integer.
Moreover, $|P\rangle$ and $|\cE(P)\rangle$ could have different mass because $\cE$ involves Lorentz
boosting. Nonetheless, the point is that $v$ and $\cE(v)$ define the same moduli space. This is in the same sense as 
the statement that for given $(r,a,\delta)$ there is a unique $I$. Clearly, for specific $(r,a,\delta)$ we can construct two different 
invariant lattices, say $I_1$ and $I_2$, which lead to different spectra, but the uniqueness statement means that by boosting 
$I_1$ we can go to $I_2$.  

The actual T-duality group is a subgroup of $\cT(I)$ that leaves the spectrum invariant. From the partition function we see 
that two shifts differing by a sign or by an element of $I^*$ give the same spectrum. This gives the equivalence relation
\be\label{vequiv}
v \sim v' =\pm v\,  \mathrm{mod}\,  I^* .
\ee
Hence, the actual T-duality group, denoted $\cT(I,v)$, maps $I$ to $I$ and at the same time takes 
$v$ to $\pm v\,  \mathrm{mod}\,  I^*$. 
The question is then which of the allowed shifts can be obtained by elements of $\cT(I)/\cT(I,v)$. 

We will first tackle this question when the invariant lattice has the form $I\simeq U + K$, where $K$ is an even lattice with signature
$(r-2,0)$ and $K^*/K=\bz_2^a$. Such $I$'s occur for $r\ge a+2$, with the exception of $I\simeq U(2) + D_4$
for the triple $(6,4,0)$ \cite{Nikulin83}. Afterwards we will treat lattices comprising $U(2)$ that also appear when $r=a$.

\subsubsection{$I\simeq U + K$}\label{subsubsec_UK}

The lattice $I$ of signature $(r-1,1)$ has $(r-1)$ moduli, which can be
taken to be a radius $R$ plus a \mbox{$(r-2)$-dimensional} Wilson line $A$,
in analogy with the $\Gamma_{\!\!(17,1)}$ lattice for heterotic compactification on a circle. Moreover, we can write the 
lattice vectors in a similar way, i.e. $P \in I$ has components $(\rho, p_L;p_R)$ given by
\be\label{momuk}
\begin{split}
\rho &= \gamma + 2 j A, \\
p_L &=  \frac1{R}\left(\frac{k}{2} + (R^2 - A^2) j - \gamma\cdot A\right), \\
p_R &= \frac{1}{R}\left (\frac{k}2 - (R^2 + A^2) j - \gamma\cdot A\right),
\end{split}
\ee
where $j, k \in \bz$, and $\gamma \in K$. It is easy to check that $P^2=2 j k + \gamma^2$.

The automorphisms of $I$ include circle T-duality, i.e. exchange of winding and quantized momentum $j \leftrightarrow k$
while $\gamma \to \gamma$. Another element $\cE_\beta$ of $\cT(I)$ is the translation of the Wilson line $A \to A +\frac12 \beta$, 
where $\beta \in K$. Its action on the quantum numbers $(j, k, \gamma)$ is 
\be\label{ebeta}
\cE_\beta:\ (j, k, \gamma) \to (j',k',\gamma')=(j, k+\gamma\cdot \beta - \frac12 \beta^2 j, \gamma - j\beta) .
\ee
Transformations $(j, k, \gamma) \to (j,k,{\mathcal W}_K(\gamma))$, where ${\mathcal W}_K$ belongs to the Weyl group of $K$,
are also in $\cT(I)$. An extended discussion of the automorphism group of $I$ is beyond the scope of this paper.
This problem has been addressed in \cite{Nikulin83, AlexeevNikulin} and below we will refer to some known results.

For simplicity we will work at a point in moduli space with $A=0$. This means that $I$ is a direct sum, i.e. $I=U+K$.
Following the convention we set in eq. \eqref{momuk}, the shift $v$ with components $(V, v_L;v_R)$ can be written as
\be\label{shiftuk}
v =\frac14\left(\alpha,\frac{n}{2R} + m R; \frac{n}{2R} - m R\right) ,
\ee
where $\alpha \in K$ and $n, m$ are integers mod 4, i.e. they can be chosen as $-1,0,1,2$. 
We will write $v=\frac14(m,n,\alpha)$  in shorthand. 
The level matching condition \eqref{LMC} reads
\begin{equation}\label{lmc_mn}
 m n+\frac{\alpha^2}{2} = -s~{\rm mod}~ 4 .
\end{equation}
We want to classify the solutions. To this purpose we notice that the choices 
\be\label{listmn}
(m,n) = (0,0), (1,0), (1,1), (1,-1), (0,2),(1,2),(2,2) 
\ee
are exhaustive. To obtain this minimal list we used that $(m,n)$ is equivalent to $(-m,-n)$ because $I$ is a direct sum of
$U$ and $K$ and $v$ is equivalent to $-v$.  Furthermore by $R \rightarrow 1/(2R)$ we can exchange $n$ and $m$.
Often we will also resort to acting on $v$ with the automorphism $\cE_\beta$, for some suitably chosen $\beta$. 
According to \eqref{ebeta} $v$ transforms as
\begin{equation}\label{vbeta}
 v=\frac{1}{4}(m,n,\alpha) \rightarrow v'=\frac{1}{4}(m',n',\alpha') = 
 \frac{1}{4}(m,n+\alpha.\beta-\frac{1}{2} \beta^2 m, \alpha-m \beta)
\end{equation}
under $\cE_\beta$. 
In general, clearly $v'$ is not in the same equivalence class as $v$. For example if $m$ is odd, and $\beta $ is a root 
(i.e. $\beta^2=2$) then $\frac14 \beta$ is obviously not in $K^*$. It is convenient to distinguish two cases depending on the 
values of $(m,n)$ as discussed below.
We will assume that $v$ is order $4$ and not order $2$, i.e. $2v \notin U+K$, nor order $1$. 
\vskip 0.1in
\noindent{\bf Case 1}
\vskip 0.1in
\noindent Suppose both $n$ and $m$ are not simultaneously even, i.e. at least one of them is odd. 
This means that the part of $v$ which is in $U$ is already of order 4.
As explained above in this case we can always take $m=1$, if necessary exchanging $n$ and $m$. 
Then the level matching condition (\ref{lmc_mn}) says that $n=-\frac12\alpha^2 -s\, {\mathrm{mod}}\, 4$. Now,
under the action of $\cE_\beta$, with $\beta= \alpha-\alpha'$, the shift $v=\frac{1}{4}(1,n,\alpha)$ goes over to 
$v'=\frac{1}{4}(1,n',\alpha')$ with $n'=-\frac12\alpha'^2 - s\, {\mathrm{mod}}\, 4$.
Thus, all the $v$'s of the form $\frac{1}{4}(1,n,\alpha)$ for arbitrary $\alpha \in K$ and $n$ satisfying level matching 
are in the same moduli space.
\vskip 0.1in
\noindent{\bf Case 2}
\vskip 0.1in
\noindent Suppose both $m$ and $n$ are even. Now the part of $v$ in $U$ is of order 1 or 2.  This can happen only when
$s+\frac{\alpha^2}{2} =0~ {\mathrm{mod}}~ 4$. By exchanging $m$ and $n$ if necessary
we can always set $(m,n)$ to be equal to $(0,0)$ or $(0,2)$ or $(2,2)$. Since we are assuming that $v$ is of order $4$, 
we have that $\frac12 \alpha \notin K$. Let us again consider the transformation induced by $\cE_\beta$, cf. eq.~\eqref{vbeta}. 
Now there are two possibilities:
\begin{trivlist}
\item[a)]There exists a $\beta \in K$ such that $\alpha\cdot\beta$ is odd (this is not always the case, for example if $K$ is 
a direct sum of several $A_1$'s such $\beta$ does not exist).
Since $n$ and $\frac{m}{2}\beta^2$ are even
and $\alpha\cdot\beta$ is odd, the new $n'$ is odd and therefore equal to $\pm 1$ mod 4. This means that we have moved to Case 1.
\item[b)] There is no $\beta \in K$ such that $\alpha\cdot \beta$ is odd. Although we cannot get to Case 1 at once, 
we can still move among $(m,n)=(0,0), (0,2), (2,2)$. For example, suppose $m=0$ and 
$\alpha\cdot\beta=2\,{\mathrm{mod}}\, 4$, then $\frac14(0,n,\alpha) \rightarrow \frac14(0,n+2,\alpha)$.
If $m=2$ and $K$ admits a $\beta$ such that $\alpha\cdot\beta-\beta^2 =2\,{\mathrm{mod}}\, 4$ then we can get 
$\frac14(2,n,\alpha) \rightarrow \frac14(2,n+2,\alpha-2\beta)$. However this has to be studied case by case, also
allowing for generic products of T-duality and Wilson line translations.
We have verified that when $\beta \in K$ with $\alpha\cdot \beta \in 2\bz +1$ does not exist,
even such generic transformations cannot connect solutions $\frac14(m,n,\alpha)$ 
in which both $m$ and $n$ are even, to solutions in which one of them is odd.
We expect this result to be valid when considering more general automorphisms.
In any event, the fact that we cannot go to Case 1 means that there exist shifts which are 
not in the same moduli space.

\end{trivlist}

Let us illustrate the above discussion with some examples. 

\begin{trivlist}
\item[1.]  $I= U + A_1$, $s=17$. 
\vskip0.1in
\noindent
The generic shift is $v=\frac14(m,n,\sqrt2 \ell)$. It suffices to consider $\ell=0,1,2$, and the values of $(m,n)$ in
the list \eqref{listmn}. The level matching condition (\ref{lmc_mn}) is $(nm + \ell^2)=-1\, {\mathrm{mod}}\, 4$. The solutions are
\be\label{vsolua1}
v_1=\frac14(1,2,\sqrt2), \quad  v_2=\frac14(1,-1,0), \quad v_3=\frac14(1,-1,2\sqrt2). 
\ee
All belong to Case 1. Thus, they are in the same moduli space. Moreover, $v_3$ and $v_2$ are equivalent because they 
differ by a vector in $I^*$. 
\vskip0.1in
\item[2.] $I = U + A_1^3$, $s=15$.
\vskip0.1in
\noindent
With $v=\frac14(m,n,\sqrt2 \ell_1, \sqrt2 \ell_2, \sqrt2 \ell_3)$, the level matching condition is
$(nm+\ell_1^2 + \ell_2^2 + \ell_3^2)=1\, {\mathrm{mod}}\, 4$.
If $n$ and $m$ are both even, a solution is to take one of the $\ell_i=1$ and the remaining zero, say
$\alpha=(\sqrt2, 0,0)$. Clearly we are in case 2b because there is no $\beta \in A_1^3$ such that $\beta \cdot \alpha$ is odd. 
The statement is that by turning on discrete Wilson lines it is not possible to connect shifts 
with at least one of $n$ or $m$ odd, e.g. $v_1=\frac14(1,1,0,0,0)$,
to shifts with both $n$ and $m$ even such as $v_2=\frac14(2,2,\sqrt2,0,0)$. 

The full automorphism group of $I=U+ A_1^3$ is generated by Weyl reflections about a fundamental set of 
2- and 4-roots found in \cite{AlexeevNikulin}. In particular, T-duality is one such reflection while Wilson line 
translations are products of even numbers of them. Looking at the fundamental reflections in this example
reveals that indeed it is not possible to connect a shift in which one of $m$ and $n$ is odd, to another in which
both are even.

\vskip0.1in
\item[3.] $I= U + E_8$, $ s=10$.
\vskip0.1in
\noindent
When $\alpha^2 \neq 4\, {\mathrm{mod}}\, 8$, level matching requires that at least one of $n$ or $m$ be odd. 
Therefore, case~1 applies and the corresponding shifts are all connected to each other by turning on suitable discrete Wilson lines
that do not change $I$. In this situation we can take $m=1$ and start from 
\be\label{vsolue8}
v_1=\frac{1}{4}(1,2,0) .
\ee
Turning on $A=-\alpha/2$, i.e. applying $\cE_\beta$ with $\beta=-\alpha$, we can then obtain all shifts $\frac14(1,n,\alpha)$, 
$n=0,\pm1,2$, satisfying level matching. Notice that this also includes shifts $\frac14(1,0,\alpha^2)$ with 
$\alpha^2 = 4\, {\mathrm{mod}}\, 8$.

The question is whether we can also connect shifts with both $n$ and $m$ even and $\alpha^2 = 4\, {\mathrm{mod}}\, 8$
to  $v_1=\frac{1}{4}(1,2,0)$.  From the analysis of case 2a we know that it is sufficient to find a $\beta \in E_8$ such that 
$\alpha \cdot \beta$ is odd. It is enough to consider $\alpha^2=4$ and $\alpha^2=12$, since larger values can be reduced by 
substracting a vector in $E_8$.
For $\alpha^2= 4$, we can take $\alpha= (0^7,2)$, in shorthand notation for the $E_8$ vector $(0,0,0,0,0,0,0,2)$.
This choice is unique up to transformations in the Weyl group of $E_8$.
We can then pick $\beta=((\frac{1}{2})^8)$,  which satisfies $\alpha \cdot \beta=1$ and hence can go to
$\frac{1}{4}(1,2,0)$ as explained in case 2a. Similarly, for $\alpha^2=12$ we can take $\alpha=(0^4,(-1)^3,3)$. 
There are several $\beta$'s fulfilling that $\alpha\cdot \beta$ is odd. For instance, $\beta=(1,0^3,1,0^3)$ or 
$((\frac{1}{2})^3,(-\frac{1}{2})^2,(\frac{1}{2})^3)$. 

In conclusion, all the shifts for $I=U+E_8$ are connected. 
It can be shown in detail that any $v$ complying with level matching can be transformed into the shift in eq.~\eqref{vsolue8}
by combining elements of $\cT(I)$. 

\vskip0.1in
\item[4.] $I= U + E_7$, $ s=11$.
\vskip0.1in
\noindent
In this case it can be shown that all shifts can be connected to $v_1=\frac{1}{4}(1,1,0)$.
We skip the details of the analysis, which is analogous to the preceeding one for $I=U+E_8$.

\vskip0.1in
\item[5.]
$I = U + D_6$, $s=12$.
\vskip0.1in
\noindent
Level matching implies that at least one of $n$ or $m$ must be odd when $\alpha^2 \neq 0\, {\mathrm{mod}}\, 8$. 
The corresponding shifts are then covered by case 1. They can all be connected to  
\be\label{vsolud6}
v_1=\frac{1}{4}(1,-1,(1^2, 0^4)) .
\ee
Here we used that $\alpha=(1^2,0^4)$,with $\alpha^2=2$, is unique modulo the Weyl group of $SO(12)$. This
$v_1$ is also connected to $\frac14(1,0,\alpha^2)$ with  $\alpha^2 = 0\, {\mathrm{mod}}\, 8$.

We next study shifts with both $(m,n)$ even, and $\alpha^2 = 0\, {\mathrm{mod}}\, 8$. Requiring
$\frac12 \alpha \notin D_6$ to avoid order 2, we can always find $\beta$ such that  $\alpha \cdot \beta$ is odd.
This is case 2a. For instance, from $v_1$ in eq.~\eqref{vsolud6} we can reach $\frac14(m,n,(2,1^4,0))$,
with $(m,n)=(0,0), (2,0), (2,2)$, via transformations in $\cT(I)$. 
The upshot is that all shifts of order 4 are connected to $v_1$.

\end{trivlist}

Shifts for other lattices of the form $I=U+K$ can be analyzed following the same procedure as above.
We will shortly discuss the resulting spectrum in some selected examples.

\vspace{-8pt}

\subsubsection{$I\simeq U(2) + K$}\label{subsubsec_U2K}

The moduli can again be taken to be a radius $R$ and a \mbox{$(r-2)$-dimensional} Wilson line $A$.
The components $(\rho,p_L; p_R)$ of $P \in I$ are now 
\be\label{momu2k}
\begin{split}
\rho &= \gamma + 4 j A,  \\
p_L &=  \frac{\sqrt2}{R}\left(\frac{k}{2} + (R^2 -2 A^2) j - \gamma\cdot A\right), \\
p_R &= \frac{\sqrt2}{R}\left (\frac{k}2 - (R^2 + 2A^2) j - \gamma\cdot A\right) \, ,
\end{split}
\ee
with $j, k \in \bz$, and $\gamma \in K$.  The squared norm reads $P^2=4 j k + \gamma^2$.

It is useful to observe \cite{Mikhailov:1998si} that
$U(2)$ can be thought of as a sublattice of $U$ by rescaling the radius. Indeed, upon $R \to R'/\sqrt2$
we see that $U(2)$ is the sublattice of $U$ defined by the condition that quantized momenta are even 
numbers while the windings are arbitrary integers. Of course one can also exchange winding and momenta.

Knowing $P\in I$ in terms of moduli also allows to find equivalent forms of the lattice.
For an interesting example mentioned before, consider $K=A_1^7$. Taking $A=0$ and
generic $R$ yields the direct sum $I=U(2) + A_1^7$. Further special moduli points are
\be\label{equivUA17}
\begin{split}
A&=\tfrac{1}{2\sqrt2}(1,1,1,1,1,1,1), \qquad R=\tfrac12, \qquad I=A_1(-1) + A_1^8,\\
A&=\tfrac{1}{4\sqrt2}(1,1,1,1,1,1,1), \qquad R=\tfrac14, \qquad I=A_1(-1) + E_8(2).
\end{split}
\ee
Therefore, $U(2) + A_1^7 \simeq A_1(-1) + A_1^8 \simeq A_1(-1) + E_8(2)$.

Other known isomorphisms involving $U(2)$ are
\be\label{extrau2}
U(2) + E_8 \simeq U + D_8; \quad U(2) + D_8 \simeq U + D_4 + D_4; \quad U(2) + D_4 + D_4 \simeq U + D^*_8(2),
\ee
where $D^*_8$ is the $SO(16)$ weight lattice. These relations follow from uniqueness of even Lorentzian lattices 
with prescribed discriminant form, as explained e.g. in \cite{Mikhailov:1998si}. By the same reasoning,
\mbox{$U(2) + D_{16} \simeq U + D_8 + D_8$}.

For future purposes we now give the explicit map from $U+D_8$ to $U(2)+E_8$.
Starting with eq.~\eqref{momuk} we do a sequence of transformations on $U + D_8$: similar to eq.~\eqref{ebeta}
first turn on a Wilson line $A_2=\beta_2/2$, then exchange the resulting winding/momentum and then 
again turn on another Wilson line $A_1 =\beta_1/2$. Choosing $\beta_1=(2,0^7)$ and $\beta_2=(\frac12^8)$ 
we then obtain that 
\be\label{mapud8}
 (j,k,\gamma) \to (j',k',\gamma')=(k-j -\tfrac12 \textstyle{\sum_i }\! \gamma_i, - 2 j' - 2 \gamma_1, 
 \gamma +j(\tfrac12^8) + j'(2,0^7)).
 \ee
Since $\gamma \in D_8$, $\gamma_1$ is integer and $\sum_i \! \gamma_i$ is even. Thus, $j'$ is
an arbitrary integer whereas $k'$ is even and $(j',k')$ just gives a generic vector of $U(2)$. Moreover,
it can be checked that $\gamma'$ spans the full $E_8$. We conclude that eq. \eqref{mapud8}
sends $U+D_8$ into $U(2) + E_8$. The above map is analogous to the map relating the $E_8\times E_8$ and 
the $Spin(32)/\bz_2$ heterotic theories in 9 dimensions \cite{Ginsparg:1986bx, Keurentjes:2006cw,Font:2020rsk}.

When $I\simeq U(2) +K$, the automorphism group $\cT(I)$ again includes exchange of winding and quantized momentum, i.e. 
$(j,k,\gamma)  \to (k,j,\gamma)$.
of the Wilson line, but now we have to distinguish
whether the scalar product of two vectors in $K$ is always even, as in $K=E_8(2)$, or just an integer, as in
$K=D_4$ or a product of $A_1$'s. In both cases the translation $A \to A +\frac14 \beta$ induces the transformation
\be\label{ebeta2}
\cE^{(2)}_\beta:\ (j, k, \gamma) \to  \to (j',k',\gamma')=(j, k+\frac12 \gamma\cdot \beta - \frac14 \beta^2 j, \gamma - j\beta) .
\ee
For $K=E_8(2)$, its is enough to take $\beta \in K$. However, for other $K$'s, additional constraints are needed to guarantee
that $k'$ is an integer. First, setting $j=0$ requires that $\gamma \cdot \beta$ be an even integer which implies that
$\beta = 2\tilde \beta$, with $\tilde \beta \in K^*$. Recall that by hypothesis $2 K^* \subset K$. Turning on $j$,
the extra term $\frac14 \beta^2 j$ is integer provided $2\tilde\beta^2$ is even, which just amounts to
$\tilde \beta \in K_e^*$.

To examine the possible shifts $v$ we set the Wilson line $A$ to zero and consider
\be\label{shiftuk2}
v =\frac14\left(\alpha,\sqrt2\left(\frac{n}{2R} + m R\right); \sqrt2\left(\frac{n}{2R} - m R\right)\right) ,
\ee
where $n, m$ are integers mod 4 and $\alpha \in K$. As before we will use the shorthand notation $v=\frac14(m,n,\alpha)$.
The level-matching condition \eqref{LMC} just translates into eq.~\eqref{lmc_mn}, but with $mn$ 
replaced by $2mn$.
To classify the solutions we can again choose $(m,n)$ to take values in the list \eqref{listmn} and allow for exchanges of
$m$ and $n$. We also apply transformations $\cE^{(2)}_\beta$, for some judicious $\beta$. Below we discuss two examples.

\begin{trivlist}
\item[1.]  $I= U(2) + A_1$, $s=17$. 
\vskip0.1in
\noindent
In general, $v=\frac14(m,n,\sqrt2 \ell)$. The level-matching condition is solved by
\be\label{vsolu2a1}
v_1=\frac14(1,1,\sqrt2).
\ee
Other solutions, e.g. with $(m,n)=(1,-1)$ and/or $\ell=3$, differ from $v_1$ by a vector in $I^*$.
\vskip0.1in
\item[2.] $I= U(2) + E_8$, $ s=10$.
\vskip0.1in
\noindent
Consider first $\alpha^2 =0\, {\mathrm{mod}}\, 8$. Level-matching leads to $mn=\pm 1$ and the two choices, with $\alpha$ fixed,
give shifts differing by an element of $I^*$. Other possibilities can be related by the transformation $\cE^{(2)}_\beta$,
taking care that $\beta \in 2 E_8$, since in this example $K^*=K_e^*=K=E_8$. Concretely,
choosing $m=n=1$, applying $\cE^{(2)}_\beta$ with $\beta=-\alpha$ and $2\alpha \in E_8$, we then find
that all $\frac14(1,1,\alpha)$ can be obtained from
\be\label{vsolu2e81}
v_1=\frac14(1,1,(0^8)).
\ee
However, there exist $E_8$ elements with $\alpha^2=8$ but $2\alpha \notin E_8$ which cannot be obtained in 
the same fashion. In particular, 
\be\label{vsolu2e82}
v_2=\frac14(1,1,(\tfrac52, \tfrac12^7))
\ee
cannot be connected to $v_1$.
Level-matching is also satisfied with $\alpha^2 =4\, {\mathrm{mod}}\, 8$ and $mn=0\, {\mathrm{mod}}\, 2$
but one can check that the solutions can be connected to either $v_1$ or $v_2$.
In summary, all shifts verifying level-matching are obtained from $v_1$ and $v_2$.

It is also interesting to determine the possible shifts for the point $(r,a,\delta)=(10,2,0)$, but starting with
$I'=U+D_8$, which is also a valid choice. In this case the shifts have the form in eq.~\eqref{shiftuk} and must 
fulfill the level-matching condition \eqref{lmc_mn}. The analysis has similarities with that in the example $I=U+E_8$
discussed before. In particular, all shifts $\frac14(1,n,\alpha)$, $n=0,\pm1,2$, which solve level-matching,
can be connected to 
\be\label{vsolud81}
v'_1=\frac14(1,2,(0^8)).
\ee
However, now it is not true that shifts with both $n$ and $m$ even and $\alpha^2 = 4\, {\mathrm{mod}}\, 8$
can also be connected to $v'_1$. The reason is that it is not guaranteed that there exists a $\beta \in D_8$ such that 
$\alpha \cdot \beta$ is odd. For example, for $\alpha=(2,0^7)$ or $\alpha=(2^3,0^5)$ there is no such $\beta$,
while for $\alpha=(1^4,0^4)$ there are many. 
We find that the solutions that cannot be connected to $v'_1$ can instead be obtained from  
\be\label{vsolud82}
v'_2=\frac14(0,0,(2,0^7)). 
\ee
These results can be neatly matched with the previous ones for $I=U(2)+E_8$.
Indeed, applying the map in eq.~\eqref{mapud8} it can be verified that the images of the shift vectors $v'_1$ and $v'_2$ 
in $U+D_8$  precisely correspond to the shifts $v_2$ and $v_1$ in $U(2)+E_8$. 

\end{trivlist}

\subsection{Spectrum}\label{subsec:spectrum}

We will now use the partition function to extract some generic features of the spectrum, 
focusing on tachyonic and massless states. The discussion will be illustrated with 
some representative examples. 

For a given sector $g^m$ (with $m=0,1,2,3$), the masses for left and right movers 
can be read off from the exponents of $q$ and  $\bar q$, respectively, 
upon expanding the terms $Z_{m,n}$ in the partition function constructed in subsection \ref{subsec_pf}. 
The left components $P_L$ of  the $\Gamma=\Gamma_{\!\!(19,3)}$ lattice of momenta,  as well as  
the oscillator numbers encoded in $\vartheta$ and $ \eta$ functions, contribute to $q$ exponents
that give $m_L^2$.
Similarly, $m_R^2$ is determined by the $\bar q$ exponents arising from right $P_R$ momenta in $\Gamma$, 
right oscillators in  $\bar\vartheta$ and $ \bar \eta$ functions, and  the weights in vector $V$ and spinor classes $Sp$ of $SO(8)$.
Besides, the level-matching condition $m_L^2=m_R^2$ must be satisfied.

In each sector we further have to take into account how the involution acting on 
$\Gamma$ restricts the lattice momenta. As explained before, in the
subsectors  $(1,g)$ and $(1,g^3)$, as well as  $(g^2,g)$ and $(g^2,g^3)$,
only momenta in the invariant lattice $I$ are selected. In
$(g,g^n)$ and  $(g^3,g^n)$, it must be instead $P\in I^*$. In the remaining sectors
general momenta $P=(P_N,P_I)\in \Gamma$ are allowed. 
The projection onto orbifold invariant states is completed by summing over
$n$ (with $n=0,1,2,3$) and keeping track of the $P$ dependent phases.
Since these phases depend on the shift vector $v$, non equivalent shifts, in the 
sense of \eqref{vequiv}, will lead to different spectra.

We will first survey the untwisted and twisted sectors separately, focusing on generic features. 
Two noteworthy results are that massless gauge vectors arise only in the untwisted sector whereas tachyons 
emerge only in twisted sectors. Afterwards we will work out the spectrum of tachyonic and massless states in examples 
with specific invariant lattice for $s=10, 15, 17, 4M$, ($M=0,\ldots,4$), which are representative values of $s$ modulo 4.
In each model we will contrast inequivalent shifts satisfying the level-matching condition $2 v^2 +\frac{s}4 \in \bz$.
We will also provide the specific  $w \in I^*/I$  satisfying
\begin{equation}
w^2= 1,\frac12,0,\frac32, \ \text{for} \ s=0,1,2,3 \, \rm{mod}\, 4,
 \label{wn}
\end{equation}
in agreement with \eqref{level2}.

\subsubsection{Untwisted sector}
\label{subsec:Untwisted} 

The expressions for the masses and projections in the untwisted sector ($m=0$) can be read from 
the terms \eqref{Z00},\eqref{Z01} and \eqref{Z02} in the partition function. We find
\be
m_L^2= \frac12 P_L^2 + N_L -1, \qquad 
m_R^2=\frac12 p^2 + \frac12 P_R^2 + N_R -\frac12,
\label{leftrightmassessgeneral}
\ee
where $N_L$ and $N_R$ are oscillator numbers.
Since $p^2\ge 1$ , for  $p\in V,Sp$, it follows that there are  no 
untwisted tachyons. Notice also that massless states must have
$p^2=1$ with $p\in V,Sp$,  $N_R=0$, and $P_R=0$.

To implement the projection on invariant states we also need the phases and lattice momenta in
each $(1,g^n)$ subsector. From the partition function we find 
\begin{alignat}{3}\label{projuntwistedgeneric}
(1,1)&:\ 1 \qquad\qquad &&P\in \Gamma, \nonumber \\
(1,g)&: \ e^{-2i\pi p \cdot v_f} e^{2i\pi P\cdot v}  \qquad\qquad  && P\in I, \nonumber \\
(1,g^2)&:  \ e^{-2i\pi p\cdot 2v_f} e^{2i\pi P\cdot(2v+w)} \qquad\qquad && P\in \Gamma, \\
(1,g^3)&: \ e^{2i\pi p\cdot v_f}e^{-2i\pi P\cdot v}\qquad\qquad  && P\in I, \nonumber
\end{alignat}
where we used that $p\cdot 3v_f=-p\cdot v_f$ and $P\cdot3 v= -P\cdot v$.
For oscillators in the reflected directions there is also an extra $-1$ in the the $(1,g)$ and $(1,g^3)$ sectors, 
which can be seen for instance in the $q$ expansion of $(\eta/\vartheta_2)^{\frac{s}2}$.
Given the momenta and oscillator numbers of states satisfying level matching we then consider
the sum over $n$ to see if they are allowed by the orbifold projection.
Below we discuss the case of massless states.

Let us first look at massless bosonic states. For right movers, we can have 
$p_{\text{o}}=(\pm1, 0,0,0)$ with $e^{2i\pi p_{\text{o}} \cdot v_f} =-1$, or 
$p_{\text{v}}=(0,\underline{\pm 1,0,0})$ with $e^{2i\pi p_{\text{v}} \cdot v_f} =1$,
where underlining means to take all permutations.
For left movers, for generic moduli, the only solution of $m_L^2=0$ is $P_L=0$ and $N_L=1$.
Among the 24 left oscillators in the light-cone, the 5 along space-time directions are invariant under the  
orbifold action and will survive when combined with the right-moving $p_{\text{v}}$ to give rise to
the 7-dimensional graviton, dilaton, and Kalb-Ramond field, plus a ``graviphoton'' leading to a $U(1)_R$ 
gauge group factor. There are also $(19-s)$ oscillators along the internal invariant directions
that survive when combined with $p_{\text{v}}$ to produce gauge vectors of $U(1)^{19-s}$, together with
$(19-s)$ scalars. Finally, there are additional scalars from the remaining $s$ oscillators along the reflected directions, 
which give invariant states when combined with $p_{\text{o}}$.

As explained above, for generic moduli the gauge group is $U(1)^{19-s}$, excepting the $U(1)_R$ from
the graviphoton. For special moduli there can be additional massless vectors of an enhanced gauge group.
To elaborate on this we consider an invariant lattice of the form $I \simeq U+K$, where  
 $K$ is an even lattice with signature $(18-s,0)$ and $K^*/K=\bz_2^a$. For such $I$ the momenta $P \in I$ are
 given in \eqref{momuk}. We go to the special point in moduli space where the Wilson line is zero, i.e. $A=0$. At this point
$P \in I$ takes the form  $P=(\gamma,p_ {L};p_{R})$, with $\gamma \in K$ and 
\begin{equation}
 p_{L}=\left(\frac{k}{2R}+jR\right),\qquad p_{R}=\left(\frac{k}{2R}-jR\right).
\label{pua1}
\end{equation}
In our conventions the self-dual radius is $R=\frac{1}{\sqrt2}$. In the following we will work
at generic $R$, unless it is otherwise stated.

In general, additional massless states must satisfy $P^2_L=P^2_{NL} + P^2_{IL}=2$.  To be more specific,  we assume first
$P_N=0$. Hence, $P_I \in I$ and
\be\label{leftgeneralpg}
P_L^2= \gamma^2+ \left(\frac{k}{2R}+jR\right)^{\!\!2}.
\ee
For $m_R^2=0$ we are picking $p=(0,\underline{\pm 1,0,0})$, and it must be $p_R=0$ which implies $k=j=0$ 
for generic radius. 
At the self-dual radius, both $p_R=0$ and $P^2_L=2$ are achieved  with $j=k=\pm1$ and the extra states enhance one of 
$(19-s)$ $U(1)$'s to $SU(2)_U$, associated to the $U$ component of the invariant lattice. Similarly, states with $\gamma^2=2$ will enhance the remaining $U(1)$'s to some group $G_K$ of rank $(18-s)$, associated to the lattice $K$. 
Recall that in general $K$ is a direct sum of $A_1$, $SO(4N)$, $E_7$ and $E_8$ root lattices

To analyze the possibilities for $G_K$ in more detail we need to specify the shift $v$ because 
for $P\not=0$ the invariant states are determined by the $v$-dependent phases given in \eqref{projuntwistedgeneric}.
For instance, all these phases are just 1 when $v$ has no components  along $K$ directions, namely when $v=\frac14(m,n,0)$.
The reason is that $P\cdot v=0$ for $j=k=0$. Moreover, $P\cdot w \in \bz$ 
since $w \in I^*$ and $P\in I$. Hence,  all $\gamma$ with $\gamma^2=2$, i.e. all the roots of $K$, are allowed
and $G_K$ is the group whose Lie algebra is $K$, which we will also call $K$ abusing notation. The conclusion is that
$G_K=U(1) \otimes K$ when the invariant lattice is the direct sum $I=U+K$ and $v=\frac14(m,n,0)$.
On the contrary, when the shift vector $v$ has a component along $K$, i.e. $v=\frac14(m,n,\alpha)$, not all roots of $K$ 
give states invariant under the orbifold projection and the group will be broken. Concretely, in this case
$P\cdot v= \frac14 \gamma \cdot \alpha$ which implies $e^{8i\pi P\cdot v}=1$ because $\gamma \cdot \alpha \in \bz$ since
both $\alpha$ and $\gamma$ belong to the lattice $K$. Thus, depending on $\alpha$ there could be roots $\gamma$ with
$P \cdot v \in \bz$ that survive the orbiflod projection, whereas those roots with $P \cdot v \notin \bz$
are projected out.

Extra charged gauge vectors can also arise from momenta with components along reflected directions, namely from 
$P\in \Gamma$ but $P \notin I$, which enter only in $(1,1)$ and $(1,g^2)$ sectors. 
For instance, if $P_I=0$ then $P_N \in N$ and for special moduli in $N$ there can be vectors with $P^2_{NL}=2$.
These states  are allowed by the orbifold projection,   
but due to the $1/4$ factor in the orbifold partition function, they will appear with a $1/2$ multiplicity. 
Indeed, this is the correct multiplicity because only states that are invariant under $g$ remain in the spectrum. In other words,
since $P_N \to -P_N$ under $g$, the invariant combinations are $|P_N\rangle + |-P_N\rangle$ and vertex operators are of the
form 
\begin{equation}
 \frac12\left(e^{iP \cdot Y}+e^{-iP\cdot  Y}\right)
 \label{refectedvectors}
 \end{equation}
with $Y$ the coordinates along the normal lattice directions.
Moreover, these invariant combinations must include both charged and Cartan generators since all oscillators along the left 
directions in $N$ are projected out when combined with right-moving $p_\tv=(0,\underline{\pm 1,0,0})$.
More generally, for special moduli of the $I$ and $N$ lattices, the gauge group could be enhanced when both 
$P_N$ and $P_I$ are different from zero. Such situations are considered in the examples presented later.
For generic moduli, space-time vectors arising from $P_N \not =0$ will be massive.

We now turn to massless fermions. In this case $p \in Sp$ and $e^{2i\pi p \cdot v_f}  =\pm i$. For states with 
$P=0$ and $N_L=1$, from the phases in \eqref{projuntwistedgeneric} we see that they are projected out, no matter if
the oscillators are along non-reflected directions or not. This means in particular that there are no massless gravitini,
consistent with absence of supersymmetry. 
For special moduli there could be massless fermions if there are solutions of $P_R=0$, $P_L^2=2$, and
$e^{2i\pi P\cdot(2v+w)}=-1$. 
In fact, there always exist $P \in \Gamma$ such that $e^{2i\pi P\cdot(2v+w)}=-1$,
as follows because otherwise $(4v^2- w^2)$ would be even in contradiction with eqs. \eqref{LMC} and \eqref{21operatorint}.
Hence, there are two independent combinations of states $|P_I,P_N\rangle$ and 
$|P_I,-P_N\rangle$ that are eigenstates of $g$ with eigenvalues $\pm i$ and therefore will couple to right-moving spinor weights to give space-time fermions. They are generically massive but at special points in the moduli space can also be massless.
We will shortly consider some examples. Notice that these fermions are necessarily charged since they have momenta
along $\Gamma$.

\subsubsection{Twisted sectors}
\label{subsec:Twisted} 

Since the $g$- and $g^3$-twisted sectors are analogous we will focus on the former.
The generic expressions for the masses are
 \be\label{leftrightmassesn1n3}
 \begin{split}
m_L^2&= \frac12 (P+ v)_L^2 + N_L -1+\frac {s}{16},\\
m_R^2&=\frac12 (p+v_f)^2 + \frac12 (P+ v)_R^2 + N_R -\frac12+\frac18,
\end{split}
\ee
where $P\in I^*$ and $N_R$, $N_L$ are oscillator numbers that can be half-integers.
It is easy to see that space-time vectors, which have $p=(0,\underline{\pm 1,0,0})$, will be massive.
On the other hand, tachyons could arise when $N_R=0$ and $p=(-1,0,0,0)$, so that $(p+v_f)^2=\frac14$.
However, level-matching prevents tachyons when $s\ge 16$. For $s < 16$ tachyons could still be avoided
depending on the shift $v$ and the particular moduli. For instance,  when $I=U + K$ and $v$ is given by
\eqref{shiftuk}, tachyons would only be present for  bounded values of the radius $R$.
In the $g$-sector it can be shown that all states with $m^2_L=m^2_R$ are allowed by the
orbifold projection, in other words,  the phases that appear in  $Z_{1,n}$ are all equal to 1 once level-matching
is imposed.

In the $g^2$-sector the masses deduced from \eqref{Z22n} are
 \be\label{leftrightmasses}
\begin{split}
m_L^2&= \frac12 (P+2 v+w)_L^2 + N_L -1,\\
m_R^2&=\frac12 (p+2 v_f)^2 + \frac12 (P+2 v+w)_R^2 + N_R -\frac12,
\end{split}
\ee
where now the oscillator numbers take only integer values.
Notice that $(p+2v_f)$ belongs either in the scalar class $Sc$ or in the conjugate spinor class $Sp'$ of $SO(8)$, since $p \in V, Sp$
and $2v_f \in V$.
It is then evident that there will not be gauge vectors in this sector and that there are
potential tachyons coming from the $Sc$ of $SO(8)$.  

We now analyze the possibilities for tachyons in the $g^2$-sector in more detail. 
Taking the ground state of the right movers to have $(p+2 v_f)=0$,  and setting all the oscillator modes to zero, leads
to the level-matching constraint
\begin{equation}
m_L^2- m_R^2 = \frac{1}{2}(P+2v+w)^2-\frac{1}{2}=0.
\label{genericmassmatch}
\end{equation}
A way to avoid tachyons would be to choose the shift $v$ such that $(2v+w) \in I$, because then $(P+2v+w)^2$ would be even
and level-matching could not be fulfilled.
However, it can be shown that $(2v+w)$ cannot lie in $I$ if the level-matching condition \eqref{level1} holds.
On the contrary, if for some $P\in  \Gamma$, $(P+2v+w)^2=1$, 
we get a level-matched state with mass square proportional to $\frac{1}{2}(-1+(P+2v+w)_R^2)$. By, if
necessary, moduli deformations corresponding to $(s,2)$ and $(19-s,1)$ boosts\footnote{In fact exactly at  $(P+2v+w)_L^2 = 1$, the would be tachyon becomes massless and one might ask what happens if one starts
giving vev to it.} we can always go to a region in the moduli space where $(P+2v+w)_R^2 < 1$
resulting in a tachyonic state if it is also $g$-invariant. Let us then look at the action of $g$ on this state. Under $g^2$, clearly the state 
is invariant since it is level-matched and the $(g^2,g^2)$ sector is obtained by $T$-transforming the $(g^2,1)$ sector. This of course means that, after diagonalizing $g$, the eigenvalues of $g$ on this level-matched state can only be $\pm 1$. There are two cases:

\noindent 1) $P_N=0$ and therefore $P_I \in I$. In this case we can obtain the $g^{-1}$-action from $Z_{2,-1}$. 
Equivalently, from \eqref{ging2} we read off the phase 
\begin{equation}
e^{-2i\pi\Phi}=e^{-2i\pi( v^2+v\cdot w-\frac{s-4}{8}+P\cdot v)} , 
\label{projphase}
\end{equation}
taking into account that we are looking at a state with $p+2v_f=0$. 
This phase (which is necessarily $\pm 1$) will determine whether this state is $g$-invariant.

\noindent 2) $P_N \neq 0$.  Now $g | P_N, P_I+2v+w\rangle \rightarrow |- P_N, P_I+2v+w\rangle$ up to some coefficients. 
Since the trace of $g$ over these two states is zero, and since the eigenvalues of $g$, as seen above, can only be $\pm 1$, we conclude that one combination of these two states will have $g$ eigenvalue $+1$ and the other $-1$. Thus, 
there will be necessarily a $g$-invariant tachyonic state.

\subsubsection{$s=17$ example}
\label{ss:s17}

There are two Nikulin points with $s=17$. Here we consider $(r,a,\delta)=(3,1,1)$ and take the invariant lattice
to be the direct sum
\be\label{Is17}
I=U+A_1.
\ee
Depending on the starting choice of moduli for the self-dual lattice $\Gamma =\Gamma_{\!\!(19,3)}$  there are different ways to obtain this invariant lattice by performing an adequate involution.    
For instance we can start at
\begin{equation}
\Gamma\simeq \Gamma_{0}=U+ E_{8}+\Gamma_{\!\!(10,2)}, \label{Gsds17E7} 
\end{equation}
where $\Gamma_{\!\!(10,2)}$ is even self-dual.
The involution then acts on $\Gamma_{\!\!(10,2)}$ by reflecting all its left and right directions, and on $E_8$ by leaving an $A_1$ 
sublattice invariant{\footnote{The action on $E_8$ vectors is $|P_1,\dots,P_6,P_7,P_8 \rangle \to 
|-P_1,\dots, -P_6, -P_8,-P_7 \rangle$.}}.
The lattice normal to $I$ is then given by
\be\label{NGzero}
N=\Gamma_{\!\!(10,2)} +E_{7}.
\ee
Another possibility is to start from
\begin{equation}
\Gamma \simeq \Gamma_{0}'=U+ U_{\text{sd}} +\Gamma_{\!\!(17,1)},
\label{Gsds17generic} 
\end{equation}
where $\Gamma_{\!\!(17,1)}$ is even self-dual and
$U_{\text{sd}}$ is the $U$ lattice at the self-dual point, i.e. the lattice vectors are those in  \eqref{pua1}
with $R=\frac1{\sqrt2}$.
The involution now acts as a reflection of all left and right directions in $\Gamma_{\!\!(17,1)}$, plus $(p_L, p_R) \to (p_L,-p_R)$
for the momenta in $U_{\text{sd}}$. Notice that $p_R=0$ implies $p_L=\sqrt2 k$, which are the momenta in the
$A_1$ component of the invariant lattice. The normal lattice is instead $N=A_{1}(-1)+\Gamma_{\!\!(17,1)}$.

As discussed previously, a generic shift vector on $I$  can be written as 
$v=\frac14(m,n,\sqrt2 \ell_1)$ and leads to the three representative solutions $v_1, v_2, v_3$ presented in \eqref{vsolua1}.  
Since $v_2$ and $v_3$ differ by a  weight in the $A_1^*$ lattice they will lead to the same spectrum. Therefore  we will just concentrate
in $v_1$ and $v_2$ corresponding to $(m,n,\ell_1)=(1,2,1)$, $(1,-1,0)$, respectively.
We also choose the vector  $w$, introduced in \eqref{wcond}, to be 
$w=(\frac{\sqrt2}{2},0;0)$, namely,  a fundamental weight in $ A_1^*/A_1$, with $w^2=\frac12$, as stated in \eqref{wn}. 

We will now examine the spectrum in the untwisted and twisted sectors.

\bigskip
\noindent
\underline{Gauge group and charged massless scalars}

Since the momenta in the $A_1$ root lattice are $\gamma=\sqrt2 \ell$, with $\ell\in \mathbb{Z}$, we see from 
\eqref{leftgeneralpg} that $P_L^2=2$ can occur for $\ell=\pm1$, and $j=k=0$ (for generic $R$). 
These solutions would lead to gauge vectors when combined with $p_\tv=(0, \underline{\pm1,0,0})$ and to massless scalars when combined with $p_{\text{o}}=(\pm 1,0,0,0)$.
As occurs in general when the shift is purely along $U$, for $v_2$ all the states are allowed and there is a gauge factor  
$U(1)\times  SU(2)$.
On the other hand, for the shift $v_1$ the states are projected out because in the sectors $(1,g)$ and ($1,g^3)$ 
the phases in \eqref{projuntwistedgeneric} include $e^{i\pi \ell}$. Thus, for $v_1$ the gauge factor is $U(1)\times  U(1)$, 
where the first $U(1)$ originates in the $U$ lattice and the second one in $A_1$.
Notice that due to the phase $e^{2i\pi p_{\text{o}} \cdot v_f} =-1$, charged scalars are projected out for $v_2$ whereas two 
massless scalars with opposite $U(1)$ charge appear for the $v_1$ shift.

Let us now search for gauge enhancements due to momenta with components along the $N$ lattice, which enter 
in $(1,1)$ and $(1,g^2)$ sectors where the lattice sum is over the full $\Gamma$.
In the case where we start from \eqref{Gsds17E7}, and $N=\Gamma_{\!\!(10,2)} +E_{7}$, the right movers with
$p_\tv$ can give massless charged vectors when combined with $P_L^2=2$ coming only from $ E_7$ roots. 
As explained before, the states appear with multiplicity $1/2$, meaning that the combinations invariant under $g$ are 
$|P_N\rangle + |-P_N\rangle$. 
Besides, these invariant combinations must include both charged and Cartan generators. 
Thus, out of the original 126 roots in $E_7$ just $63$ combinations are kept and an 
extra $SU(8)$ gauge factor emerges, for both shifts $v_1$ and $v_2$.

The counting of massless scalars proceeds in a similar manner. 
Now there are $63$ states coming from the anti-symmetric combinations of  $E_7$ roots 
that pick up a minus sign under $g$, just like the 7 oscillators along the left reflected directions associated to $E_7$.
Under the $SU(8)$ gauge group the states transform in the 70-dimensional representation (rank-four antisymmetric tensor). 
These states can then couple to the right-moving sector oscillators along ${\bt^2}$ in a \mbox{$g$-invariant way}. 
Thus, altogether there are two (one for each torus direction) massless scalars in the ${\bf{70}}$ representation of $SU(8)$.

It remains to examine the possibility of having both $P_I$ and $P_N$ not zero, thus $P_N \in N^*$ and $P_I \in I*$.
With the lattices in \eqref{Is17} and \eqref {NGzero}, we can take the components of $P=(P_N,P_I)$ to be fundamental 
weights of $E_7$ and $A_1$ respectively, namely $P=(\tw_{56}, \tw_2)$, with $\tw^2_{56}=\frac32$ and $\tw^2_2=\frac12$. 
Clearly $P^2=2$ and $P\cdot(2v+w)=\frac12(\ell_1+1)$. Hence, combining with $p_\tv$ gives states that are projected
out for $v_2$ whereas for $v_1$ they will appear with multiplicity $1/2$, meaning again that we must form invariant combinations
under the orbifold action. 
The upshot is that for $v_1$ there are 56 extra states. Recall now that for $v_1$,
we had already found a factor $U(1)$ from $A_1$ with $P_N=0$, and a factor $SU(8)$ from $E_7$ with $P_I=0$.
The extra 56 states, which transform under $U(1)$ and $SU(8)$, indeed complete the adjoint of $SO(16)$.
Altogether the gauge group for $v_1$ is $U(1)\otimes SO(16)$, with the $U(1)$ from the lattice $U$. 
Besides, for $v_1$ there are massless scalars that accommodate into two ${\bf{128}}$ spinor representations of $SO(16)$.
For $v_2$ the full group is  $U(1)\otimes SU(2) \otimes SU(8)$ and there are two massless scalars in the 
${\bf{70}}$ representation of $SU(8)$. 

It is interesting to compare with the results obtained had we started
with the decomposition in \eqref{Gsds17generic}, leading to $N=A_{1}(-1)+\Gamma_{\!\!(17,1)}$.
In this case there would be no enhancement from $P_N \not = 0$ as long as the moduli in the self-dual $\Gamma_{\!\!(17,1)}$ 
are generic. To see what can happen at a special point, we take $\Gamma_{\!\!(17,1)} = \Gamma_{\!\!(9,1)} + E_8$,
where  $\Gamma_{\!\!(9,1)}$ is even self-dual. At this point there will emerge an additional $SO(16)$ from invariant
combinations  $|P_N\rangle + |-P_N\rangle$, with $P_N$ one of the 240 roots of $E_8$.

\bigskip
\noindent
\underline{Massless fermions}

As mentioned before,  there exist massless fermions in the untwisted sector if $e^{2i\pi P\cdot(2v+w)}=-1$
and $P^2=2$ (with $P_R=0$). In the example at hand we already know that 
$P=(\tw_{56}, \tw_2)$ has $P^2=2$ and $P\cdot(2v+w)=\frac12(\ell_1+1)$. Thus, when $\ell_1=0$, i.e. for the shift $v_2$,
there are massless fermions. Forming invariant combinations we see that they transform as $({\bf 2}, {\bf 28})$ under
the  $SU(2) \otimes SU(8)$ group found for the shift $v_2$.

\bigskip
\noindent
\underline{Tachyons}

For $s=17$ all states in the $g$-sector are massive. We then
concentrate in the $g^2$-sector, where tachyonic states can appear if the level-matching condition \eqref{genericmassmatch} is 
fulfilled, and the states happen to be invariant under the orbifold action. There are two possible cases, depending on whether
$P_N=0$ or not.

We first examine the situation when $P_N=0$ and therefore $P_I=(\sqrt2\ell, p_L; p_{R}) \in I$, with $p_L$ and $p_R$ given in \eqref{pua1}. It is easy to check that the values
\begin{alignat}{3}\label{ljkvaluesmassmatching12}
(\ell,j,k)&= (-1,0, 0),(-1,-1,-2)\qquad && {\text{for}} \ \ v_1,  \nonumber \\
(\ell,j,k)&= (0,-1,0),(0,0,1)\qquad \qquad &&  {\text{for}}\ \ v_2,
\end{alignat}
solve the constraint \eqref{genericmassmatch} and could lead to tachyons depending on the radius $R$.
For instance, replacing the values for the $v_1$ shift 
in \eqref{leftrightmasses} we find that  tachyons could appear for radius such that $(\sqrt2-1) <R<(\sqrt2+1)$.
However, in all cases, the phase \eqref{projphase} in $(g^2,g)$ and $(g^2, g^3)$ turns out to be $e^{-2i\pi\Phi}=-1$.
Hence, when adding the contributions from $(g^2,1)$ and $(g^2, g^2)$, which have phase $+1$, possible tachyons with $P\in I$ 
will be projected out.

We now turn to the case with $P_{N}\ne 0$, in which tachyons will appear at specific moduli points as argued before.
For a concrete example, when $N=E_7 + \Gamma_{\!\!(10,2)}$, we choose moduli such that
\be
\Gamma_{\!\!(10,2)}= \Gamma_{\!\!(9,1)} + U_{\text{sd}},
\label{gamma102}
\ee
where  $\Gamma_{\!\!(9,1)}$ is even self-dual and $U_{\text{sd}}$ is the $U$ lattice at the self-dual radius.
We then take $P_N$ to be purely along $U_{\text{sd}}$, concretely $P_N=(0^7, 0^9, \frac{1}{\sqrt2};0, \frac{1}{\sqrt2})$.
We also choose $P_I=(\sqrt2\ell, p_L; p_{R})$,  where $p_L$ and $p_R$ are given in \eqref{pua1}.
Since $P_N^2=0$, the same values of $(\ell, j,k)$ in \eqref{ljkvaluesmassmatching12} again satisfy 
the level-matching condition \eqref{genericmassmatch}. 
For the $v_1$ shift we then find
 \be
m^2_R= m_L^2= \frac12\Big
(\frac1{2R}+\frac12R\Big)^2-\frac34,
\ee
which is negative for $(\sqrt3-1) <\sqrt 2 R< (\sqrt3+1)$.
Similarly, for the shift $v_2$ 
 \be
m^2_R=m_L^2=\frac12\Big
(\frac1{4R}+\frac12R\Big)^2-\frac12,
\label{rightmassestachyonv2}
\ee
which is negative in the range $(2-\sqrt2)< 2 R<(2+\sqrt2)$.
Thus, for $R$ in the above regions, tachyons will appear from $g$-invariant combinations.

\subsubsection{$s=4M$ examples }
\label{sec:s4m}

Let us first describe the invariant and normal lattices of the examples.
They are obtained by starting at a point in moduli space where
\be
\Gamma \simeq \Gamma_0 = U + \widehat{\Gamma}_{(18,2)}, \qquad
\widehat{\Gamma}_{(18,2)} = (D_{18};D_2)[ (Sp_{18}; Sp_2),(V_8;V_2)] .
\ee
As explained in Appendix \ref{app_lattices}, $[(Sp_{18};Sp_2),(V_8;V_2)]$ are glue vectors such that the
even self-dual $\widehat{\Gamma}_{(18,2)}$ consists of the correlated classes
\be
(Sp_{18}; Sp_2) + (Sc_{18}; Sc_2) +  (V_{18}; V_2) + (Sp'_{18}; Sp'_2).
\label{c182}
\ee 
Now we define the $\bz_2$ involution on $\Gamma$ to act only on $\widehat{\Gamma}_{(18,2)}$ by reflecting
the 2 right-movers along $D_2$ as well as $s$ directions along $D_{18}$.
The structure of the resulting invariant and normal lattices can be understood from the decomposition under
$D_{18-s} + D_s$ of the $D_{18}$ classes, namely from
\be\label{c182decomp}
\begin{split}
 Sp_{18}&=(Sp_{18-s}, Sp_{s} )+(Sp'_{18-s}, Sp'_{s}),\\
 Sc_{18}&=(Sc_{18-s}, Sc_{s} )+(V_{18-s}, V_{s}), \\
 V_{18}&=(V_{18-s}, Sc_{s} )+(Sc_{18-s}, V_{s}), \\
 Sp'_{18}&=(Sp'_{18-s}, Sp_{s} )+(Sp_{18-s}, Sp'_{s}).
\end{split}
\ee
In the invariant lattice we have to set the components along $D_{s}$ and $D_2$ to zero, which is an element of the
respective scalar classes. Thus, besides the inert $U$, $I$ just descends from  $(Sc_{18}; Sc_2)$ and only contains
the scalar class, i.e. the root lattice, of $D_{18-s}$. On the other hand, in the normal lattice the components along
$D_{18-s}$ must be set to zero. From eqs. \eqref{c182} and \eqref{c182decomp} we then read that $N$ is
$(D_s;D_2)$ with correlated classes $(Sc_s; Sc_2)$ and  $(V_s; V_2)$, or equivalently, with glue vectors
$[(V_s;V_2)]$.  Summarizing, the resulting invariant and normal lattices are
 \be
I=U+ D_{18-s}, \qquad N=(D_s;D_2)[(V_s;V_2)],
\label{I(18-s)}
\ee
where $s=4M$, $M=0,1,2,3,4$. The corresponding Nikulin point is $(r,a,\delta)=(20-4M,2,1)$.

For $s=4M$ the level-matching condition \eqref{LMC} requires $2v^2\in  \bz$ and there are several
choices for the shift $v$, as discussed in section \ref{subsubsec_UK}.
We will limit ourselves to the simple choice $(m,n,\alpha)=(0,1,0)$ so that
\begin{equation}
 v=\frac14\left(0, \frac{1}{2R}; \frac{1}{2R}\right).
 \label{vu140}
\end{equation}
For $w \in I^*/I$ we select  $w=({\overbrace{ 1,0,\dots,0}^{18-s},0;0})$, belonging to the vector class  $V_{18-s}$. Notice that $w^2=1$,
in agreement with \eqref{wn}.

\bigskip
\noindent
\underline{Gauge group}
 
In these examples the invariant lattice is of the form $I=U+K$, with $K=D_{18-s}$.
Since the shift $v$ has no components along $K$, the gauge group from $P\in I$ is 
then $U(1)\otimes SO(36-2s)$, as follows from the general analysis in section \ref{subsec:Untwisted}.

To deduce the gauge factors arising from momenta with $P_N\not =0$, we need to find the possibilities with 
$P_L^2=P_{IL}^2+P_{NL}^2=2$, and $P_R=0$, which can be done by looking at the classes in eq. \eqref{c182decomp}.
When $P_I=0$, it must be that $P_N \in N$ belongs either to $(Sc_s; Sc_2)$  or to $(V_s; V_2)$, as explained before.
Since $P_R=0$, necessarily $P_{NL} \in Sc_{s}$ and the solutions of $P_{NL}^2=2$ are the $2s(s-1)$ non-trivial
roots of $D_s$. Forming $g$-invariant combinations then gives $s(s-1)$ states, which correspond to the adjoint of
$SO(s)\otimes SO(s)$.
When both $P_I$ and $P_N$ are not zero, but again $P_R=0$, besides the scalar classes already considered, we are left
with $P_{IL} \in V_{18-s}$ and $P_{NL} \in V_s$. Together they give $(36-2s)\times(2s)$ additional solutions
of $P_L^2=2$, and therefore $s(36-2s)$ $g$-invariant states, which can be thought to transform as vectors
under $SO(s)\otimes SO(36-2s)$. In fact, the total number of states, with both $P_N=0$ and 
$P_N \not =0$, turns out to be $\left[\tfrac12 s(s-1) + \tfrac12 (36-s)(35-s)\right]$, which is precisely the
dimension of $SO(s) \times SO(36-s)$. The number of invariant vector states, and their transformation properties, 
indicate that the full gauge group is $U(1)\otimes SO(36-4M)\otimes SO(4M)$, adding the $U(1)$ from the $U$ lattice. 
As mentioned in the general discussion, when moving along moduli space to 
generic moduli points we expect that the normal lattice contributions 
become massive and the gauge group left will be $U(1) \otimes SO(36-8M)$. 

\bigskip
\noindent
\underline{Massless fermions}

Massless fermions are absent in these examples. The candidates with $P_{IL} \in V_{18-s}$ and $P_{NL} \in V_s$
are projected out because they have $e^{2i\pi P\cdot(2v+w)}=1$.

\bigskip
\noindent
\underline{Tachyons}

In the $g$-twisted sector there are generically tachyons, arising for instance from states with
$N_L=N_R=0$ and $p=(-1,0,0,0)$, as seen from the masses given in \eqref{leftrightmassesn1n3}. 
Further setting $P=(P_{18-s},\frac{k}{2R}+jR;\frac{k}{2R}-jR)$, with $P_{18-s}$ along the $(18-s)$
invariant directions, and taking  $P_{18-s}=0$, yields
 \be
m_L^2= \tfrac12\big(\!\left(k+\tfrac14\right)\tfrac{1}{2R}+jR\, \big)^{\!2}-\tfrac {4-M}{4},
\label{leftrightmassesg1sectors4m}
\ee 
which can be negative for specific values of $R$ when $M<4$. Thus, tachyons will be present provided there is level-matching
$m_L^2=m_R^2$. Indeed, this condition has 
integer solutions \mbox{$j=\frac{3-M}{4k+1}\in \bz$}, for each value of $M$, e.g. $k=0$, $j=3-M$.

Let us now inspect the $g^2$-twisted sector. For $P=(P_N,P_I)$ we take $P_N=(P_{s};P_{2R})$ and
$P_I=(P_{18-s},\frac{k}{2R}+jR;\frac{k}{2R}-jR)$, where 
$P_{s}$ and $P_{2R}$ are respectively along the $s$ left and the two right reflected directions,
and $P_{18-s}$ was defined before.
Recall that in the $(g^2,1)$ and $(g^2,g^2)$ sectors $P\in \Gamma$, while $P\in I$ in $(g^2,g)$ and $(g^2,g^3)$.
The level-matching condition \eqref{genericmassmatch} leads to
 \be
(k+\tfrac12)j + \tfrac12 (P_{18-s}+w)^2 -\tfrac12=
 -\tfrac12 P_ {s}^2+\tfrac12 P_{2R}^2,
 \label{massmatchings4m}
\ee
which can be satisfied taking $P_s=P_{2R}=0$, 
$P_{18-s}=0$ or $P_{18-s}=(-1,\underline{\pm1,0\dots,0})$,  and $j=0$.
This data implies
$m_L^2=\frac12 \big((k+\frac12)\frac1{2R}\big)^{\!2}\!-\!\frac12$, therefore tachyons  could appear
for $R>\frac14$. The projection phase \eqref{projphase}, present in $(g^2,g)$ and $(g^2,g^3)$ sectors, 
turns out to be $e^{-2i\pi\Phi}=-(-1)^M$, for $j=0$.
Hence, tachyons with $P \in I$ are invariant for $M$ even but will be projected out for $M$ odd.

Again, it is easy to see that tachyons appear when momenta have non vanishing components along the normal lattice.  
For instance, since classes are correlated we can choose momenta $P_{18-s} \in V_{18-s}$ and $P_s \in V_s$.
Now we can take $(P_{18-s}+w)=0$ but $P_s^2=1$ and the level-matching condition \label{massmatchings4m}
is still verified with $j=0$ and $P_{2R}=0$. Thus, the same values as above for negative masses are obtained.
Since $P_N\not=0$, there will be necessarily $g$-invariant tachyonic states.

\subsubsection{$s=10$ example}
\label{sec:s10}

In this example the invariant lattice is chosen to be
\begin{equation}
 I=U+E_8,
\end{equation}
corresponding to $(r,a,\delta)=(10,0,0)$.
It can be reached starting from a moduli point where $\Gamma$ takes the form in eq. \eqref{Gsds17E7}. 
The involution now acts by reflecting the 2 right-movers as well as the  $s=10$ left directions along  
 $\Gamma_{\!\!(10,2)}$. Thus, $N=\Gamma_{\!\!(10,2)}$. 
As shown in  section \ref{subsubsec_UK}, item 3, all shifts of order 4 are connected to $v$ given in
\eqref{vsolue8}, with $(m,n,\alpha)=(1,2,0)$. Besides, we can set $w=0$, according to eq.~\eqref{wn} 
since $s=2\,{\rm{mod}}\, 4$.

\bigskip
\noindent
\underline{Gauge group}

From the general analysis in section \ref{subsec:Untwisted}, we conclude that for $P\in I$, since the shift 
$v$ has no components along the $K$ lattice, the gauge group is simply the one associated to the root lattice of $K$, namely 
$U(1)\otimes E_8$. For generic moduli in $N$ this will be the full group, but there are extra factors for special moduli.
For instance, since $\Gamma_{\!\!(10,2)}$ is even self-dual, there exist moduli such that $N=U_{\text{sd}} + U_{\text{sd}} + E_8$, 
where an additional $U(1)^2\otimes SO(16)$ will appear from normal directions. The $U(1)$ factors originate in the $g$-invariant combinations of charged vectors in $U_{\text{sd}}$, whereas the  $SO(16)$ comes from symmetrized charged $E_8$ roots.  

\bigskip
\noindent
\underline{Tachyons}

In the $g$-twisted sector tachyons arise for particular values of $R$. For instance, from eq.~\eqref{leftrightmassesn1n3}
we find  that states with $P=0$, $p=(-1,0,0,0)$, and $N_L=N_R=0$, are tachyonic when 
\mbox{$(5-2\sqrt6)< R^2 <(5 + 2\sqrt6)$}.  

In the  $g^2$-twisted sector it is easy to see that tachyons appear when $P_N \not=0$.
Otherwise, for $P_N=0$ and $P_I=(\gamma, \frac{k}{2R}+jR; \frac{k}{2R}-jR)$, $\gamma \in E_8$,
the level-matching condition \eqref{genericmassmatch} has solutions with $\gamma=0$ and $(k,j)=(0,0),(-2,-1)$. 
The resulting mass is negative for $(3-2\sqrt2)< R ^2 <(3+2\sqrt2)$. Tachyons will  be  present 
in this range of $R$, because the projection phase \eqref{projphase} can be checked to be 1.

\subsubsection{$s=15$ example}

The invariant lattice is taken to be $I = U + A_1^3$, corresponding to $(r,a,\delta)=(5,3,1)$. 
The shift $w$ can be chosen as a fundamental weight in each $A_1$ component, thus $w^2=\frac32$,
in agreement with eq.~\eqref{wn} for $s=3\,{\rm{mod}}\, 4$.
From the general analysis we know that for a shift vector with no components along $K= A_1^3$, the resulting gauge group 
associated to $I$ is $U(1)\times SU(2)^3$. A possible such shift is $v_1=\frac14(1,1,0,0,0)$.
As discussed in section \ref{subsubsec_UK}, item 2, this $v_1$ cannot be connected by boosts to other allowed shifts
such as $v_2=\frac14(2,2,\sqrt2,0,0)$.
We will not go into a detailed analysis of the spectrum, which proceeds as in the previous examples.

\section{Moving in the moduli space of orbifold theories}\label{sec_motion}

A remarkable feature of the Nikulin orbifolds that we have constructed is the existence of a web of transitions between models with different $(r,a,\delta)$. 
The fundamental link is a decrease of $s$ by $1$, or equivalently an increase of $r$ by $1$, and it is realised through shuffling an $A_1$ component in the normal lattice $N$ to the invariant lattice $I$.
The question we would like to pose is: are all the points in figure \ref{figureNikulin} connected in this way?
Namely, starting from one point in the figure, can one make a sequence of transitions to reach any other point?

The answer to this question turns out to be positive.
We will first prove in subsection \ref{subsec_changing} that the partition functions of two models with $s$ differing by one --- through moving an $A_1$ lattice from $N$ to $I$ --- indeed agree.
This follows our conformal field theory analysis of the $SU(2)$ flat connections in section \ref{section_cft}.
We will see that upon turning on particular moduli one reaches one point or another.
This corresponds to a motion between the Higgs and Coulomb branches of the moduli space of the theory and will be analysed in detail.

In subsection \ref{sec_U3E8} we will perform a systematic analysis of transitions between models.
We start from $\ktl\simeq U^3 + E_8^2$ and apply the $\bz_2$ involution such that one or both $E_8$ factors are in the normal lattice.
We then shuffle the $A_1$ components in the $E_8$ lattice(s) from $N$ to $I$.
This changes the value of $s$ by $-1$ and so is referred to as an $s$-transition. 
The action of the orbifold also applies to the $U^3$ part and will play an important role in our analysis.
However, the $U^3$ components come for the ride and are not affected by the shuffling of the $A_1$ factors.
We will end this subsection by working out explicit examples of transitions.

It turns out that all but four points in figure \ref{figureNikulin} can be constructed using the method in subsection \ref{sec_U3E8}.
The remaining four points correspond to the triples $(10,8,0)$, $(10,10,0)$, $(14,6,0)$ and $(18,4,0)$.
The triples $(10,8,0)$ and $(10,10,0)$ have $E_8(2)$ components in their normal and invariant lattices.
The $E_8(2)$ lattice is obtained by exchanging the two $E_8$ factors.
The triples $(14,6,0)$ and $(18,4,0)$ have $U(2)^2$ components in their normal lattice which is obtained by starting from $\ktl\simeq U^2+\Gamma_{(9,1)} + E_8$ and defining the proper action of the $\bz_2$ involution on the even self-dual lattice $\Gamma_{(9,1)}$.
In subsection \ref{subsec_exm} we construct these four models and show how they are reached by making $s$-transitions from other points.

We therefore conclude that all the 75 models are connected.
This seems to suggest that these theories sit in a big moduli space where each triple $(r,a,\delta)$ and its associated moduli space form a subspace of it.
Furthermore, these subspaces are connected through $s$-transitions between the models.
There is however a caveat here: as discussed in subsection \ref{subsec_LMC}, there are particular models for which not {\it all} the shift vectors can be connected through automorphisms of the lattice --- see case 2, part (b), in subsection \ref{subsubsec_UK}.
This therefore raises the question whether the big moduli space is unique.

Let us consider the model with $I=U+A_1^3$ which was studied in subsection \ref{subsubsec_UK}.
We showed that there are two families of shift vectors in this theory which cannot be connected through turning on a Wilson line.
Furthermore, we showed that for this theory the two families of shift vectors cannot be connected by any automorphism of the lattice which may in general contain applying T-dualities and turning on any number of Wilson lines.
At this point one might conclude that the $I=U+A_1^3$ model sits in two disconnected big moduli spaces associated with the two families of the shift vectors.
There is however yet another interesting aspect of the $I=U+A_1^3$ model: using the methods we develop in this section, we can make a transition from this model to the theory with $I=U+E_7$ and map the two disconnected families of the shift vectors in $I=U+A_1^3$ to the shift vectors in the $I=U+E_7$ theory.
In the latter theory, however, {\it all} shift vectors are connected! (See item $4.$ in subsection \ref{subsubsec_UK}). 
Therefore, the disconnected families of shift vectors of $I=U+A_1^3$ are connected through making the $s$-transition to $I=U+E_7$.
This means that the existence of a unique big moduli space is still conceivable: while in a particular subspace of it there may be regions which are disconnected in that subspace, these regions are individually connected to another subspace within which all theories are connected.
Thus in this way the two families of the shift vectors are indeed connected in the big moduli space.

This reasoning however applies only to one model.
In order to draw a definite conclusion on the existence of a unique big moduli space one needs to analyse all theories with disconnected families of shift vectors. 
We plan to study this further in future work.

Table \ref{tab_IN}
summarises the invariant and normal lattices that we will construct in this section for all the triples in figure \ref{figureNikulin}.
We stress that the lattices in this table are found at a particular point in moduli space.

\afterpage{
{\footnotesize
\begin{center}
\setlength{\extrarowheight}{.15pt}
\begin{longtable}{|c||c|c|c|c|}
\hline
 $\#$ & $s$ & $(r,a,\delta)$ & $I$ & $N$ \\ 
\hline\hline
1 & 19 & $(1,1,1)$ & $A_1(-1)$ & $U^2+A_1+E_8^2$\\[1mm]\hline
2 & \multirow{3}{*}{18} &  $(2,0,0)$ & $U $& $U^2+E_8^2$\\
3 &  &  $(2,2,0)$ & $U(2)$ & $U(2)+U+E_8^2$\\
4 &  &  $(2,2,1)$ &  $A_1(-1)+A_1$ & $A_1(-1)+A_1+U+E_8^2$\\[1mm]\hline
5 &  \multirow{2}{*}{17}  & $(3,1,1)$ & $U+A_1$ & $U^2+E_7+E_8$\\
6 &  &  $(3,3,1)$ & $U(2)+A_1$ & $U(2)+U+E_7+E_8$\\[1mm]\hline
7 &  \multirow{2}{*}{16}  &  $(4,2,1)$ & $U+D_2$ & $U^2+D_6+E_8$\\
8 &  &  $(4,4,1)$ & $U(2)+D_2$ &$ U(2)+U+D_6+E_8$\\[1mm]\hline
9 &  \multirow{2}{*}{15}  &  $(5,3,1)$ & $U+D_2+A_1$ & $U^2+D_4+A_1+E_8$\\
10 &  &  $(5,5,1)$ & $U(2)+D_2+A_1$ & $U(2)+U+D_4+A_1+E_8$\\[1mm]\hline
11 &  \multirow{4}{*}{14}  &  $(6,2,0)$ & $U+D_4$ & $U^2+D_4+E_8$\\
12 &  &  $(6,4,0)$ & $U(2)+D_4$ & $U(2)+U+D_4+E_8$\\
13 &  &  $(6,4,1)$ & $U+D_2^2$ & $U^2+D_2^2+E_8$\\
14 &  &  $(6,6,1)$ & $U(2)+D_2^2$ & $U(2)+U+D_2^2+E_8$\\[1mm]\hline
15 &  \multirow{3}{*}{13}  &  $(7,3,1)$ & $U+D_4+A_1$ & $U^2+D_2+A_1+E_8$\\
16 &  &  $(7,5,1)$ & $U+D_2^2+A_1$ & $U^2+D_2^2+E_7$\\
17 &  &  $(7,7,1)$ & $U(2)+D_2^2+A_1$ & $U(2)+U + D_2^2 + E_7$\\[1mm]\hline
18 &  \multirow{4}{*}{12}  &  $(8,2,1)$ & $U+D_6$ & $U^2+D_2+E_8$\\
19 &  &  $(8,4,1)$ & $U+D_2+D_4$ &  $U^2+D_6+D_4$\\
20 &  &  $(8,6,1)$ & $U+D_2^3$ & $U^2+D_6 + D_2^2$\\
21 &  &  $(8,8,1)$ & $U(2)+D_2^3$ & $U(2)+U+D_6+D_2^2$\\[1mm]\hline
22 &  \multirow{5}{*}{11}  &  $(9,1,1)$ & $U+E_7$ & $U^2+A_1+E_8$\\
23 & &  $(9,3,1)$ & $U+D_6+A_1$ & $U^2+D_2+E_7$\\
24 & &  $(9,5,1)$ & $U+D_2+D_4+A_1$ & $U^2+D_6+D_2+A_1$\\
25 & &  $(9,7,1)$ & $U+D_2^3+A_1$ & $U^2+D_2^2+D_4+A_1$\\
26 & &  $(9,9,1)$ & $U(2)+D_2^3+A_1$ & $U(2)+U+D_2^2+D_4+A_1$\\[1mm]\hline
27 &  \multirow{11}{*}{10}  &  $(10,0,0)$ & $U+E_8$ & $U^2+E_8$\\
28 & &  $(10,2,0)$ & $U(2)+E_8$ & $U(2)+U+E_8$\\
29 & &  $(10,2,1)$ & $U+E_7+A_1$ & $U^2+A_1+E_7$ \\
30 & &  $(10,4,0)$ & $U+D_4^2$ & $U^2+D_4^2$ \\
31 & &  $(10,4,1)$ & $U+D_2+D_6$ & $U^2+D_6+D_2$\\
32 & &  $(10,6,0)$ & $U(2)+D_4^2$ & $U(2)+U+D_4^2$\\
33 & &  $(10,6,1)$ & $U+D_2^2+D_4$ & $U^2+D_2^2+D_4$\\
{34} & &  ${(10,8,0)}$ & $U+E_8(2)$ & $U^2+E_8(2)$\\
35 & &  $(10,8,1)$ & $U+D_2^4$ & $U^2+D_2^4$\\
{36} & &  ${(10,10,0)}$ & $U(2)+E_8(2)$ & $U(2)+U+E_8(2)$\\
37 & &  $(10,10,1)$ &  $U(2)+D_2^4$ & $U(2)+U+D_2^4$\\[1mm]
\hline
38 &  \multirow{6}{*}{9}  &  (11,1,1) & $U+E_8+A_1$ &$U^2+E_7$ \\*
39 & &  (11,3,1) & $U+D_2+E_7$ & $U^2+D_6+A_1$\\
40 & &  (11,5,1) & $U+D_6+D_2+A_1$ & $U^2+D_2+D_4+A_1$\\
41 & &  (11,7,1) & $U+D_2^2+D_4+A_1$ & $U^2+D_2^3+A_1$\\
42 & &  (11,9,1) & $U(2)+D_2^2+D_4+A_1$ & $U(2)+U+D_2^3+A_1$\\
43 & &  (11,11,1) & $U(2)+A_1+D_2^4$& $U(2)+A_1(-1)+D_2^4$\\[1mm]\hline
44 &  \multirow{5}{*}{8}  &  (12,2,1) & $U+E_8+D_2$ & $U^2+D_6$\\
45 & &  (12,4,1) & $U(2)+D_2+E_8$ & $U(2)+U+D_6$\\
46 & &  (12,6,1) & $U+D_2^2+D_6$ & $U^2+D_2^3$\\
47 & &  (12,8,1) & $U(2)+D_2^2+D_6$ & $U(2)+U+D_2^3$\\
48 & &  (12,10,1) & $U+A_1^2+D_2^4$ & $A_1^2(-1)+D_2^4$\\[1mm]\hline
49 &  \multirow{4}{*}{7}  &  (13,3,1) & $U+D_4+E_7$ & $U^2+D_4+A_1$\\
50 & & (13,5,1) & $U+D_2^2+E_7$ & $U^2+D_2^2+A_1$\\
51 & &  (13,7,1) & $U(2)+D_6+D_4+A_1$ & $U(2)+U+D_2^2+A_1$\\
52 & &  (13,9,1) & $U+A_1^2+D_2^2+D_4+A_1$ & $A_1^2(-1)+D_2^3+A_1$\\[1mm]\hline
53 &  \multirow{6}{*}{6}  &  (14,2,0) & $U+D_4+E_8$ & $U^2+D_4$\\
54 & &  (14,4,0) & $U(2)+D_4+E_8$ & $U(2)+U+D_4$\\
55 & &  (14,4,1) & $U+D_6^2$ & $U^2+D_2^2$\\
{56} & &  {(14,6,0)} & $U(2)+D_8+D_4$ & $U(2)^2+D_4$\\
57 & &  (14,6,1) & $U(2)+D_6^2$ & $U(2)+U+D_2^2$\\
58 & &  (14,8,1) & $U+A_1^2+D_2^2+D_6$ & $A_1^2(-1)+D_2^3$\\[1mm]\hline
59 &  \multirow{3}{*}{5}  &  (15,3,1) & $U+D_6+E_7$ & $U^2+D_2+A_1$\\
60 & &  (15,5,1) & $U(2)+D_6+E_7$ & $U(2)+U+D_2+A_1$\\
61 & &  (15,7,1) & $U+A_1^2+D_6+D_4+A_1$ & $A_1^2(-1)+D_2^2+A_1$\\[1mm]\hline
62 &  \multirow{3}{*}{4}  &  (16,2,1) & $U+D_6+E_8$ & $U^2+D_2$\\
63 & &  (16,4,1) & $U(2)+D_6+E_{8}$ & $U(2)+U+D_2$\\
64 & &  (16,6,1) & $U+A_1^2+D_6^2$ & $A_1^2(-1)+D_2^2$\\[1mm]\hline
65 &  \multirow{3}{*}{3}  &  (17,1,1) & $U+E_7+E_8$ & $U^2+A_1$\\
66 & &  (17,3,1) & $U(2)+E_7+E_8$ & $U(2)+U+A_1$\\
67 & &  (17,5,1) & $U+A_1^2+D_6+E_7$ & $A_1^2(-1)+D_2+A_1$\\[1mm]\hline
68 &  \multirow{5}{*}{2}  &  (18,0,0) & $U+E_8^2$ & $U^2$\\
69 & &  (18,2,0) & $U(2)+E_8^2$ & $U(2)+U$\\
70 & &  (18,2,1) & $U+A_1+E_7+E_8$ & $U+A_1(-1)+A_1$\\
{71} & &  {(18,4,0)} & $U(2)+D_{8}+E_8$ & $U(2)^2$\\
72 & &  (18,4,1) & $U+A_1^2+D_{6}+E_8$ & $A_1^2(-1)+D_2$\\[1mm]\hline
73 &  \multirow{2}{*}{1}  &  (19,1,1) & $U+A_1+E_8^2$ & $U+A_1(-1)$\\
74 & &  (19,3,1) & $U+A_1^2+E_7+E_8$ & $A_1^2(-1)+A_1$\\[1mm]\hline
75 & 0 &  (20,2,1) & $U+A_1^2+E_8^2$ & $A_1(-1)^2$\\[1mm]
\hline
\caption{Summary of invariant and normal lattices corresponding to each point in Figure \ref{figureNikulin}.}
\label{tab_IN}
\end{longtable}
\end{center}
}
}

\subsection{Changing $s$}\label{subsec_changing}

To clarify the main idea let us first consider the $(1,g)$ sector. In the partition function $Z_{0,1}$, given in eq.~\eqref{Z01},
we can identify separate contributions from non-compact bosons, fermions, the lattice $I$ and the lattice $N$.
In the $N$ contribution, denoted $Z_N(1,g)$, the lattice sum is absent and only
oscillator terms enter because the $s$ left-moving and the two right-moving normal directions are reflected. More precisely,
\be\label{zn1g}
Z_N(1,g) =  \left(\frac{2\eta}{\vartheta_2}\right)^{\!\! \frac{s}2}  \!\! \left(\frac{2\bar{\eta}}{\bar{\vartheta}_2}\right)
= \left(\frac{2\eta}{\vartheta_2}\right)^{\!\!\frac{s-1}2} \!\! \left(\frac{2\bar{\eta}}{\bar{\vartheta}_2}\right) \,
\frac{1}{\eta} \sum_{P \in A_1} q^{\frac12 P^2} e^{2\pi i P\cdot v_1}\ ,
\ee
where $v_1$ has components $(v_{1L};v_{1R})=(\frac1{2\sqrt2};0)$ and $2v^2_1=\frac14$. 
In the second equality we used the identities
\be\label{identitya1}
\left(\frac{2{\eta}}{{\vartheta}_2}\right)^{\!\!\frac{1}2}  
= \frac{1}{ q^{\frac1{24}} \prod\limits_n (1+  q^n)} =
\frac{1}{\eta} \sum_{n}  q^{n^2} e^{i\pi n} = 
\frac{1}{\eta} \sum_{P \in A_1} q^{\frac12 P^2} e^{2\pi i P\cdot v_1}\ .
\ee
We also used the fact that $P \in A_1$ means $P=\sqrt2 n$. 
The main take from the second form of $Z_N(1,g)$ in eq.~\eqref{zn1g} is that
a decrease of $s$ by 1 is accompanied by the emergence of a lattice sum over $A_1$. In turn this lattice sum can be
absorbed in the contribution of the invariant lattice thereby increasing $r$ by 1. The analysis further shows that 
if the mother theory characterized by $s$ has shift $v$, then the daughter theory with $(s-1)$ will have shift
$(v+v_1)$. A consistency check is that $2(v+v_1)^2 +\frac{s-1}4 = 2v^2 +\frac{s}4$, implying that
if the mother theory satisfies the level-matching condition (\ref{LMC}) so does its daughter.

So far we have argued that the partition functions of the mother and daughter theories do match in the $(1,g)$ sector.
To match the full partition functions of the two theories, it will be enough to consider the $(1,g)$ and $(1,g^2)$ sectors since all other sectors are obtained by modular transformations of these two sectors. 
Matching in the $(1,g^2)$ sector, however, is more involved because we need to specify how an explicit $A_1$ arises in the original lattice $N$.
We will shortly elaborate on this using particular examples in which we will also provide an interpretation of the underlying physics.

As explained in section \ref{subsec_LMC}, the smallest value of $r$ for which a shift $v$ satisfying level matching exists is $r=2$, 
with two possibilities for $a$. 
We will first show how models with $r \ge (a+2)$ and $\delta=1$ can be obtained starting from $(2,0,0)$.
The models with $r=a$ and/or $\delta=0$ will be discussed later.

Let us now examine the transitions starting from $(2,0,0)$. At this point we have $I=U$, whereas $N$ 
is a generic even self-dual lattice{\footnote{ We denote even self-dual lattices of signature $(8j+d,d)$ by $\Gamma_{\!\!(8j+d,d)}$.}
$\Gamma_{\!\!(18,2)}$. 
We take $N=\Gamma_{\!\!(10,2)} + E_8$, but other choices such as $N=U+\Gamma_{\!\!(17,1)}$ can also be made.
The next step is to find an $A_1$ inside $N$ in order to go from $r=2$ to $r=3$. The natural candidate in this example is 
an $A_1$ inside $E_8$. To proceed we then decompose $E_8$ in terms of $E_7 + A_1$. Vectors in the $E_8$ lattice split
as $(R_7,R_1)+(F_7,F_1)$. Here $R_7$ and $R_1$ denote the root lattices themselves, renamed for clarity, whereas $F_7$ and $F_1$ 
refer to weights in the conjugacy classes of ${\bf{56}}$ of $E_7$ and ${\bf{2}}$ of $A_1$ respectively. 

In the $r=2$ theory the partition function in the $(1,g)$ sector can be written as
\begin{equation}\label{z01r2}
Z_{0,1} = \hat{Z}_{0,1} \left(\frac{2{\eta}}{{\vartheta}_2}\right)^{\!\!\frac{1}2} ,
\end{equation}
where $\hat{Z}_{0,1}$ represents the partition function of all the remaining CFT, i.e. non-compact bosons, fermions, 
lattice $I=U$, and the remaining pieces $\Gamma_{\!\!(10,2)}$ and $E_7$ in $N$.
The $I$ part includes the dependence on the shift $v$,  which is assumed to fulfill level matching
$2 v^2 + \frac{s}4 \in \bz$ for $s=18$. 
The detailed form of $\hat{Z}_{0,1}$, which can be read from \eqref{Z01}, is not needed because it 
does not change in our method to go from $r=2$ to $r=3$.
The remaining term $\left(2{\eta}/{\vartheta}_2\right)^{\!\!\frac12}$ 
is precisely the partition function of the oscillator along $A_1$, which being part of $N$, does not come with a lattice sum.
We now use the identity \eqref{identitya1} to obtain
\begin{equation}\label{z01r3}
 Z_{0,1} = \hat{Z}_{0,1}\, \times \,  \frac{1}{\eta}  \sum_{P \in A_1} \bar{q}^{\frac{1}{2}P^2} e^{2\pi i P \cdot v_1} .
\end{equation}
Here $v_1$ has right component zero and left component equal to half the fundamental weight of $A_1$, i.e. as before
$(v_{1L}, v_{1R})=(\frac1{2\sqrt2},0)$ and $2v_1^2=\frac14$. 
Equation \eqref{z01r3} is exactly the partition function for the $(3,1,1)$ model where $I= U+A_1$ and $N=\Gamma_{\!\!(10,2)}+E_7$.
The new shift is $v+v_1$. As proven in general before (see below eq. (\ref{identitya1})), since $v$ satisfies level matching for $s = 18$ so does $v +v_1$ for $s = 17$.

We now look at the $(1,g^2)$ sector for $(2,0,0)$.
In this sector we must take into account the decomposition of $E_8 \subset N$ in terms of $E_7 + A_1$
because the partition function $Z_{0,2}$ effectively involves a sum over $N$. We then write
\begin{equation}\label{z02r2}
 Z_{0,2}=\hat{Z}_{0,2}^{R_7} \times \frac{1}{\eta} \sum_{P\in R_1} {q}^{\frac{1}{2}P^2} + 
\hat{Z}_{0,2}^{F_7} \times \frac{1}{\eta}  \sum_{P\in F_1} {q}^{\frac{1}{2}P^2} .
\end{equation}
Here $\hat{Z}_{0,2}^{R_7}$ is the partition function of all other degrees of freedom in the CFT, including the piece
from $E_7$ in the class $R_7$ that is correlated with the $A_1$ class $R_1$. Analogously, $\hat{Z}_{0,2}^{F_7}$
contains the part from $E_7$ in the class $F_7$ correlated with the $A_1$ class $F_1$. Explicit expressions for
$\hat{Z}_{0,2}^{R_7}$ and $\hat{Z}_{0,2}^{F_7}$ are not necessary as they remain unchanged in the 
transition from $r = 2$ to $r = 3$. Note also that for the $(2, 0, 0)$  model $w$ is trivial since $I^*_o$ is null for $I = U$ --- see above eq. (\ref{wcond}).

In the $(3, 1, 1)$ model with $I = U + A_1$ and $N = \Gamma_{\!\!(10,2)} + E_7$, $Z_{0,2}$ also includes sums over the
two classes in the weight lattice of $A_1$, which are correlated with the two classes in the weight lattice of $E_7$. 
In fact, in $(3, 1, 1)$ we have
\begin{equation}\label{z02r3}
  Z_{0,2}= \hat{Z}_{0,2}^{R_7} \times \frac{1}{\eta}\sum_{P\in R_1} {q}^{\frac{1}{2}P^2}e^{2\pi i P\cdot(2v_1+w_1)} 
  + \hat{Z}_{0,2}^{F_7}  \times \frac{1}{\eta}\sum_{P\in F_1} {q}^{\frac{1}{2}P^2}e^{2\pi i P\cdot(2v_1+w_1)}
\end{equation}
where $v_1$ was introduced before and $w_1$ is defined as $e^{2 i\pi P^2} = e^{2i\pi P\cdot w_1}$ for all $P$
in the weight lattice of $A_1$. It follows that $w_1$ is the fundamental weight of $A_1$ and that $2v_1 +w_1$
is a root of $A_1$. Therefore, $e^{2\pi i P\cdot(2v_1+w_1)}= 1$ for all $P$ in the weight lattice of $A_1$. 
We thus conclude that equations \eqref{z02r2} and \eqref{z02r3} are the same. This proves that in the $(1, g^2)$ sector 
the partition functions of the $(2, 0, 0)$ and $(3, 1, 1)$ models are exactly equal.

After showing that the $(2, 0, 0)$ model gives the same partition function as the $(3, 1, 1)$ model let us discuss the physical interpretation.
In the $(3, 1, 1)$ model, with $I = U + A_1$, the $A_1$ can be realized by a left-moving boson denoted $Y$.
The Kac-Moody currents are $J_3=\partial Y$ and $J_\pm=e^{\pm i \sqrt{2}Y}$.
In this case $g$ acts as $Y \to Y + 2\pi v_1$, where the shift $v_1$ has components $v_{1L} = \frac1{2\sqrt2}$ and  $v_{1R} = 0$.
Thus, $J_3 \to J_3$ and $J_\pm \to - J_\pm$ under $g$.
The exactly marginal operators of the theory are $J_3\bar\partial\bar X_3$ and $J_\pm\bar\partial\bar X_{1,2}$, where $\bar\partial\bar X_i$ are the right-moving bosons with $i=3$ being the direction in $I$ and $i=1, 2$ being the two right-moving directions in $N$ that are reflected by $g$.
The theory also has $U(1)$ massless gauge fields $J_3\bar\partial\bar X^\mu$ where $\mu$ refers to noncompact space-time directions.{\footnote{In the $-1$ ghost picture the operators correspond to $U(1)$ massless gauge fields $J_3\bar\psi^\mu$ as well as massless scalars $J_3\bar\psi^3$ and $J_\pm\bar\psi^{1,2}$, where $\bar\psi$ are right-moving fermions.}}

There are two possible deformations that can be made at this point --- see subsection \ref{subsec_mod} for related discussion.
One possibility is to give a vev to $J_3\bar\partial\bar X_3$.
This deformation is along what we call the Coulomb branch because it leaves the
$U(1)$ gauge symmetry unbroken. However, this vev will make $J_\pm\bar\partial\bar X_{1,2}$ massive. This is all part of
the $(3, 1, 1)$ moduli space. The second possibility is to give a vev to  e.g. $( J_+ + J_-)\bar\partial\bar X_1$.  This will
break the $U(1)$ gauge symmetry and in particular $J_3\bar\partial\bar X_3$  will become massive. We say that this
deformation is along the Higgs branch. In order to solve the Higgs branch at arbitrary points
it is convenient to use rebosonisation to write $( J_+ + J_-)=J'_3;=\partial Y'$.
On $Y'$, $g$ acts as reflection and we get to the moduli space of the $(2,0,0)$ model, where the $A_1$ is part of the $N$ lattice.

Note that in all the above discussion the details of the rest of the CFT did not play a role. All we needed in the  
$(2,0,0)$ model, was the appearance  of an $A_1$ point in the Higgs
branch (the moduli space of $(18,2)$ boosts in the $N$ direction) so that there is a massless $U(1)$ gauge field and
the corresponding scalar. Once that happens, we can give vev to the latter scalar. Some of the Higgs branch fields become 
massive now and we end up with the moduli space of the $(3,1,1)$ model.

The procedure to go from $(2,0,0$) to $(3,1,1)$ can be repeated to reach models with larger value of $r$.
For example, $(4,2,1)$ can be obtained from $(3,1,1)$, where $I=U+A_1$ and \mbox{$N=\Gamma_{\!\!(10,2)} + E_7$},
by decomposing $E_7$ in terms of $D_6+A'_1$, where the prime is put to distinguish from the $A_1$ in the previous step.
The two $E_7$ classes decompose as $R_7 =(Sc_{6},R'_1)+ (Sp_{6},F'_1)$ and \mbox{$F_7=(V_{6},F'_1)+(Sp'_{6},R'_1)$}. 
In the $(1,g)$ sector we now split $\hat{Z}_{0,1}$ in eq.~\eqref{z01r3} into an $A'_1$ piece and the rest of the CFT and write
\begin{equation}
  \hat{Z}_{0,1}= \hat{Z'}_{0,1}  \left(\frac{2{\eta}}{{\vartheta}_2}\right)^{\!\!\frac{1}2} .
 \end{equation}
This can be recast in the form of the $(4,2,1)$ model as
\begin{equation}
  \hat{Z}_{0,1}= \hat{Z'}_{0,1} \times \frac{1}{\eta} \sum_{P \in R'_1} {q}^{\frac{1}{2}P^2} e^{2\pi i P\cdot v_2}
\end{equation}
where $v_2$ is half the fundamental weight of $A'_1$. The total shift becomes $v+v_1+v_2$, which satisfies the level matching
condition for $r=4$.
Substituting back in eq.~\eqref{z01r3} yields
\begin{equation}
  Z_{0,1}= \hat{\zp}_{0,1} \times \frac{1}{\eta^2} 
  \sum_{P \in D_2} {q}^{\frac{1}{2}P^2} e^{2\pi i P \cdot (v_1+v_2)}
\end{equation}
where we have used the fact that the root lattice of $D_2$ is the scalar class $Sc_2$ of $SO(4)$, which is the same as $(R_1,R'_1)$. 
This is exactly the $Z_{0,1}$ for the $(4,2,1)$ model in which $I=U+D_2$ and $N=\Gamma_{\!\!(10,2)} + D_6$.

In the $(1,g^2)$ sector,  the $\hat{Z}_{0,2}^{R_7}$ and $\hat{Z}_{0,2}^{F_7}$  in eq.~\eqref{z02r3} become
\begin{equation}\label{z02r4aux}
\begin{split}
  \hat{Z}_{0,2}^{R_7}&=\hat{\zp}_{0,2}^{Sc_{6}}  \times \frac{1}{\eta} 
  \sum_{P\in R'_1} {q}^{\frac{1}{2}P^2}e^{2\pi i P\cdot (2v_2+w_2)} 
  + \hat{\zp}_{0,2}^{Sp_{6}}  \times \frac{1}{\eta}  
  \sum_{P\in F'_1} {q}^{\frac{1}{2}P^2}e^{2\pi i P\cdot (2v_2+w_2)}
\\[2mm]
  \hat{Z}_{0,2}^{F_7}&=\hat{\zp}_{0,2}^{Sp'_{6}} \times \frac{1}{\eta}  \sum_{P\in R'_1} {q}^{\frac{1}{2}P^2}e^{2\pi i P\cdot (2v_2+w_2)} +
  \hat{\zp}_{0,2}^{V_{6}}  \times \frac{1}{\eta}  \sum_{P\in F'_1} {q}^{\frac{1}{2}P^2}e^{2\pi i P\cdot (2v_2+w_2)}
\end{split}
\end{equation}
Here $w_2$ is the fundamental weight in $A'_1$.
Substituting this in eq.~\eqref{z02r3} then gives
\begin{equation}\label{z02r4}
\begin{split}
 & Z_{0,2}=\frac{1}{\eta^2} \Big[\hat{\zp}_{0,2}^{Sc_{6}} \sum_{P\in Sc_{2}} {q}^{\frac{1}{2}P^2}
 e^{2\pi i P\cdot (2(v_1+v_2)+(w_1+w_2))}
+  \hat{\zp}_{0,2}^{Sp_{6}}  \sum_{P\in Sp_{2}} {q}^{\frac{1}{2}P^2}e^{2\pi i P\cdot (2(v_1+v_2)+(w_1+w_2))}
  \\[2mm] 
  &\qquad\quad\;\; +
  \hat{\zp}_{0,2}^{V_{6}}  \sum_{P\in V_{2}} {q}^{\frac{1}{2}P^2}e^{2\pi i P\cdot (2(v_1+v_2)+(w_1+w_2))} 
  +  \hat{\zp}_{0,2}^{Sp'_{6}}  \sum_{P\in Sp'_{2}} {q}^{\frac{1}{2}P^2}e^{2\pi i P\cdot (2(v_1+v_2)+(w_1+w_2))} \Big]
  \end{split}
\end{equation}
where we have used the decompositions $Sc_{2}=(R'_1,R_1)$, $V_2=(F'_1,F_1)$, $Sp_2=(F'_1,R_1)$ and $Sp'_2= (R'_1,F_1)$
for the classes of $D_2$ in $A'_1+A_1$.
Since $w_1$ and $w_2$ are fundamental weights of  $A_1$ and $A'_1$, for $D_2$ now in $I$, it turns out that 
$w =w_1+w_2$ is in the $V_2$ class, in agreement with our general analysis for $w$ when $I=U+D_{4n+2}$ --- see the discussion below eq. (\ref{wcond}). 
Furthermore, all the phases above are trivial since $2v_1+w_1$ and $2v_2+w_2$ are roots of  $A_1$ and $A'_1$. 
Hence, $Z_{0,2}$ is clearly unchanged except that eq.~\eqref{z02r4} is the expression corresponding to the $(4,2,1)$ model.

Actually it is clear that at every step when we convert an $A_1$ which is in the $N$ directions, to an $A_1$ along $I$ directions  
there will be additional contributions to $v$ equal to half fundamental and to $w$ equal to the fundamental of the latter $A_1$. 
Therefore this will automatically ensure the level-matching condition (\ref{LMC}) and the condition \eqref{21operatorint} on $w$. Moreover,  $2v+w$ will be trivial and the form of $Z_{0,2}$ will remain unchanged.

So far we discussed connecting the first three points on the line $r=a+2$ in Figure \ref{figureNikulin}. 
We can continue the procedure to obtain various models with $I\simeq U + K$ for other values of $r$. 
However, at first sight it is not clear whether all models can be connected.
As one increases $r$, the first question arises at the transition from $r=5$ to $r=6$.
Concretely, we have to check whether $(6,4,1)$ and $(6,2,0)$, with $I\simeq U+D_2+D_2$ and $I\simeq U+D_4$ respectively, can 
both be reached starting from $(5,3,1)$.
It is clear that the first model can be obtained by taking the two $D_2=A_1+A_1$ in two different $E_8$ lattices in 
$N=\Gamma_{\!\!(2,2)}+E_8^2$. This can be shown as outlined above. 
For the second case where $D_4$ is embedded in one $E_8$, one might suspect that the above procedure could fail for the following reason. In the decomposition $D_4 \rightarrow A_1+A_1+A_1+A_1$,  the scalar class $Sc_4$ of $D_4$ splits 
as $(R_1,R_1,R_1,R_1)+(F_1,F_1,F_1,F_1)$ . Of course $Z_{0,2}$ will always work, as $2v+w$ in each $A_1$ factor is trivial. 
The problem would be with $Z_{0,1}$ where from the point of view of the  $(6,2,0)$ model the lattice sum must be in $I$ which 
includes $(R_1,R_1,R_1,R_1)+(F_1,F_1,F_1,F_1)$ in the $D_4$ part decomposed in terms of $ A_1+A_1+A_1+A_1$. However,
in the above procedure for every $A_1$ that we rebosonize only $R_1$ contributes to $Z_{0,1}$. Of course the solution is trivial. 
For our procedure of obtaining $v$, the part of $v$ in $D_4$, in the above decomposition, will be $\frac{1}{2}(F_1,F_1,F_1,F_1)$, 
so the contribution of $(F_1,F_1,F_1,F_1)$ to $Z_{0,1}$ will vanish because for each $A_1$ it involves the Jacobi function 
$\vartheta_1$ that is identically zero.

In the above we did not do anything to $U$. In the whole process of changing $s$ step by step, $U$ was a spectator.
Thus, we can proceed in the same manner with models that contain $U(2)$ in their invariant lattice without touching $U(2)$ --- see table \ref{tab_IN}.
Hence, it seems that all the models in the $U(2)$ series are also connected, except the one involving $E_8(2)$ that appears only for $r=10$
and that will be analyzed in subsection \ref{subsec_exm}.

\subsection{A systematic way of changing $s$ in $E_8$}\label{sec_U3E8}

Can all orbifold models associated with the triples $(r,a,\delta)$ in figure \ref{figureNikulin} be connected using the method of shuffling $A_1$ lattices?
To answer this question we start with the even self-dual lattice $\ktl \simeq\Gamma_0= U^3 + E_8^2$.
The idea is to focus on the $E_8$ components  inside $\Gamma_0$.
Since changing $s$ involves shuffling $A_1$'s, it is best to decompose $E_8$ in terms of $A_1^8$.
The form of $E_8$ vectors in the $A^8_1$ basis can be deduced in steps starting from the decompositions of $E_8$ under $D_4 + D_4$ with standard correlated conjugacy classes, namely 
\be\label{e8d4d4}
(Sc_4,Sc_4)+(V_4,V_4)+(Sp_4,Sp_4)+(Sp'_4,Sp'_4)\ .
\ee
Next each $D_4$ lattice is decomposed into $D_2 + D_2$ with correlated classes
\begin{alignat}{3}\label{d4d2d2}
Sc_4&=(Sc_2,Sc_2)+(V_2,V_2)\ ,\qquad\quad\; &V_4&=(Sc_2,V_2)+(V_2,Sc_2)\ , \nonumber \\
Sp_4&=(Sp_2,Sp_2)+(Sp'_2,Sp'_2)\ ,\qquad\quad  &Sp'_4&=(Sp_2,Sp'_2)+(Sp_2,Sp'_2)\ .
\end{alignat}
Finally, the $D_2$ lattice is decomposed into $A_1+A_1$ with correlations
\be\label{d2a1a1}
Sc_2 = (R,R), \quad V_2=(F,F), \quad Sp_2=(R,F), \quad \quad Sp'_2=(F,R)\ ,
\ee
where $R$ and $F$ are the root lattice and the fundamental class of $A_1$ (we are dropping a subscript 1 to simplify expressions).
The end result is that the groupings of classes are such that the number of $F$'s is multiple of 4, as required also
by the fact that $E_8$ is even. More precisely, there is one each of $R^8$ and $F^8$, plus 14 possible orderings
of $R^4 F^4$. 
The latter are such that in any two groupings there is an even number of $F$ overlaps, also needed because $E_8$ is integral.
One can also check that the total number of vectors with length square 2 is 240,  equal to the number of $E_8$ roots. 

Moving $A_1$'s from normal to invariant directions is straightforward in terms of the decomposition under $A^8_1$.
Suppose that we start from a model where all the directions of one $E_8$, say the first, are reflected. 
Then $I_1$ coming from this $E_8$ is null and $N_1=E_8$. By moving one $A_1$ of this normal lattice to invariant directions
we land at $I_1=A_1$ and $N_1=E_7$, which can be seen by decomposing $E_7$ in terms of $A_1^7$. Besides,
the contribution to $a$ from this $E_8$, denoted $a_1$, is 1. Thus, after the decrease $\Delta_{s_1}=-1$ 
we have data $(I_1,N_1,a_1)=(A_1,E_7,1)$. We can further move two and three $A_1$'s in a similar way.
If we move four $A_1$'s to invariant directions there are two possiblities depending on the relative positions.
Moving $s_1$  $A_1$'s is the same as moving $(8-s_1)$ $A_1$'s but exchanging $I_1$ and $N_1$
with $a_1$ remaining unchanged. The results are shown in table \ref{tab_e8}.
The second column in this table can also be understood as giving the corresponding invariant and normal lattices
of the possible $\bz_2$ automorphisms of $E_8$. Indeed, the results can be obtained applying the same theorem \cite{Nikulin83} 
that produces Figure \ref{figureNikulin}.
\begin{table}[h!]\begin{center}
\renewcommand{\arraystretch}{1.05}
\setlength\tabcolsep{5pt}
{\normalsize{
\begin{tabular}{|c|c|}
\hline
$\phantom{-}\Delta_{s_1}$ & $(I_1,N_1,a_1)$ \\
\hline
$\phantom{-}0$ &  $(\text{null},E_8,0)$  \\ \hline
$-1$ &  $(A_1,E_7,1)$  \\ \hline
$-2$ &  $(A_1^2,D_6,2)$  \\ \hline
$-3$ &  $(A_1^3,A_1+D_4,3)$  \\ \hline
$-4$ &  $(A_1^4,A_1^4,4)$  \\ \hline
$-4$ &  $(D_4,D_4,2)$  \\ \hline
$-5$ &  $(A_1+D_4,A_1^3,3)$  \\ \hline
$-6$ &  $(D_6,A_1^2,2)$  \\ \hline
$-7$ &  $(E_7,A_1,1)$  \\ \hline
$-8$ &  $(E_8,\text{null},0)$ 
 \\ \hline
\end{tabular}
}}
\caption{Invariant and normal lattices obtained from moving $A_1$'s in $E_8$. }
  \label{tab_e8}\end{center}\end{table}

The same analysis works for the second $E_8$ and yields $(I_2,N_2,a_2)$ as in table \ref{tab_e8}.
Suppose we start from the $(2,0,0)$ model with $I=U$ and $N=U + E_8^2$. Using the transitions in table \ref{tab_e8} in the two $E_8$'s 
we obtain models with $I=U+K=U+I_1+I_2$ up to a maximum value of $a=8$.
These models are located inside the dashed blue lines in figure \ref{fig_nik_2}.
There are 41 points inside this region and all but 5 of them are connected in this way.
The 5 points are: $(6,4,0)$, $(10,2,0)$, $(10,6,0)$, $(10,8,0)$ and $(14,4,0)$ and are denoted by blue circles in the figure.
Except for $(10,8,0)$, the remaining four points can be reached from $I=U(2)$ and $N=U(2) + U + E_8^2$.
Note that the $(10,2,0)$ model has $I\simeq U(2)+E_8\simeq U+D_8$ and can be obtained by replacing $E^2_8$ in $\ktl$ by the $Spin(32)/\bz_2$ lattice. 

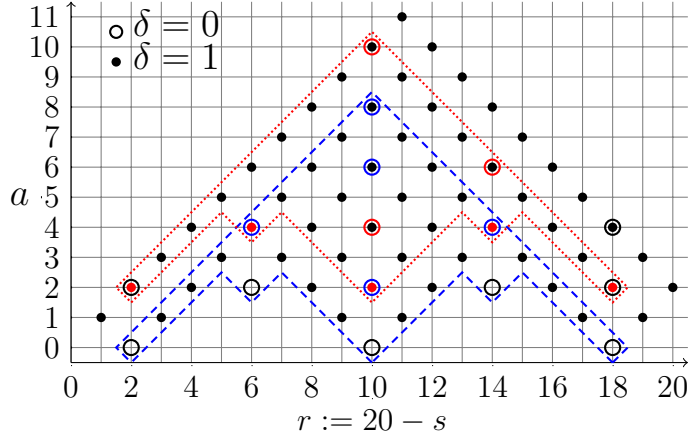
\begin{figure}[h!]
\centering
\begin{tikzpicture}[scale=0.4]
    \draw[gray] (0, -0.5) grid (20.5, 11.5);    \draw[thin, black, ->] (0, -0.5) -- (20.5, -0.5);
      \foreach \x in {0,2,4,6,8,10,12,14,16,18,20}
      \draw[thin] (\x, -0.6) -- (\x, -0.6)        node[anchor=north] {\(\x\)};
    \foreach \x in {10}      \draw[thin] (\x, -1.7) -- (\x, -1.7)        node[anchor=north] {$r:=20-s$};    \draw[thin, black, ->] (0,-0.5) -- (0, 11.5);
    \foreach \y in {0,1,2,3,4,5,6,7,8,9,10,11}      \draw[thick] (-0.0, \y) -- (0.0, \y)        node[anchor=east] {\(\y\)};
    \foreach \y in {5}      \draw[thick] (-1.0, \y) -- (-1.0, \y)        node[anchor=east] {\large$a$};
    
    \draw[thick, black, fill=black] (1,1) circle (1.2mm);    
    \draw[thick, black] (2,0) circle (2.5mm);
    \draw[thick, black] (2,2) circle (2.5mm);
    \draw[thick, red, fill=red] (2,2) circle (1.2mm);

    \draw[thick, black, fill=black] (3,1) circle (1.2mm);
    \draw[thick, black, fill=black] (3,3) circle (1.2mm);

    \draw[thick, black, fill=black] (4,2) circle (1.2mm);
    \draw[thick, black, fill=black] (4,4) circle (1.2mm);

    \draw[thick, black, fill=black] (5,3) circle (1.2mm);
    \draw[thick, black, fill=black] (5,5) circle (1.2mm);

    \draw[thick, black] (6,2) circle (2.5mm);
    \draw[thick, blue] (6,4) circle (2.5mm);
    \draw[thick, red, fill=red] (6,4) circle (1.2mm);
    \draw[thick, black, fill=black] (6,6) circle (1.2mm);

    \draw[thick, black, fill=black] (7,3) circle (1.2mm);
    \draw[thick, black, fill=black] (7,5) circle (1.2mm);
    \draw[thick, black, fill=black] (7,7) circle (1.2mm);

    \draw[thick, black, fill=black] (8,2) circle (1.2mm);
    \draw[thick, black, fill=black] (8,4) circle (1.2mm);
    \draw[thick, black, fill=black] (8,6) circle (1.2mm);
    \draw[thick, black, fill=black] (8,8) circle (1.2mm);

    \draw[thick, black, fill=black] (9,1) circle (1.2mm);
    \draw[thick, black, fill=black] (9,3) circle (1.2mm);
    \draw[thick, black, fill=black] (9,5) circle (1.2mm);
    \draw[thick, black, fill=black] (9,7) circle (1.2mm);
    \draw[thick, black, fill=black] (9,9) circle (1.2mm);

    \draw[thick, black] (10,0) circle (2.5mm);
    \draw[thick, blue] (10,2) circle (2.5mm);
    \draw[thick, red, fill=red] (10,2) circle (1.2mm);
    \draw[thick, red] (10,4) circle (2.5mm);
    \draw[thick, black, fill=black] (10,4) circle (1.2mm);
    \draw[thick, blue] (10,6) circle (2.5mm);
    \draw[thick, black, fill=black] (10,6) circle (1.2mm);
    \draw[thick, blue] (10,8) circle (2.5mm);
    \draw[thick, black, fill=black] (10,8) circle (1.2mm);
    \draw[thick, red] (10,10) circle (2.5mm);
    \draw[thick, black, fill=black] (10,10) circle (1.2mm);

    \draw[thick, black, fill=black] (11,1) circle (1.2mm);
    \draw[thick, black, fill=black] (11,3) circle (1.2mm);
    \draw[thick, black, fill=black] (11,5) circle (1.2mm);
    \draw[thick, black, fill=black] (11,7) circle (1.2mm);
    \draw[thick, black, fill=black] (11,9) circle (1.2mm);
    \draw[thick, black, fill=black] (11,11) circle (1.2mm);
    
     \draw[thick, black, fill=black] (12,2) circle (1.2mm);
     \draw[thick, black, fill=black] (12,4) circle (1.2mm);
    \draw[thick, black, fill=black] (12,6) circle (1.2mm);
    \draw[thick, black, fill=black] (12,8) circle (1.2mm);
    \draw[thick, black, fill=black] (12,10) circle (1.2mm);

     \draw[thick, black, fill=black] (13,3) circle (1.2mm);
    \draw[thick, black, fill=black] (13,5) circle (1.2mm);
    \draw[thick, black, fill=black] (13,7) circle (1.2mm);
    \draw[thick, black, fill=black] (13,9) circle (1.2mm);

    \draw[thick, black] (14,2) circle (2.5mm);
    \draw[thick, blue] (14,4) circle (2.5mm);
    \draw[thick, red, fill=red] (14,4) circle (1.2mm);
    \draw[thick, red] (14,6) circle (2.5mm);
    \draw[thick, black, fill=black] (14,6) circle (1.2mm);
    \draw[thick, black, fill=black] (14,8) circle (1.2mm);

     \draw[thick, black, fill=black] (15,3) circle (1.2mm);
    \draw[thick, black, fill=black] (15,5) circle (1.2mm);
    \draw[thick, black, fill=black] (15,7) circle (1.2mm);

     \draw[thick, black, fill=black] (16,2) circle (1.2mm);
    \draw[thick, black, fill=black] (16,4) circle (1.2mm);
    \draw[thick, black, fill=black] (16,6) circle (1.2mm);

     \draw[thick, black, fill=black] (17,1) circle (1.2mm);
    \draw[thick, black, fill=black] (17,3) circle (1.2mm);
    \draw[thick, black, fill=black] (17,5) circle (1.2mm);

    \draw[thick, black] (18,0) circle (2.5mm);
    \draw[thick, black] (18,2) circle (2.5mm);
    \draw[thick, red, fill=red] (18,2) circle (1.2mm);
    \draw[thick, black] (18,4) circle (2.5mm);
    \draw[thick, black, fill=black] (18,4) circle (1.2mm);

     \draw[thick, black, fill=black] (19,1) circle (1.2mm);
    \draw[thick, black, fill=black] (19,3) circle (1.2mm);

     \draw[thick, black, fill=black] (20,2) circle (1.2mm);

\node[above,black] at (3.5,8.9) {\large ${\delta=1}$};
\node[above,black] at (3.5,10) {\large$\delta=0$};
    \draw[thick, black] (1.5,10.5) circle (2mm);
    \draw[thick, black, fill=black] (1.5,9.5) circle (1.2mm);
\draw[densely dashed,thick,blue] (18.0,-0.5)--(18.5,0)--(10,8.5)--(1.5,0)--(2,-0.5) (10,-0.5)--(7,2.5)--(6,1.5)--(5,2.5)--(2,-0.5) (18,-0.5)--(15,2.5)--(14,1.5)--(13,2.5)--(10,-0.5);
\draw[densely dotted,thick,red] (2,1.5)--(5,4.5)--(6,3.5)--(7,4.5)--(10,1.5)--(13,4.5)--(14,3.5)--(15,4.5)--(18,1.5)--(18.5,2)--(10,10.5)--(1.5,2)--(2,1.5);
  \end{tikzpicture}
\caption{Starting from the $(2,0,0)$ model and shuffling $A_1$ lattices in the two $E_8$ components of $\ktl\simeq\Gamma_0=U^3+E_8^2$ as in table \ref{tab_e8}, we connect to models with $I=U+K$ and $a\le8$. These models are located inside the dashed blue lines. Five points depicted as blue circles are not obtained in this way. Starting from $(2,2,0)$ model and moving the $A_1$ lattices we connect to models with $I=U(2)+K$ and $a\le10$ located inside the dotted red lines. The red dots and circles in this region may not be obtained in this way.}
\label{fig_nik_2}
\end{figure}

Larger values of $a$ (i.e. $a>8$) can be obtained by changing the action of the $\bz_2$ involution on the $U^3$ block in $\ktl$.
The contribution from this block to $I$ and $N$ lattices are denoted as $I_3$ and $N_3$, of signature $(3-s_3, 1)$ and $(s_3,2)$, respectively.
In the previous paragraph we considered an involution that reflects two of the $U$ factors and hence obtained $I_3=U$, $N_3=U^2$, $s_3=2$ and $a_3=0$.
Higher values of $a_3$ may be obtained by exchanging two $U$'s which yields a $U(2)$ component in both $I_3$ and $N_3$.
Another possibility is to take the $U$ lattice at the $SU(2)$ point, i.e. $(A_1;A_1)[(F_1;F_1)]$, and reflect only the left or the right part of it.
Combining these actions we find that $(I_3,N_3,a_3)$ are of the form given in table \ref{tab_u3} (recall that $U(2) + A_1 \simeq A_1(-1) + A^2_1$).
\begin{table}[h!]\begin{center}
\renewcommand{\arraystretch}{1.15}
\setlength\tabcolsep{5pt}
{\normalsize{
\begin{tabular}{|c|c|}
\hline
$s_3$ & $(I_3,N_3,a_3)$ \\
\hline
$3$ & $(A_1(-1), U^2 + A_1, 1)$ \\ \hline
$2$ & $(U, U^2, 0)$,\quad $(A_1(-1)+ A_1, A_1(-1)+ A_1+U, 2)$,\quad $(U(2), U(2)+U, 2)$ \\ \hline
$1$ & $(U+A_1, U+ A_1(-1),1)$,\quad $(U(2) + A_1, U(2)+A_1(-1), 3)$ \\ \hline
$0$ & $(U+A_1^2, A^2_1(-1), 2)$
 \\ \hline
\end{tabular}
}}
\caption{Invariant and normal lattices embedded in $U^3$. }
  \label{tab_u3}\end{center}\end{table}

Suppose we take $I_3=U(2)$ and start from $I=U(2)$ and $N=U(2)+U+E_8^2$.
Moving the $A_1$ components in the two $E_8$'s we connect to models with a maximum value of $a = 10$.
These models are located inside the region surrounded by dotted red lines in figure \ref{fig_nik_2}.
There are again some models inside this domain which may not be obtained in this way and are denoted by red dots and circles in the figure.
Some of them may be reached from other choices of $I_3$ in table \ref{tab_u3}.
The remaining possible forms of the $I_3$ lattice yield similar domains as the ones shown in figure \ref{fig_nik_2} but with values of $r$ shifted.
We will not go through all the details here but only emphasise the important results.
It turns out that {\it all} but four models $(r,a,\delta)$ in figure \ref{fig_nik_2} have invariant and normal lattices that can be written as $I=I_1+I_2+I_3$, $N=N_1+N_2+N_3$, $a=a_1+a_2+a_3$, where $(I_3,N_3,a_3)$ are given in table \ref{tab_u3} and $(I_1,N_1,a_1)$ and $(I_2,N_2,a_2)$ are given in table \ref{tab_e8}.
We have collected the list of $I$ and $N$ lattices which are constructed in this way in table \ref{tab_IN}.

The four exceptions are the entries 34, 36, 56 and 71 in the table.
The former two have $E_8(2)$ components in their $I$ and $N$ lattices and the latter two have $U(2)^2$ components in their normal lattice.
As such, they cannot be constructed using the method developed in this subsection.
We will construct these four models in subsection \ref{subsec_exm} and show how they are connected to the models we constructed here.

So far we have shown that the models inside each $I_3$ domain are connected to each other through moving the $A_1$ components in $E_8^2$. {\footnote{Recall that the asymmetric orbifold construction may not be realised for the triples $(1,1,1)$ and $(2,2,1)$ --- see the discussion at the beginning of section \ref{subsec_LMC}.}}
There are seven such domains corresponding to the entries in table \ref{tab_u3}.
Are the different $I_3$ domains connected to each other?
The answer is yes.
This is because there are common triples in the overlap of any two domains.
Consider two domains $I_3$ and $I_3'$.
The models located in the overlap of these two domains may be described in two ways: $I\simeq I_3+K\simeq I_3'+K'$ where the two descriptions are related by a boost transformation in $SO(r-1,1)$.
The former description is connected to the $I_3$ domain.
Upon performing a boost transformation we obtain the latter description which is connected to all points in the $I_3'$ domain.
Therefore the two domains are indeed connected to each other.

As an example consider the two domains with $I_3=U$ and $I_3'=U(2)$ in figure \ref{fig_nik_2}.
Start from the triple $(6,6,1)$ with $I=U(2)+A_1^3+A_1$ and $N=U(2)+U+A_1+D_4+E_7$.
In the invariant lattice the $A_1^3$ components come from the same $D_4$ whereas the remaining $A_1$ has been moved from a different $D_4$. 
Upon moving the remaining $A_1$ in the former $D_4$, we make a transition to the triple $(7,5,1)$ with $I=U(2)+D_4+A_1$ and $N=U(2)+U+D_4+E_7$.
Next we make a boost transformation in $SO(6,1)$ to obtain $I=U+D_2^2+A_1$.
The associated normal lattice is $N=U^2+D_2^2+E_7$.
We can now use $A_1$ shifts to reach other points within the $I_3=U$ domain.

We end this subsection by providing an explicit example of making transitions between models in oder to explain the ideas and the method developed so far.

\subsubsection{Example: transitions from $(12,8,1)$}\label{subs12_8_1}

This example serves to illustrate transitions from the $r=(a+4)$ line in figure \ref{figureNikulin}.
We begin by describing the invariant and normal lattices corresponding to a $\bz_2$ involution acting on 
$\ktl\simeq\Gamma_0=U^3 + E_8^2$, now with the second and third $U$'s at the $SU(2)$ point.
Under $\bz_2$ the first $U$ is untouched, while in the second and third only the right part is reflected.
Thus, from $U^3$ there is a contribution  $I_3=U+A_1^2$ to $I$ and $N_3=A_1(-1)^2$ to $N$, with $A_1^2$ classes in $I^*$
correlated with $A^2_1(-1)$ classes in $N^*$. The action on $E^2_8$ is defined below.

One of the $E_8$ lattices, say the first, is decomposed as $D_4+D_4$, with standard correlated conjugacy classes, see eq. \eqref{e8d4d4}.
We then reflect one $D_4$ to obtain further contributions $I\supset D_4$ and $N \supset D_4$, with their classes correlated in $I^*$ and $N^*$.
The second $E_8$ is also decomposed under $D_4+D_4$ but we resort to triality in one factor to exchange $V_4$ and $Sp_4$. 
The classes are then correlated as
\be\label{e8d4d4triality}
(Sc_4,Sc_4)+(V_4,Sp_4)+(Sp_4,V_4)+(Sp'_4,Sp'_4).
\ee
In this second $E_8$ we then split each $D_4$ factor as $D_2 + D_2$, each with correlations given in eq. \eqref{d4d2d2}, and then reflect one $D_2$ from each of the $D_4$ factors. 
As explained before, the invariant piece from the second $E_8$ must be $D_2 + D_2 =A_1^4$. Moreover,  the normal piece from the
second $E_8$ will also be $D_2+D_2$. The conjugacy classes of $I^*$ and $N^*$ will be correlated according to the decompositions 
made in the process.

 All in all, we find $I =U + A_1^2 + D_4 + D_2^2 $, $N = A_1(-1)^2 + D_4 + D_2^2$ with definite correlations between $I^*$ and $N^*$ classes. 
While the correlation of the classes of $A_1(-1)$ with $A_1$ and $D_4$ with $D_4$ is standard, the correlation of classes of $D_2^2$ of $N$ with  $D_2^2$ of $I$, which is a bit non-standard, is given by
\begin{align}\label{D2_2}
 &~ (D_2^*/D_2,D_2^*/D_2;D_2^*/D_2,D_2^*/D_2)  \\
&= (Sc_2,Sc_2;Sc_2,Sc_2)+(V_2,Sc_2;V_2,Sc_2)+(Sc_2,V_2;Sc_2,V_2)+(V_2,V_2;V_2,V_2)  \nonumber \\
&+ (Sc_2,Sp_2;V_2,Sp_2)+(V_2,Sp_2;Sc_2,Sp_2)+(Sc_2,Sp'_2;V_2,Sp'_2)+(V_2,Sp'_2;Sc_2,Sp'_2) \nonumber \\
&+ (Sp_2,Sc_2;Sp_2,V_2)+(Sp'_2,Sc_2;Sp'_2,V_2)+(Sp_2,V_2;Sp_2,Sc_2)+(Sp'_2,V_2;Sp'_2,Sc_2) \nonumber  \\
  &+ (Sp_2,Sp_2;Sp'_2,Sp'_2)+(Sp'_2,Sp_2;Sp_2,Sp'_2)+(Sp_2,Sp'_2;Sp'_2,Sp_2)+(Sp'_2,Sp'_2;Sp_2,Sp_2). \nonumber
\end{align}
where we used the ordering $(N^*/N;I^*/I)$.

Now we can take one of the $A_1$ in $N$ (it could be either from $D_2^2$ or an $A_1$ inside $D_4$) and rebosonize it so that it 
becomes part of $I$ with a shift equal to half the fundamental in that $A_1$. 
To find the new $I$, we need to figure out how the fundamental of that $A_1$ that we have picked in $N$ is correlated to the rest of the conjugacy classes.

First consider $A_1$ in one of the $D_2$ in $N$ and decompose that  $D_2$ in terms of $A_1^2$. 
Under $A^2_1$ the various conjugacy classes of $D_2^*/D_2$ are given in eq. \eqref {d2a1a1}.
Finally we move the second $A_1$ to $I$. Then the right hand side of eq. \eqref{D2_2} becomes 
\begin{align}
  &\quad\  (Sc_2,R;R,Sc_2,Sc_2)+(V_2,R;R,V_2,Sc_2)+(Sc_2,F;F,Sc_2,V_2)+(V_2,F;F,V_2,V_2) \\
                                           &+ (Sc_2,R;F,V_2,Sp_2)+(V_2,R;F,Sc_2,Sp_2)+(Sc_2,F;R,V_2,Sp'_2)+(V_2,F;R,Sc_2,Sp'_2)  \nonumber  \\
                                           &+ (Sp_2,R;R,Sp_2,V_2)+(Sp'_2,R;R,Sp'_2,V_2)+(Sp_2,F;F,Sp_2,Sc_2)+(Sp'_2,F;F,Sp'_2,Sc_2)  \nonumber\\
  &+ (Sp_2,R;F,Sp'_2,Sp'_2)+(Sp'_2,R;F,Sp_2,Sp'_2)+(Sp_2,F;R,Sp'_2,Sp_2)+(Sp'_2,F;R,Sp_2,Sp_2)\ .
      \nonumber
\end{align}
In the above we have moved the second $A_1$ to the right of the semicolon because we have rebosonized that $A_1$ so that it is 
along the new $I$ directions. 
The new invariant piece from the second $E_8$  will be obtained by taking all the vectors that have zero entry on the left of the semicolon which means that the left of the semicolon must be $(Sc_2,R)$. Hence, this new invariant piece  
will be $(R,Sc_2,Sc_2) + (F,V_2,Sp_2)$. 
Note that the first class is just $A_1+D_2+D_2$, while the second one admits 16 vectors of square length 2.
Thus, the two terms altogether give 31 vectors of square length 2, which are actually the roots of $D_4 + A_1$. 
This can be seen by writing the last $D_2$ in  $A_1+D_2+D_2$ as $A_1+A_1$, and noting that $Sp_2$ is neutral with respect to the second $A_1$. Combining all the remaining parts of $I$ we find that the new full invariant lattice is $I'=U+A_1^3+D_4^2$,
which corresponds to the $(13,7,1)$ model. 
The new normal lattice is $N'= A_1(-1)^2+D_4+D_2+A_1$.

What if we had moved an $A_1$ inside $N$ from the first $E_8$, namely from $D_4$ which we write as $D_2+A_1+A_1$ ? 
The $E_8$ conjugacy classes before rebosonizing an $A_1$ are
\begin{align}
  &\quad\ (Sc_2,R,R;Sc_4)+(V_2,F,F;Sc_4)+(Sc_2,F,F;V_4)+(V_2,R,R;V_4)\nonumber\\
  &+(Sp_2,R,F;Sp_4)+ (Sp'_2,F,R;Sp_4)+(Sp'_2,R,F;Sp'_4)+ (Sp_2,F,R;Sp'_4).
\end{align}
Next we move the second $A_1$ to the $I$ direction which means moving the semicolon one place to the left. 
The new $I'$ can come only from states that have zero in the first two entries, i.e. they must be $(Sc_2,R)$ in the normal direction. 
Thus, the invariant piece coming from this $E_8$ will be $A_1+D_4$. Combining with the remaining parts of $I$ we get $I'=U+A_1^3+D_4+D_2^2$ and $N'=A_1(-1)^2+D_2^2+A_1+D_2$, which give the $(13,9,1)$ model. 
Hence, the end point depends on which $A_1$ is moved from $N$ to $I$.

We can continue the procedure. For instance, suppose we move two $A_1$'s, one from the first $E_8$ and another from the second 
$E_8$. The new invariant lattice can be found combining the results in the above paragraphs for the 2 $E_8$'s, together with $U+A_1^2$. In this way we obtain $I'=U+A_1^2+(A_1+D_4)+(A_1+D_4)$, which gives the $(14,8,1)$ model with $N'=A_1(-1)^2+A_1^2+D_2^2$.

\subsection{The remaining triples}\label{subsec_exm}
In the previous subsections we developed a method for making transitions between orbifold models.
We started from $\ktl\simeq\Gamma_0=U^3+E_8^2$, defined the action of the $\bz_2$ involution to obtain $I$ and $N$, and moved $A_1$ components from the $N$ to $I$ to change the value of $s$.
We observed that there are four triples in figure \ref{figureNikulin} which cannot be constructed in this way.
These points are $(10,8,0)$, $(10,10,0)$, $(14,6,0)$ and $(18,4,0)$.
Triples $(10,8,0)$ and $(10,10,0)$ contain $E_8(2)$ in their $I$ and $N$ lattices which is obtained by exchanging the two $E_8$'s.
Triples $(14,6,0)$ and $(18,4,0)$ have $U(2)^2$ in $N$ which cannot not be obtained from the involutions of the $U^3$ block as can be seen in table \ref{tab_u3}.
In this subsection we construct these four models and show how to connect to them by making $s$-transitions from other points inside the $I_3$ domains.
We therefore conclude that all the 75 triples in figure \ref{figureNikulin} are connected to each other through $s$-transitions.

\subsubsection{$(10,8,0)$} \label{subs10_8_0}

For the theory with $(r,a,\delta)=(9,9,1)$, the invariant lattice is generically $I \simeq U(2)+A_1^7$. 
As shown in eq. \eqref{equivUA17} there is a point in moduli space where
$I= A_{1}(-1)+E_8(2)$, which happens to be more convenient to start to survey possible transitions.
This point can also be obtained directly from $\Gamma_{\!\!(19,3)}\simeq\Gamma_{\!\!(2,2)}+U+E_8^2$, 
taking the $\bz_2$ involution to act by reflecting $\Gamma_{\!\!(2,2)}$, exchanging the two $E_8$'s, and reflecting the left part of 
$U$ at the $SU(2)$ point, i.e. at the self-dual radius. 
In this case $N=\Gamma_{\!\!(2,2)}+A_1+E_8(2)$. Besides, the conjugacy classes of $I^*/I$ and $N^*/N$ are correlated 
between $A_{1}(-1)^*/A_{1}(-1)$ and $A_1^*/A_1$, as well as between the $E_8(2)^*/E_8(2)$'s in $I$ and $N$. 
Now, in the $A_1$ appearing in $N$ we can rebosonize $J_1\rightarrow J'_3$ as discussed in subsection \ref{subsec_changing}. 
The result is that the $A_1$ of $N$ will move to $I$ and since the $A_{1}(-1)^*/A_{1}(-1)$ and $A_1^*/A_1$ classes were correlated, the new $I$ will be $U+E_8(2)$ which is the $(10,8,0)$ model.

Instead of using this procedure on the $A_1$ of $N$, we could have taken $\Gamma_{\!\!(2,2)}=U+U$, with one of the $U$'s at 
the $SU(2)$ point. 
We could then perform the rebosonization of the $J_1$ current on the left-moving $A_1$ of this $U$ so that it becomes part of $I$. 
However, now the classes of this left-moving $A_1$ is \emph{not} correlated with the already existing $A_{1}(-1)$ in $I$.  
Therefore, after this transition, $I=A_{1}(-1)+A_1+E_8(2)$ and we arrive at the $(10,10,1)$ model,  
since this $I$ is equivalent to $U(2)+A_1^8$. 
Indeed, $U(2)+A_1^7+A_1 \simeq A_{1}(-1)+E_8(2)+A_1$.

\subsubsection{$(10,10,0)$}\label{subs10_10_0}

With our prescribed method of turning an $A_1$ lattice in the normal direction to the invariant direction, 
it is not obvious how we can connect the point $(10,10,0)$ with $I=U(2)+E_8(2)$, to another point with $r=9$. 
However, we can try to connect it to a point with $r=11$. 
Let us then start from $(11,11,1)$  and take $I=U(2)+A_1+D_2^4$. 
This can be realized starting from the self dual lattice $\ktl\simeq U^3+E_8^2$, and writing each $E_8$ as $D_4^2$ with conjugacy classes given in eq. \eqref{e8d4d4triality} (observe that we have used triality in one $D_4$ to exchange $V_4$ and $Sp_4$). 
We then split each $D_4$ factor as $D_2 + D_2$, each with correlations given in eq. \eqref{d4d2d2}.
The involution acts by reflection of one $D_2$ factor in each $D_4$, exchange of the first two $U$'s, and reflection of
the right movers in the third $U$ at the $SU(2)$ point. This gives the desired $I=U(2)+A_1+D_2^4$, and
$N= U(2)+A_{1}(-1)+D_2^4$, with correlated classes implied by the defining involution. 
This result follows because in $I$ each of the reflected $D_2$ lattice vectors must be set to zero, which is an element of $Sc_2$.
Thus, from eqs. \eqref{e8d4d4triality} and \eqref{d4d2d2} we see that the invariant piece from each
$E_8$ can only come from $(Sc_4,Sc_4)$ and must be $D_2 + D_2=A_1^4$.

As discussed in subsection \ref{subsubsec_U2K}, it can be shown that $I=U(2)+A_1^9$ is in the same moduli space as $I=A_{1}(-1)+A_1^2+E_8(2)$ --- see eq. \eqref{equivUA17}.
In fact, $A_{1}(-1)+A_1^2 \simeq U(2)+A_1$. 
Thus, we conclude that  $I=U(2)+A_1^9$ is equivalent to $I=U(2)+A_1+E_8(2)$.
We now use the method of rebosonization on the $A_1$ factor of the latter and convert it to an $A_1$ in the normal direction.
Note that we are \emph{increasing} $s$ by one unit and therefore this method will work only if the shift vector $v$ of the $(11,11,1)$ model has a part in this $A_1$ equal to half the fundamental.
After this transition we attain $I=U(2)+E_8(2)$, which is indeed the $(10,10,0)$ model.

Note that we could have used the method of transition one step earlier, namely taking $I=A_{1}(-1)+A_1^2+E_8(2)$. 
If there is a part of the shift equal to half the fundamental in one of the $A_1$ factors, we could convert it into an $N$ direction 
resulting in $I =A_{1}(-1)+A_1+E_8(2)$ which is the $(10,10,1)$ model. 
Thus, even though we were not able to go from $r=9$ directly to $(10,10,0)$, we could go from $(9,9,1)$ to $(10,10,1)$ as described in subsection \ref{subs10_10_0}.
Then, inverting the process just described, we could go to the $(11,11,1)$ model, and then as explained in the previous paragraph we can transition to the $(10,10,0)$ model.

Note that while moving from $s$ to $(s-1)$ we do not have to worry about the shift vector: the $A_1$ which moves from $N$ to $I$ automatically comes with a consistent shift.
However, when we move from $s$ to $s+1$, an $A_1$ must move from $I$ to $N$, and that is possible only if the shift vector has a component along this $A_1$ which is equal to half the fundamental.

\subsubsection{$(18,4,0)$}\label{subs18_4_0}

Let us start with $\ktl\simeq U^2+\Gamma_{(9,1)}+E_8$ where $\Gamma_{(9,1)}$ is an even self-dual lattice.
We will choose a particular point in the moduli space of $\Gamma_{(9,1)}$ such that it can be expressed as $(D_9;D_1)[(V_9;V_1),(Sp_9;Sp_1)]$ with correlated classes $(Sc_9;Sc_1)$, $(V_9;V_1)$, $(Sp_9;Sp_1)$ and $(Sp'_9;Sp'_1)$.
Furthermore we decompose $D_9$ in terms of $(D_8,D_1)$.
Various conjugacy classes in this decomposition are $Sc_9=(Sc_8,Sc_1)+(V_8,V_1)$, $V_9=(Sc_8,V_1)+(V_8,Sc_1)$, $Sp_9 =(Sp_8,Sp_1)+(Sp'_8,Sp'_1)$ and $Sp'_9 =(Sp_8, Sp'_1) + (Sp'_8,Sp_1)$.
Thus $\Gamma_{(9,1)}$ at this point can be expressed in the decomposition $(D_8,D_1;D_1)$ as 
\bea\label{18_4_0}
&&\Gamma_{(9,1)}=(Sc_8,Sc_1;Sc_1)+(V_8,V_1;Sc_1)+(Sc_8,V_1;V_1)+(V_8,Sc_1;V_1)+\\
&&\qquad\;\,+\;(Sp_8,Sp_1;Sp_1)+(Sp'_8,Sp'_1,Sp_1)+(Sp_8, Sp'_1;Sp'_1)+,(Sp'_8, Sp_1;Sp'_1)\ .\nn
\eea
The involution in $\ktl$ we take to be exchange of two $U$'s and reflection along the $D_1(-1)$ and $D_1$ directions in $\Gamma_{(9,1)}$.
The contribution to $I$ and $N$ coming from the exchange of two $U$'s is $U(2)$ for each lattice.
The contribution to $I$ coming from the reflection in $\Gamma_{(9,1)}$ is obtained by setting momenta along $D_1(-1)$ and $D_1$ directions to zero and as a result, it is just scalar class of $D_8$ --- see eq. \eqref{18_4_0}. 
The normal lattice is obtained by setting the momenta along the directions of $D_8$ to zero and hence, the contribution to $N$ from this part is $(Sc_1;Sc_1)+(V_1;V_1)$.
This is indeed the $U(2)$ lattice as can be seen by setting $R=1/\sqrt2$ in \eqref{momu2k} and observing that the lattice has the from $(p_L;p_R)=(k+w,k-w)$ where the two entries are either both even or both odd, i.e. it is the lattice $(D_1;D_1)[(V_1,V_1)]$.
Altogether we find $I=U(2)+D_8+E_8$ and $N=U(2)^2$.

In the above construction we further decompose $E_8$ in terms of $(E_7,A_1)$ and reflect $A_1$ --- see eqs. \eqref{alphatwistE82E7} and \eqref{E7roots} in appendix \ref{app_lattices} for this decomposition.
The result is $I=U(2)+D_8+E_7$ and $N=U(2)^2+A_1$, i.e. the $(17,5,1)$ model.
Now the $s$-transition can be performed by moving the $A_1$ from $N$ to $I$, resulting in the above $(18, 4, 0)$ model.
Note that the $(17,5,1)$ model has $I\simeq U(2)+D_8+E_7 \simeq U+A_1^2+E_8+A_1+D_4$ and $N\simeq U(2)^2+A_1 \simeq A_1^2(-1)+A_1^2+A_1$ where the second descriptions may be obtained from our construction in section \ref{subsec_changing} with $I_3=U+A_1^2$ in table \ref{tab_u3}.
Therefore the $(18,4,0)$ model can be connected to other triples in this domain.

\subsubsection{$(14,6,0)$}\label{subs14_6_0}
We proceed as for the $(18,4,0)$ model just discussed.
In addition, we decompose $E_8$ in terms of $(D_4,D_4)$ as in \eqref{e8d4d4} and reflect one $D_4$.
The result is $I=U(2)+D_8+D_4$ and $N=U(2)^2+D_4$, which gives the $(14,6,0)$ model.
Next we want to make an $s$-transition to reach this point.
To do so, consider the above construction but do one more reflection inside the $D_8$ lattice appearing in $\Gamma_{(9,1)}$ as follows.
Decompose $D_8$ in terms of $(D_6,D_2)$ and furthermore decompose $D_2 =(A_1,A_1)$ --- see eq. \eqref{d2a1a1}.
The correlated classes are given by $Sc_8=(Sc_6,R,R)+(V_6,F,F,)$, $V_8=+(Sc_6,F,F)+(V_6,R,R)$, $Sp_8=(Sp_6,R,F)+(Sp'_6,F,R)$, $Sp'_8=(Sp'_6,R,F)+(Sp_6,F,R)$.
Now in the decomposition $(D_6,A_1,A_1,D_1;D_1)$, reflect one $A_1$ as well as $(D_1;D_1)$.
The contribution  coming from $\Gamma_{(9,1)}$ to $I$ and $N$ is thus $D_6+A_1$ and $U(2)+A_1$.
Combining everything together we find $I = U(2)+D_6+A_1+D_4$ and $N=U(2)^2+A_1+D_4$ which is the $(13,7,1)$model.
Now we make an $s$-transition by moving the $A_1$ from $N$ to $I$.
By construction above, the $I^*/I$ classes are correlated with the $N^*/N$ classes in such a way that upon moving the $A_1$ from $N$ to $I$ we get back to the $(14, 6, 0)$ model.

\section{Final remarks}
\label{sec_final}

In this work we constructed asymmetric orbifolds describing heterotic strings on $\bt^3/\bz_2$ at the
worldsheet level. In the context of M-theory/heterotic duality, these theories are dual to M-theory on K3 surfaces 
quotiented by a non-symplectic involution.
Such involutions are uniquely characterised by the three parameters $(r,a,\delta)$ of the sublattice of 
the K3 lattice which is invariant under the involution.
For a given $(r,a,\delta)$, the heterotic orbifold is additionally characterised by the shift vector $v$ constrained by level matching.
We found that for the two triples with \mbox{$(r,a,\delta)=(1,1,1)$} and $(2,2,1)$, the asymmetric orbifold is not viable due 
to the small rank of the invariant lattices, which makes it impossible to find a shift vector satisfying the level matching condition.
One possible way to remedy this problem is to compactify the theory further on a circle, i.e. move to six dimensions.
The direction along the circle is chosen to be left invariant under the involution and as such the circle is referred to as the spectator circle. One would do the analysis in six dimensions and then take the limit where the radius of the spectator circle goes to infinity to recover the seven-dimensional theory.
The rank of the invariant lattice is bigger in six dimensions because of the presence of the spectator circle.
One then asks whether it is now possible to define shift vectors that satisfy the level matching condition in the new theory.
If so, one would classify the inequivalent families of such shift vectors and analyse how they emerge in the original theory as we take the radius of the spectator circle to infinity.

We proved that all the asymmetric orbifold models are connected to each other through \mbox{$s$-transitions.}
This raised the question whether there exists a unique big moduli space where each model associated with the triple $(r,a,\delta)$ is a subspace of it. The answer is not clear because we also saw that some models contain several inequivalent shift vectors.
However, the analysis of one such model, namely the one associated with $(r,a,\delta)=(5,3,1)$, showed that the inequivalent shift 
vectors are connected by making an $s$-transition to the $(9,1,1)$ theory.
Whether the same conclusion holds for all models with inequivalent shifts remains to be studied.
It will be interesting to study the existence of inequivalent shift vectors and the connection between the asymmetric orbifolds in one lower dimension, upon compactifying the theory on an spectator circle. It could be that all models are connected, as it
occurs for the 10-dimensional supersymmetric and non-supersymmetric heterotic theories compactified on a circle
\cite{Ginsparg:1986bx, Ginsparg:1986wr}.

Although tachyons are not allowed in the untwisted sector of our perturbative  models, they generically appear in the twisted 
sectors in some regions of moduli space.
For instance, in some models tachyons occur only for values of the circle radius such that $R_{min}<R<R_{max}$.
They become massless at the endpoints and massive outside this interval.
When present, tachyons only signal that the vacuum is unstable. 
An appealing scenario is that tachyon condensation produces a stable lower dimensional vacuum,
as shown to occur in some non-supersymmetric string setups \cite{Horava:2007hg, Hellerman:2007zz, Basile:2018irz, Kaidi:2020jla, Basile:2021vxh}. It would be worthwhile to examine if such dynamics could take place in the non-supersymmetric 
heterotic orbifolds that we have studied.

We have seen that in the heterotic on $\bt^3/\bz_2$ there always exist untwisted fermions, massless or massive,
which are necessarily charged. On the M-theory or type IIA side, charged fermionic states are non-perturbative, coming
from membranes or D2-branes wrapped on various 2-cycles.
This result prompts the question whether the K3 quotient by the given Nikulin involution supports fermions or not,
i.e. whether it is a spin manifold or not.
As discussed at the end of section \ref{section_nikulin}, 
for most $(r,a,\delta)$ the answer is negative except for the two cases, $r=2$, $a=2,0$, and $\delta=0$.
Now, in M-theory or type IIA on a spin manifold one would expect fermions that are not charged, for instance
massive KK modes of the gravitino. If the quotient is spin,
absence of uncharged fermions in the heterotic could still be
explained arguing that KK modes of M-theory or type IIA in the strong coupling limit (i.e. heterotic theory)
may disappear or become infinitely massive.
Another point is that, even if the 4-dimensional  K3 quotient is not spin,
such as the Enriques case, the branes are localized on 2-cycles that do not see the whole 4-dimensional space. 
It could very well be that on these 2-cycles there exists a spin structure. Perhaps that is all that is needed to obtain 
7-dimensional or 6-dimensional fermions from these brane states. 
For quantum consistency, the involution must necessarily act with some phases on the membrane or D2-brane states, 
since it does on the heterotic states that have momentum along the $\ktl$ lattice and feel the effect of the
shifts $v$ and $w$. What are these quantum consistency conditions in M-theory or IIA theory ?
A way to examine this problem would be to study type IIA on an Enriques surface. The idea would be
to look at D2-branes wrapped on some 2-cycles of K3 and see how the involution
modifies the analysis. Consistency conditions might have to be imposed on non-trivial RR backgrounds
necessary for the heterotic/type IIA duality, as found in supersymmetric compactifications to four and six
dimensions where similar issues also arise \cite{Ferrara:1995yx, Schwarz:1995bj, Aspinwall:1995mh, deBoer:2001wca}.
We plan to come back to these matters in the future.

Finally, another possible extension of this work is to construct heterotic $\bt^3$ asymmetric orbifolds 
in which the orbifold generator includes other non-symplectic automorphisms of the $\ktl$ lattice such as those
analized e.g. in \cite{Artebani08, Artebani11, Garbagnati13}. It is also conceivable to consider higher dimensional 
tori $\bt^d$ and quotient by automorphisms of the even self-dual lattice $\Gamma_{\!\!16+d, d}$ that break 
supersymmetry.

\section*{Acknowledgments}

We are grateful to D. Allcock, E. Andr\'es,  L. da Rold, C. Mayrhofer, H. Parra De Freitas, S. Theisen and G. Thompson
for interesting comments and valuable insights. 
A.~Font and G. Aldazabal  acknowledge hospitality and support from IFT UAM-CSIC via the Centro de Excelencia Severo Ochoa 
Program under Grant SEV-2016-0597.
The work of G. Aldazabal is partially supported  by the CONICET grant PIP-11220200100981CO.
The work of B.S. Acharya is supported by a grant from the Simons Foundation (\#488569, Bobby Acharya).

\appendix

\section{Construction of $I$ and $N$ lattices for all $(r,a,\delta)$}
\label{app_lattices}
The key player in our analysis is the even self-dual lattice $\Gamma_{(19,3)}$ which is unique up to $SO(19,3)$
transformations.
We study $\bz_2$ involutions of $\Gamma_{(19,3)}$, denoted by $\theta$, that act by reflecting $s$ left-moving and 
2 right-moving directions. Such involutions have been classified by Nikulin \cite{Nikulin80, Nikulin83, Nikulin86}. Each
$\theta$ leaves invariant a lattice, denoted $I$, of rank $r$ and signature $(r-1,1)$, where $r=20-s$. 
The normal lattice in $\ktl$, denoted $N$, has rank $(2+s)$ and signature $(s,2)$. 
Both $I$ and $N$ are even sub-lattices of $\ktl$ and satisfy $I^*/I =N^*/N = (\bz_2)^a$ where $a\in\bz_{\ge0}$.
Lattices which satisfy the latter condition are called 2-elementary lattices.
All possible involutions, as well as the corresponding lattices $I$ and $N$, can be completely characterized by the
triple $(r,a,\delta)$, where the invariant $\delta$ is defined in eq. \eqref{deldef}. 
Figure \ref{figureNikulin} depicts the 75 allowed triples found by Nikulin.

In the main body of the paper we developed the asymmetric orbifold realisation of the involutions in the context of heterotic string theory. 
The orbifold construction is performed in its full generality, i.e. it is independent of the invariants $(r,a,\delta)$ of the models.
Nonetheless at various points we have discussed examples associated with specific triples in order to illustrate the ideas.

For each triple $(r,a,\delta)$, there exist special points in the moduli space of the corresponding lattices where $I$ and $N$ are given by an orthogonal direct sums of 2-elementary lattices.
The purpose of this appendix is to provide explicit actions of involutions on $\Gamma_{(19,3)}$ that yield $I$ and $N$ of this form for all the 75 triples.
The involutions are not in general realised uniquely and what we present is one possible action of $\theta$ for each triple.
In our constructions of even self dual lattices we shall use the method of {\it gluing} the ADE lattices which defines glue vector generators among the conjugacy classes of the lattices such that upon addition, they generate all other conjugacy classes.
We refer the reader to \cite[Chapter 4]{ConwaySloane} for details.

\underline{Notations}: We denote a lattice $\Lambda$ with glue vectors $G_1, G_2,\cdots G_n$ by $\Lambda[G_1, G_2,\cdots,G_n]$.
For instance, $D_n[V_n]$ has one glue vector generator being the vector conjugacy class, two conjugacy classes in total (together with the scalar class), and is an odd self-dual lattice.
The plain lattice $\Lambda$ with no glue vectors refers to the root lattice of $\Lambda$.
When there are lattices with correlated classes we quote them inside parenthesis, e.g.  $(D_8,D_2)[(V_8,V_2)]$.
For lattices with a Lorentzian signature, the left and right-moving components are placed on the left and right sides of a semicolon, for instance $(D_8;D_2)[(V_8;V_2)]$ denotes a lattice with signature $(8,2)$ and with correlated classes.
The squared norm of a vector $v$ in a lattice with purely left (right) directions is $||v||^2\ge0$ ($||v||^2\le0$).

We follow two steps to construct the invariant and normal lattices $I$ and $N$:\\
1. Start at a point in the moduli space of the even self-dual lattice $\Gamma:=\Gamma_{(19,3)}$ where it takes a convenient form.\\
2. Find an involution that leads to $I$ and $N$ with the desired values of $(r,a,\delta)$.\\
The invariant lattice is generically a sum of 2-elementary lattice components given in table \ref{2elebb}.
The value of $a$ for $I$ is the sum of those of the component lattices.
$I$ has $\delta=1$ if any of the component lattices has $\delta=1$ and otherwise it has $\delta=0$.

\begin{table}[h]
\centering
\begin{adjustbox}{width=\textwidth}
\begin{tabular}{ |c||c|c|c|c|c|c|c|c|c|}
\hline
 $\Lambda$ & U & U(2) & $A_1(-1)$ & A$_1$ & E$_7$ & E$_8$ & E$_8$(2) & D$_{4m}$ & D$_{4m+2}$  \\ 
\hline\hline
$(r,a,\delta)$ & (2,0,0) & (2,2,0) & (1,1,1) & (1,1,1) & (7,1,1) & (8,0,0) & (8,8,0) & ($4m$,2,0) & ($4m+2$,2,1) \\  
\hline
\end{tabular}
\end{adjustbox}
\caption{Even 2-elementary lattice building blocks for the construction of $\Gamma_{(r-1,1)}$ sublattices of K3 which are invariant under Nikulin's involutions.}
\label{2elebb}
\end{table}

The leftmost points in each row in figure \ref{figureNikulin} make the diagonal line $r=a$, namely, the value of the rank coincides with the multiplicity of the center. 
We will see that in order to reproduce these points, we need to include a contribution $U^2$  in the starting even self-dual lattice. 
Namely, $\Gamma= U^2+\Gamma'$ with $\Gamma'$ an even self-dual lattice of signature $(17,1)$. 
By permuting the two $U$ lattices, namely, 
\begin{equation}
 \theta|P_{1L},P_{2L};P_{1R},P_{2R}\rangle=|P_{1R},P_{2R};P_{1L},P_{2L}\rangle
 \label{u2}
\end{equation}
we obtain the $U(2)$ lattice where the length squares of the vectors are scaled by $2$ with respect to the original $U$. 
We thus have $(r,a,\delta)=(2,2,0)$ for $U(2)$, see table \ref{2elebb}.
To obtain the desired invariant lattice $I$, we may need to further act on $\Gamma'$.

Before we begin the construction, let us review different involutions in the lattice $E_8\equiv D_8[Sp_8]$ which lead to invariant sublattices useful for our constructions.
In particular, we consider involutions that leave invariant $A_1$ or $E_7$, both contributing with $a=1$ since $E_7^*/E_7=A_1^*/A_1=\mathbb{Z}_2$.
Note that in order to obtain an odd value of $a$, the lattice $I$ must contain one $E_7$ and/or an odd number of $A_1$ components.
The fundamental classes in $E_7^*$ (i.e. \bf56}) and in $A_1^*$ have only half-integer lengths.
Thus, lattices with odd values of $a$ will \emph{always} have $\delta=1$. 

Recall that the 240 roots of $E_8\equiv D_8 [Sp_8]$ (i.e. vectors with square norm equal to two) are
\be\label{D8basis}
(\underline{\pm1, \pm 1, 0, 0, 0, 0, 0, 0}), \quad
(\underbrace{\pm \tfrac12, \pm \tfrac12, \pm \tfrac12, \pm \tfrac12, \pm \tfrac12, \pm \tfrac12, \pm \tfrac12, \pm \tfrac12}_{\text{even number of +'s}})\ ,
\ee
where underlining means permutations. 
Under the involution
\begin{eqnarray}
\theta|P_{1},\cdots,P_{6},P_{7},P_{8} \rangle&=&|P_{1},\cdots,P_{6},P_{8},P_{7} \rangle
\label{alphatwistE82E7}
\end{eqnarray}
which permutes the last two entries (and corresponds to $s=1$), the invariant lattice corresponds to the $E_7$ root lattice with the 126 roots given by
\be\label{E7roots}
(\underline{\pm1, \pm 1, 0, 0, 0, 0}, 0, 0)\ , \quad \pm(0, 0, 0, 0, 0, 0, 1,1)\ ,\quad
\pm(\underbrace{\pm \tfrac12, \pm \tfrac12, \pm \tfrac12, \pm \tfrac12, \pm \tfrac12, \pm \tfrac12}_{\text{even number of +'s}}, \tfrac12, \tfrac12)\ .
\ee

It is also straightforward to identify the involutions that lead to $I=D_{6-m}+A_1$, $m=\{0,2,4,6\}$.
For instance, permute two momentum entries in $D_8 [Sp_8]$ and reflect the rest (hence, $s=1+6=7$):
\begin{eqnarray}
\theta|P_{1},P_{2},P_{3},\cdots,P_{8}\rangle&=&|P_{2},P_1,-P_3,\dots,-P_{8} \rangle\ .
 \label{alphatwistA1}
\end{eqnarray}
This involution leaves the states $|1,1,0,\dots,0\rangle$ and $|-1,-1,0,\dots,0\rangle$ invariant which correspond to the roots of the $A_1$ lattice. 
Likewise, 
\begin{eqnarray}
\theta|P_{1},P_2,P_3,P_4,P_5,\cdots,P_{8}\rangle&=&|P_{2},P_1,P_3,P_4,-P_5,\cdots,-P_{8} \rangle
 \label{alphatwistA1D2}
\end{eqnarray}
(with $s=1+4=5$) leads to $I= A_1+ D_2$.
The invariant lattices $I= A_1+ D_4$ and $I= A_1+D_6$ are obtained in a similar way.
We conclude that by permuting the first two entries and reflecting an $m$ number of the remaining $6$ momentum entries we find the invariant lattice $I=A_1+D_{6-m}$ with $\delta =1$, $a=3$ for $m\ne6$ and $a=1$ for $m=6$. 

In the remainder of this appendix we will provide explicit constructions of the invariant lattices $I$ (and correspondingly the normal lattices $N$) for each point in figure \ref{figureNikulin}.
For some constructions the invariant lattices are the same as the entries in table \ref{tab_IN}, and for some constructions they are isomorphic to the corresponding entries.

\subsection{Even $a$}

Let us first consider lattices with even values of $a$.
An even self-dual lattices with signature $(\gamma_+,\gamma_-)$ and $\gamma_+-\gamma_-\equiv0\mod8$ is unique up to lattice isometries \cite{Milnor58}.
Thus, $\Gamma\cong U^3+E_8^2$.
We shall start with an even self-dual lattice of interest $\Gamma$ and define the specific involutions which act on $\Gamma$ to yield the desired lattices $I$ and $N$.
For $a=2$ we perform involutions to arrive at an invariant lattice of the form $I=U+D_{2m}$. 
Since $D_{2m}^*/D_{2m}=\mathbb{Z}_2^2$, we have $a=2$ --- see table \ref{2elebb}.
Thus the associated invariant lattices correspond to the triples $(2m+2,2,0)$ for $m=2k$ and $(2m+2,2,1)$ for $m=2k+1$, $k\in\bz$. 
For $a=4$, we take involutions which yield $I=U+D_{2m_1}+D_{2m_2}$.
This gives the triples $(2+2m_1+2m_2,4,0)$ for $m_1=2k_1$, $m_2=2k_2$, and $(2+2m_1+2m_2,4,1)$ for $m_1=2k_1+1$, $m_2=2k_2+1$ where $k_1,k_2\in\bz$.
We shall continue with such a decomposition to obtain all values of $a=2,4,6,8,10$.

There are also other possibilities at hand, e.g taking involutions such that the invariant lattice contains the $E_8(2)$ lattice with the triple $(8,8,0)$ and/or the $U(2)$ lattice with $(2,2,0)$.
Moreover, an even value of $a$ with $\delta=1$ can also be obtained by taking an even number of $A_1$ components and/or 2 $E_7$ components --- {\it cf.} eqs. \eqref{alphatwistE82E7} and \eqref{alphatwistA1}.

\subsubsection{$a=0$}\label{sec:a0}
Start with the even self-dual lattice $\Gamma=U+(D_{2};D_{2})[(Sp_2;Sp_2),(V_2;V_2)] +  D_8[Sp_8] + D_8[Sp_8]$.
The lattice $D_8[Sp_8]$ is even self-dual and is equal to the $E_8$ lattice.
Consider the involution changing $s_0$ signs in the left $D_2$, $s_1$ signs in the first $E_8$, $s_2$ sings in the second $E_8$ and the signs of the two right components in $D_2$.
The total value of $s$ is $s=s_0+s_1+s_2$.
For instance, we choose $s_0=2$,  $s_1=8$, $s_2=8$ with the involution
\begin{equation}
 \theta|P\rangle=|-P_{1L},-P_{2L}, -Q_{1L},\dots -Q_{8L},-Q_{1L}',\dots, -Q_{8L}';-P_{1R},-P_{2R}  \rangle
 \label{alpha200}
 \end{equation}
where $P_L$, $Q_L$, $Q'_L$ and $P_R$ correspond to the left $D_2$, the first $E_8$, the second $E_8$ and the right $D_2$, respectively.
This involution leads to $I=U$, $(r,a,\delta)=(2,0,0)$.
Similarly, taking other values of $s_0$, $s_1$, and $s_2$ we find lattices of different ranks.
All lattices with $a=0$ in figure \ref{figureNikulin} are obtained:\\ 
$s_0=2$, $s_1=0$, $s_2=0$ gives $s=2$, $I=U+D_{8}[Sp]+D_{8}[Sp]$ with $(r,a,\delta)=(18,0,0)$.\\
$s_0=2$, $s_1=8$, $s_2=0$ gives $s=2$, $I=U+D_{8}[Sp]$ with $(r,a,\delta)=(10,0,0)$.\\
$s_0=2$, $s_1=8$, $s_2=8$ gives $s=2$, $I= U$ with $(r,a,\delta)=(2,0,0)$.

\subsubsection{$a=2$}\label{sec:a2}
Start with the even self dual lattice $\Gamma= U + (D_{18};D_{2})[ (Sp_{18};Sp_{2}),(V_{18};V_{2})]$ with conjugacy classes $(Sc_{18},Sc_{2})$, $(V_{18},V_{2})$, $(Sp_{18},Sp_{2})$ and $(Sp'_{18},Sp'_{2})$.
Besides changing the signs of 2 right components we change $s$ signs in left weights, where $s$ is even, to obtain $I=U+ D_{18-s}$.
We find that all $a=2$ points are reproduced in this way except for $(r,a,\delta)=(2,2,0)$, $(2,2,1)$, $(10,2,1)$ and $(18,2,1)$.

Now consider $\Gamma= U+(D_{2};D_{2})[(Sp;Sp),(V;V)]+ D_8[Sp_8]+ D_8[Sp_8]$. 
Reflecting first the two right movers leaves us with $(D_2,0)$ and further reflecting one left entry in $D_2$ gives $(D_{2};D_{2})[(Sp,Sp),(V,V)]\rightarrow A_1$.
Note that the corresponding component in the normal lattice is of the form $(A_{1};D_{2})[(R_1,Sp'_2)]$.
Next, performing the involution \eqref{alphatwistE82E7} on one of the $E_8$ lattices yields $E_7$ and we find  $I=U+A_1+E_7+E_8$ with  $(r,a,\delta)=(18,2,1)$.
Reflecting in addition all the directions along the $E_8$ lattice gives $I=U+A_1+E_7$ with $(r,a,\delta)=(10,2,1)$.

To obtain the triple $(2,2,1)$, consider $\Gamma=  (A_1;A_1) [(F_1;F_1)]^2 + U+E_8^2$.
Reflect the left $A_1$ in one $(A_1;A_1) [(F_1;F_1)]$ and the right $A_1$ in the other one.
Further reflect along all the 18 directions in $U+E_8^2$.
This yields $I=(A_1;A_1)$ with $(r,a,\delta)=(2,2,1)$.
 
Finally, consider $\Gamma =U^3+E_8^2$.
Performing the involution \eqref{u2} on two $U$ lattices which leads to $U(2)$, inverting the left and right directions in the remaining $U$ as well as the $16$ left directions along $E_8^2$ we find: $I=U(2)$, $(r,a,\delta)=(2,2,0)$, $s=18$.

\subsubsection{$a=4$}\label{subsection_a4}
Consider $\Gamma=U+(D_{m};D_{2})[(Sp_m;Sp_2),(V_m;V_2)] +G_{18-m}$, $m=2+8k=\{2,10,18\}$.
The value of $m=18$ leads us to the construction in subsection \ref{sec:a2}.
Here we consider $m=\{2,10\}$. 
The lattice $G_{18-m}$, i.e. $G_{16}$ and $G_{8}$, must be even self dual. 
For $m=2$ we can choose $G_{16}=D_{16}[Sp_{16}]$ or $G_{16}=D_{8}[Sp_8]+ D_{8}[Sp_8]$ and for $m=8$ we have $G_{8}=D_{8}[Sp_8]$.
The idea is to invert $s$ left weight components distributed among $D_{m}$ and $G_{18-m}$ where $s$ is an even integer.

Let us first consider $m=2$ and choose the even self dual lattice $G_{16}=\Gamma_{16}=D_{16}[Sp]$. 
We invert two right entries in $D_2$ and $s$ signs in $G_{16}$ ($s\ne 16$).
This gives $I=U+D_2+D_{16-s}$ with $(r,a,\delta)=(20-s,4,1)$. 
\footnote{Note that if we were to invert the signs of $s$ entries as 2 in $D_2$ and 
$s-2$ in $G_{16}$  then $I= D_{16-(s-2)}+U$, $(r,a,\delta)=(20-s,2, 0),(20-s,2, 
1) $ for $s=2+8k$, $s=4k$ respectively and we recover the $a=2$ result above 
.}
We thus obtain the triples $(18,4,1)$, $(16,4,1)$, $(14,4,1)$, $(12,4,1)$, $(10,4,1)$, $(8,4,1)$ and $(6,4,1)$.

Next we consider $G_{16}=D_8 [Sp]+D_8[Sp]$.
Besides the reflection on two right directions, we invert $s_0$ signs in the left $D_2$ lattice, $s_1$ signs in the first $D_8$ and $s_2$ signs in the second $D_8$ lattices. where $s_0=2$ and $s_1$ and $s_2$ are even integers.
We then obtain $I=U+D_{8-s_1}+D_{8-s_2}$ with $s=2+s_1+s_2$, $s_1,s_2\ne0$ and $r=18-s$.
This provides an alternative way of constructing some of the lattices with $(r,4,1)$ triples with $r>4$.
Notice that for $s_1=s_2=4$ the length squares of all weight vectors in $D_{8-s_1=4}$ and $D_{8-s_2=4}$ are integers and so $\delta=0$.
Thus in this case we obtain the triple $(10,4,0)$.

The triples $(4,4,1)$, $(6,4,0)$ and $(14,4,0)$ cannot be constructed in this way.
However, it appears that we can construct the invariant lattices for these models using the $U(2)$ lattice.
The special property of the $(6,4,0)$ triple is that it is the unique point with $a<r$ that necessarily contains the $U(2)$ lattice in its constituent components \cite{Nikulin83}.
Start with $\Gamma=U^3+ D_8[Sp_8]+D_8[Sp_8]$, permute two $U$ lattices to get $U(2)$, reflect along the left and right directions of the third $U$ lattice, change the sings of $s_1$ momentum entries in the first $D_{8}$ and $s_2$ entries in the second $D_8$.
We obtain:\\
$s_1=6$, $s_2=8 $ gives $I= U(2)+D_{2}$ with $(r,a,\delta)=(4,4,1)$.\\
$s_1=4$, $s_2=8 $ gives $I= U(2)+D_{4}$ with $(r,a,\delta)=(6,4,0)$.\\
$s_1=4$, $s_2=0$ gives $I= U(2)+D_{4} + E_8 $ with $(r,a,\delta)=(14,4,0)$.

Finally, we consider the triple $(18,4,0)$.
This case is slightly more subtle because the normal lattice has signature $(2,2)$ with $(r,a,\delta)=(4,4,0)$ and the only possible way to realise this is to have a lattice isomorphic to $N=U(2)^2$.
This model is constructed in section \ref{subs18_4_0}.

\subsubsection{$a=6$}\label{subsection_a6}
Consider the even self dual lattice $\Gamma=U+(D_{2};D_{2})[(Sp_2;Sp_2),(V_2;V_2)] + D_{8}[Sp_8]+ D_8[Sp_8]$.
Following our previous constructions, we consider $s=s_1+s_2$ inversions distributed among the two $D_8$ components where $s_1,s_2\ne0$.
This gives $I=U+D_2+D_{8-s_1}+ D_{8-s_2}$ with $a=6$ and we obtain the triples $(16,6,1)$, $(14,6,1)$, $(12,6,1)$, $(10,6,1)$ and $(8,6,1)$.

To construct the triple $(6,6,1)$ consider the even self dual lattice $\Gamma= U^2 + (A_1;A_1) [(F_1;F_1)]+D_{8}[Sp_8]+D_{8}[Sp_8]$.
Perform an involution that leads to $U(2)$, inverts the signs on both $A_1$ directions, and inverts 6 directions in each $D_{8}$ lattice.
This gives $s=1+1+6+6=14$ and we obtain  $I=U(2)+D_2+D_2$ with $(r,a,\delta)=(6,6,1)$.
All triples with $a=6$ points are constructed.

The two remaining points to construct are the $a=6$ triples with $\delta=0$: $(10,6,0)$ and $(14,6,0)$.
As for the former, the invariant lattice may be constructed by starting from $\Gamma=U^3+D_{8}[Sp_8]+ D_8[Sp_8]$, exchanging two $U$ lattices, and reflecting the third $U$ as well as four entries in each $E_8$.
This yields $I=U(2)+D_4+D_4$.

The construction of the triple $(14,6,0)$ is similar to the $(18,4,0)$ model and is discussed in section \ref{subs14_6_0}.

\subsubsection{$a=8$}\label{subsection_a8}
Start with $\Gamma=U+(D_{2};D_{2})[(Sp_2;Sp_2),(V_2,V_2)]+ (D_{4},D_4) [(Sp_4,V_4),(V_4,Sp_4)]+ D_8[Sp_8]$.
Consider the inversion of $s_0$ signs in $D_2$, $s_1$ in the first $D_4$, $s_2$ signs in the second $D_4$ and  $s_3$ signs in the $D_8$ lattice.
This gives $s=s_0+s_1+s_2+s_3$.
To obtain $a=8$, we set $s_0=0$, $s_1=s_2=2$ and $s_3\ne0$.
This gives $I=U+D_2+D_{2}+D_{2}+D_{8-s_3}$.
The triples $(14,8,1)$, $(12,8,1)$ and $(10,8,1)$ are constructed in this way.

Next consider the triples $(10,8,0)$ and $(8,8,1)$.
The former can be obtained from $I=U+E_8(2)$.
To construct the latter, consider $\Gamma=U^3+D_8[Sp]+  (D_{4},D_4)[Sp_4,V_4]$.
Obtain $U(2)$ under the involution as before and switch $s_1=6$ signs in $D_8$ as well as $s_2=s_3=2$ signs in the two $D_{4}$ lattices.
We find $I=U(2)+D_2+D_2+D_2$ with $(r,a,\delta)=(8,8,1)$.

\subsubsection{$a=10$}\label{subsection_a10}
Start with $\Gamma=U+(D_{2};D_{2})[(Sp_2;Sp_2),(V_2;V_2)]+(D_{4},D_4)[(Sp_4,V_4),(V_4,Sp_4)]^2$. 
Consider as before reflections $s_i$ acting on each factor with $s_0=0$ along the left $D_2$.
Choosing $s_1=s_2=s_3=s_4=2$ we find we find $I=U+D_2+D_{2}+D_{2}+D_{2}+D_{2}$ with $(r,a,\delta)=(12,10,1)$.

There remains the two points $ (10,10,0)$ and $(10,10,1)$ to be constructed.
Consider $\Gamma=  U^3  +  D_{8}[Sp_8]+D_{8}[Sp_8] $.
First, exchange two $U$'s to obtain $U(2)$.
Next, reflect along the directions of the remaining $U$.
Finally, permute the two $E_8$ lattices.
We obtain $I= U(2)+E_8(2)$ with $(r,a,\delta)=(10,10, 0)$.

To obtain the $(10,10,1)$ triple, start from $\Gamma=  (A_1;A_1) [(F_1;F_1)]^3  +  D_{8}[Sp_8]+D_{8}[Sp_8]$.
Reflect 2 right momenta in the first two $(A_1;A_1)$ factors.
Next reflect the left momenta in the second and third $(A_1;A_1)$ factors.
Finally permute the two $E_8$ momenta.
This gives $I=(A_1;A_1)+E_8(2)$ with $(r,a,\delta)=(10,10, 1)$.

\subsection{Odd $a$}
As discussed earlier, for odd values of $a$ the invariant lattice must contain an $E_7$ component and/or an odd number of $A_1$ components, see the paragraph below eq. (\ref{u2}).
We shall now construct the invariant lattices $I$ for odd values of $a$.

\subsubsection{$a=1$}\label{subsection_a1}
We start with the lattice $\Gamma=U+(D_{2};D_{2})[(Sp_2;Sp_2),(V_2;V_2)]+ D_8[Sp_8]+ D_8[Sp_8]$.
We next decompose the left $D_2$ lattice as $(A_1,A'_1)$ with correlated classes $Sc_2=(R_1,R_1)$, $V_2=(F_1,F_1)$, $Sp_2=(F_1,R_1)$ and $Sp'_2=(R_1,F_1)$.
Perform an involution that reflects the two right directions along $D_2$ as well as one left direction along $A_1'$.
This gives $I=U+A_1+E_8+E_8$ with $(r,a,\delta)=(19,1,1)$.
If in addition we reflect eight directions of one of the $E_8$ lattices, we obtain the triple $(11,1,1)$.
Similarly, reflecting eight directions of both $E_8$'s gives the triple $(3,1,1)$.
 
Start with the same even self dual lattice, reflect the two left and the two right directions along $(D_2;D_2)$, apply the involution on $E_8$ which leads to $E_7$ --- see eq. (\ref{alphatwistE82E7}).
We obtain $I=U+E_8+E_7$ with $(r,a,\delta)=(17,1,1)$.
Reflect in addition along the directions of the $E_8$ lattice to obtain $I=U+E_7$ with $(r,a,\delta)=(9,1,1)$.

In order to obtain the triple $(1,1,1)$ consider $\Gamma=  [(A_1;A_1) (F_1;F_1)] +U^2+E_8^2$.
Reflect the left $A_1$ direction, as well as all directions along $U^2$ and $E_8^2$.
We obtain $I=A_1(-1)$ with $(r,a,\delta)=(1,1,1)$.

\subsubsection{$a=3$}\label{subsection_a3}
Start with $\Gamma=U+(D_{10};D_{2})[(Sp_{10};Sp_2),(V_{10};V_2)] +D_{8}[Sp_8]$.
Reflect $s$ directions along $D_{10}$ where $s$ is even and $s_1\ne 10$.
Perform the involution \eqref{alphatwistE82E7} to obtain $ I=U+D_{10-s_1}+E_7$.
This yields the triples $(19,3,1)$, $(17,3,1)$, $(15,3,1)$, $(13,3,1)$ and $(11,3,1)$.

Alternatively, use the $D_8 [Sp_8]$ base in eq. \eqref{D8basis} and apply the involution \eqref{alphatwistA1} to obtain $I=U+D_{10-s_1}+A_1$.
We obtain the triples $(5,3,1)$, $(7,3,1)$, $(9,3,1)$, $(11,3,1)$ and $(13,3,1)$ where the last two were already constructed in the above paragraph.

Finally, to construct the $(3,3,1)$ triple start with $\Gamma=U^2+(A_1;A_1)(F_1;F_1)+D_8[Sp_8]+ D_8[Sp_8]$.
Permute the two $U$'s to obtain $U(2)$, reflect the left and the right $A_1$ components, apply the involution \eqref{alphatwistA1} on the first $D_8[Sp_8]$ lattice and reflect all eight entries in the second $D_8[Sp_8]$.
We find $I=U(2)+A_1$ with $(r,a,\delta)=(3,3,1)$.

\subsubsection{$a=5$}\label{subsection_a5}
Consider $\Gamma=U+(D_{2};D_{2})[(Sp_2;Sp_2),(V_2,V_2)] +D_8[Sp_8]+D_8[Sp_8]$.
Following the above constructions, perform involutions to obtain $I=U+D_2+D_{8-s_1}+A_1$ and $I=U+D_2+D_{8-s_1}+E_7$, where $s$ is even and $0<s<8$.
For $I=U+D_2+D_{8-s_1}+A_1$ we find the triples $(11,5,1)$, $(9,5,1)$ and $(7,5,1)$ and for $I=U+D_2+D_{8-s_1}+E_7$ we obtain $(17,5,1)$, $(15,5,1)$ and $(13,5,1)$.

To obtain the only remaining triple $(5,5,1)$ start from $\Gamma= U^2+ (A_1;A_1)(F_1;F_1)+ D_8[Sp_8]+ D_8[Sp_8]$.
Perform the permutation leading to $U(2)$, reflect the right $A_1$ direction and perform involutions on each $D_8[Sp]$ to obtain $A_1^2$.
We find $I=U(2)+A_1^3 $ with $(r,a,\delta)=(5,5,1)$.

\subsubsection{$a=7$}\label{subsection_a7}
Consider $\Gamma=U+(D_{2};D_{2})[(Sp_2;Sp_2),(V_2;V_2)] +D_8[Sp_8]+D_8[Sp_8]$.
Reflect the two right directions along $D_2$, perform $s_1$ reflections in the first $D_{8}[Sp_8]$ lattice where $s_1$ is even and $0<s_1<8$.
Next apply the involution of the type \eqref{alphatwistA1D2} to the second $D_8[Sp_8]$ to select the invariant lattice $D_{6-s_2}+A_1$, where $s_2$ is even and $0<s_2<6$.
We obtain $I=U+D_2+D_{8-s_1}+D_{6-s_2}+A_1$ corresponding to the triples $(15,7,1)$, $(13,7,1)$, $(11,7,1)$ and $(9,7,1)$.

To obtain the $(7,7,1)$ triple start with $\Gamma= U^2+(A_1;A_1)(F_1;F_1)+D_8[Sp_8]+D_8[Sp_8]$.
Permute the two $U$ components to obtain the $U(2)$ lattice, reflect the right $A_1$ direction as well as six directions along each $D_8[Sp_8]$.
We find  $I=U(2)+D_2+  D_2+ A_1 $ with $(r,a,\delta)=(7,7,1)$.

\subsubsection{$a=9$}\label{subsection_a9}
Start with $\Gamma=U+(D_{2};D_{2})[(Sp_2;Sp_2),(V_2;V_2)]+(D_{4},D_4) [(Sp_4,V_4),(V_4,Sp_4)]+D_8[Sp_8]$.
Following our approach in subsection \ref{subsection_a8}, we reflect the two right directions along $D_2$, $s_1$ left directions in the first the $D_4$, $s_2$ left directions in the second $D_4$, and perform an involution on the $D_8[Sp_8]$ component which leaves $D_{6-s_3}+ A_1$ invariant.
Here $s_1,s_2,s_3>0$ and even.
We find $I=U+D_2+D_{4-s_1}+D_{4-s_2}+D_{6-s_3}+A_1$ which yields the triples $(13,9,1)$ and $(11,9,1)$.

Next consider $\Gamma= U^2+(A_1;A_1)[(F_1;F_1)]+(D_{4},D_4)[(Sp_4,V_4),(V_4,Sp_4)]+ D_8[Sp_8]$.
Perform the involution on $U^2$ to obtain the $U(2)$ lattice, reflect the right component $A_1$, and reflect $s_1=2$, $s_2=2$ and $s_3=6$ directions along $D_4$, $D_4$ and $D_8$.
This yields $I=U(2)+A_1+D_2+D_2+D_2$ with $(r,a,\delta)=(9,9,1)$.

\subsubsection{$a=11$}\label{subsection_a11}
Start with $\Gamma= U^2+ (A_1;A_1)[(F_1;F_1)]+(D_{4},D_4)[(Sp_4,V_4),(V_4,Sp_4)]+ (D_{4},D_4)[(Sp_4,V_4),(V_4,Sp_4)]$.
Perform the involution on the $U$ lattices to obtain $U(2)$, reflect the right direction along $A_1$, reflect 2 entries in each $D_4$ component.
We find $I= U(2)+A_1+D_2+D_{2}+D_{2}+D_{2}$ with $ (r,a,\delta)=(11,11,1)$.

This concludes the construction of the $I$ and $N$ lattices for all 75 points in figure (\ref{figureNikulin}).

\small\baselineskip=.87\baselineskip
\let\bbb\bibitem\def\bibitem{\itemsep1.5pt\bbb}

\bibliographystyle{JHEP}
\bibliography{Nik_het}

\end{document}